\documentclass[symmetry,article,accept,moreauthors,pdftex]{Definitions/mdpi}

\usepackage{combelow}

\usepackage{amsmath}
\usepackage{amssymb}
\usepackage{graphicx}
\usepackage{bm}
\usepackage{braket}
\usepackage{enumitem}
\usepackage{tensor}
\usepackage{color}
\usepackage{slashed}
\usepackage{epstopdf}
\usepackage{mathrsfs}  
\usepackage{mathtools}

\numberwithin{equation}{section} 

\definecolor{victor}{rgb}{0,0.5,0.75}
\definecolor{EW}{rgb}{0.4,0,1}

\def\hatt{{\hat{t}}}
\def\hatr{{\hat{r}}}
\def\htheta{{\hat{\theta}}}
\def\hati{{\hat{\imath}}}
\def\hatj{{\hat{\jmath}}}
\def\hatk{{\hat{k}}}
\def\hatl{{\hat{\ell}}}
\def\hatx{{\hat{x}}}
\def\haty{{\hat{y}}}
\def\hatz{{\hat{z}}}

\def\hvarphi{{\hat{\varphi}}}
\def\halpha{{\hat{\alpha}}}
\def\hbeta{{\hat{\beta}}}
\def\hgamma{{\hat{\gamma}}}
\def\hsigma{{\hat{\sigma}}}

\def\hzeta{{\hat{\zeta}}}

\def\hlambda{{\hat{\lambda}}}

\def\hl#1{#1}

\def\wt{{\overline{t}}}
\def\wr{{\overline{r}}}
\def\ws{{\overline{s}}}
\def\wO{{\overline{\Omega}}}
\def\tO{{\widetilde{\Omega}}}
\def\wrho{{\overline{\rho}}}
\def\wz{{\overline{z}}}

\def\vx{\bm{x}}
\def\vgamma{\bm{\gamma}}

\renewcommand{\linenumbers}{}

\extrafloats{10}

\firstpage{1} 
\pubvolume{1}
\issuenum{1}
\articlenumber{0}
\pubyear{2021}
\copyrightyear{2021}
\externaleditor{\hl{Academic Editor: Firstname Lastname}} 
\datereceived{} 
\dateaccepted{} 
\datepublished{} 
\hreflink{https://doi.org/} 

\pdfoutput=1

\Title{Vortical Effects for Free Fermions on Anti-De Sitter Space-Time}
\TitleCitation{Vortical Effects for Free Fermions on Anti-De Sitter Space-Time}

\Author{\hl{Victor E.~Ambru}\cb{s} $^{1}$\orcidA{} and Elizabeth Winstanley $^{2,}$\hl{*}\orcidB{}}  

\AuthorNames{Victor E.~Ambrus and Elizabeth Winstanley}
\AuthorCitation{Ambru\cb{s}, V.E.; Winstanley, E.}
\address{%
$^{1}$ \quad Department of Physics, West University of Timi\cb{s}oara, 
Bd.~Vasile P\^arvan 4, 300223 Timi\cb{s}oara, Romania;  Victor.Ambrus@e-uvt.ro\\
$^{2}$ \quad Consortium for Fundamental Physics, School of Mathematics and Statistics, The University of Sheffield, Hicks Building, Hounsfield Road, Sheffield S3 7RH, UK}

\corres{\hl{Correspondence:  E.Winstanley@sheffield.ac.uk}}

\abstract{Here, we study a quantum fermion field in rigid rotation at finite temperature 
on anti-de Sitter space. We assume that the rotation rate $\Omega$ is
smaller than the inverse radius of curvature $\ell ^{-1}$,  so that there is no speed of light 
surface and the static (maximally-symmetric) and 
rotating vacua coincide. 
This assumption enables us to follow a geometric approach employing a closed-form expression for the vacuum two-point function, which 
can then be used to compute thermal expectation values (t.e.v.s). 
In the high temperature regime, we find a perfect analogy with known results on Minkowski space-time, uncovering curvature effects in the form of extra terms involving the Ricci scalar $R$. The axial vortical effect is validated and the axial flux through two-dimensional slices is found to escape to infinity for massless fermions, while for massive fermions, it is completely converted into the pseudoscalar density $-i {\bar \psi} \gamma^5 \psi$. Finally, we discuss volumetric properties such as the total scalar condensate and the total energy within the space-time and show that they diverge as $[1 - \ell^2 \Omega^2]^{-1}$ in the limit $\Omega \rightarrow \ell ^{-1}$.
}


\begin{document}

\section{Introduction}
\label{sec:intro}

One of the deepest and most fundamental results of quantum field theory in curved space-time is the Unruh effect~\cite{Fulling:1972md,Davies:1974th,Unruh:1976db,Crispino:2007eb,Fulling:2014wzx}.
Let us describe this effect in terms of the canonical quantization of a free quantum field.
In canonical quantization, the field is decomposed with respect to an orthonormal basis of mode solutions of the classical field equation. 
On a static space-time possessing a global time-like Killing vector, it is natural to employ mode solutions which have a definite frequency, in other words to perform a Fourier transform of the field with respect to the time coordinate defined by the Killing vector.
The basis of mode solutions is split into ``positive'' and ``negative'' frequency modes. 
Upon quantization, the coefficients of the modes in the field expansion are promoted to operators, the coefficients of positive frequency modes corresponding to annihilation operators and the coefficients of negative frequency modes corresponding to creation operators. 
The natural vacuum state is then that state which is annihilated by all the annihilation operators. 

On global Minkowski space-time, an inertial observer will define a vacuum state via the above process, using their proper time as the time coordinate.
The vacuum thus constructed is the Minkowski vacuum, 
being the same for all inertial observers. 
Consider instead an observer uniformly 
accelerating in Minkowski space-time.
Such an observer can follow the standard canonical quantization procedure, using their proper time as the time coordinate, and hence construct a vacuum state.
The crux of the Unruh effect is that the vacuum state constructed by the accelerating observer (which we call the Rindler vacuum) is {\em {not}} the global Minkowski vacuum. 
Since the inertial and accelerating observers do not agree on the definition of the vacuum state, they also do not agree on the notion of a~``particle''.

The properties of the Rindler and Minkowski vacua are the same for both bosonic and fermionic fields~\cite{Candelas:1978gg}.
Our interest in this paper is free quantum fields, but the effect can be proven to also apply to general interacting quantum fields~\cite{Bisognano:1975ih,Bisognano:1976za,Sewell:1982zz}.
In particular, the Minkowski vacuum contains a thermal distribution of Rindler particles, with the temperature proportional to the proper acceleration of the linearly accelerating observer. 
While the Unruh effect occurs in flat Minkowski space-time, there is a compelling analogy with quantum states on static black hole space-times~\cite{Sewell:1982zz,Kay:1985zs,Kay:1988mu}.
In this analogy, the role of a linearly accelerating observer is played by a static observer outside the black hole event horizon; the corresponding vacuum state is the Boulware state~\cite{Boulware:1974dm,Boulware:1975pe}.
In a static black hole space-time, an inertial observer is freely-falling and the corresponding vacuum state is the Hartle-Hawking state~\cite{Hartle:1976tp}. 
The Hartle-Hawking state contains a thermal distribution of particles relative to the Boulware state, in direct analogy to the properties of the Rindler and Minkowski vacua.
In the black hole case, the thermal particles are produced due to the Hawking effect~\cite{Hawking:1974rv,Hawking:1974sw}.

The Unruh effect as described above is associated with uniform linear acceleration in flat space-time. 
Alternatively, one may consider a rotating observer, having uniform circular motion about an axis in Minkowski space-time.
Such an observer is accelerating, but it is the direction of their velocity rather than its magnitude which is changing. 
A~natural question then arises: is the vacuum state defined by a rotating observer the same as the global Minkowski vacuum?  Answering this question is surprisingly subtle and depends on the nature of the quantum field under consideration.

For the simplest type of free quantum field, a quantum scalar field, the nonrotating and rotating vacua are identical~\cite{Denardo:1978dj,Letaw:1979wy}, but for a quantum fermion field, there are two inequivalent quantizations.
The usual nonrotating vacuum state can be constructed following the standard quantization procedure~\cite{Vilenkin:1980zv}. 
Employing an alternative quantization procedure~\cite{Iyer:1982ah} leads to a rotating vacuum state (which is not the same as the nonrotating vacuum state~\cite{Ambrus:2014uqa}).
This difference in the behaviour of bosonic and fermionic fields arises in the canonical quantization procedure leading to the definition of vacuum states.
For a boson field, the split of field modes into ``positive'' and ``negative'' frequencies is constrained by the fact that, in order to obtain a consistent quantization, positive frequency modes must have a positive ``norm'', while negative frequency modes must have a negative ``norm''. 
In contrast, for a fermion field, {\em {\hl{all}}} 
 field modes have positive norm, which means that there is greater freedom to define positive and negative frequency modes.
Thus, as seen above, it is possible to define quantum states for a fermion field which have no analogue for a boson~field. 

We have already discussed how the Unruh effect for linear acceleration yields insights into the definition and properties of quantum states on nonrotating black hole space-times. 
To explore the analogy between rotating states in Minkowski space-time and quantum states on rotating black holes, one first needs to consider
rigidly-rotating thermal states in flat space-time.
Considering first a quantum scalar field, rigidly-rotating thermal states are ill-defined everywhere in unbounded Minkowski space-time~\cite{Vilenkin:1980zv,Duffy:2002ss}.
For a quantum fermion field, rigidly-rotating thermal states can be defined on the unbounded space-time~\cite{Ambrus:2014uqa}.
Expectation values of observables in these states are regular on the axis of rotation and everywhere inside a cylindrical surface, known as the speed-of-light (SLS) surface.
The SLS is defined as the surface on which an observer travelling with uniform angular speed in a circle centred on the axis of rotation must travel with the speed of light.
For rigidly-rotating thermal states of fermions, expectation values diverge as the SLS is approached~\cite{Ambrus:2014uqa}.

As pointed out by Vilenkin~\cite{Vilenkin:1980zv}, the spin-orbit coupling inherent at the level of the Dirac equation leads to a nonvanishing flux of neutrinos (particles with left-handed chirality) directed along the macroscopic vorticity of the fermionic fluid. This connection between chirality flux and vorticity is now understood as the chiral vortical effect~\cite{Kharzeev:2015znc}, which states that in a system with nonvanishing local vorticity $\bm{\omega}$, an axial (chirality) charge current is generated via the constitutive relation $\bm{J}_A = \sigma^\omega_A \bm{\omega}$, where $\sigma^\omega_A$ is the axial vortical conductivity. This constitutive relation, believed to stem from the chiral and gravitational anomalies~\cite{Landsteiner:2011iq}, forms the basis of dissipationless anomalous transport and is purely quantum in nature~\cite{Son:2009tf}. 

Regular rigidly-rotating thermal states exist for both boson and fermion fields if, instead of considering the whole of Minkowski space-time, a space-time region inside a cylindrical boundary is studied. 
If the cylindrical boundary lies within the SLS, the rotating and nonrotating vacua coincide for both scalar and fermion fields 
\cite{Duffy:2002ss,Ambrus:2015lfr}.
In this case rotating and nonrotating thermal states are distinct, but expectation values of observables in both states are regular everywhere inside and on the boundary~\cite{Duffy:2002ss,Ambrus:2015lfr}. 
Similar conclusions hold for a spherical boundary inside the SLS~\cite{Zhang:2020hct}.

It may be argued that inserting a time-like boundary into Minkowski space-time is somewhat artificial. 
For this reason, in this paper we consider anti-de Sitter (adS) space-time.
Like Minkowski space-time, adS is maximally symmetric, which simplifies the study of quantum field theory on this background.
However, unlike Minkowski space-time, in adS null infinity is a time-like surface. 
This time-like surface is not directly analogous to the cylindrical boundary in Minkowski space-time; in particular, time-like geodesics do not reach the adS boundary in finite proper time. 
Some of the features of rotating systems in adS are however the same as those for Minkowski space-time.

Quantum field theory on adS has been studied for many years, starting with the seminal work in~\cite{Avis:1977yn}, where a quantum scalar field is considered and the nonrotating vacuum state constructed.
Quantum field theory on adS is complicated by the presence of the time-like boundary at infinity, on which boundary conditions must be imposed~\cite{Avis:1977yn,Benini:2017dfw,Dappiaggi:2016fwc,Dappiaggi:2017wvj,Dappiaggi:2018pju,Dappiaggi:2018xvw}.
For a quantum scalar field, applying either Dirichlet~\cite{Avis:1977yn,Allen:1986ty,Kent:2014nya,Ambrus:2018olh} or Neumann ~\cite{Avis:1977yn,Allen:1985wd,Allen:1986ty}
boundary conditions yields a global vacuum state which, 
like the global Minkowski vacuum, respects the maximal symmetry of the underlying adS space-time.
In recent work, it has been shown that this symmetry is broken if one instead imposes Robin (mixed) boundary conditions on a quantum scalar field~\cite{Dappiaggi:2018xvw,Pitelli:2019svx,Barroso:2019cwp,Morley:2020ayr}. 
Maximally-symmetric global vacuum states can also be constructed on adS for fermion and other quantum fields~\mbox{\cite{Allen:1985wd,Allen:1986qj,Ambrus:2015mfa,Belokogne:2016dvd}}. 

In Minkowski space-time, static thermal states, like the Minkowski vacuum, are maximally symmetric. This is not the situation on adS. 
Instead, thermal radiation tends to ``clump'' away from the space-time boundary~\cite{Allen:1986ty,Ambrus:2017cow,Ambrus:2018olh}. 
For fermion fields, the thermal expectation value (t.e.v.) of the stress-energy tensor (SET) is isotropic and of perfect fluid form~\cite{Ambrus:2017cow}, but for a scalar field, the t.e.v. of the SET has a nonzero pressure deviator~\cite{Ambrus:2018olh}.
Further properties of static vacuum and thermal states on adS are explored in~\cite{Burgess:1984ti,Caldarelli:1998wk,Camporesi:1991nw,Camporesi:1992tm,Camporesi:1992wn,Ambrus:2017vlf}.

What about rotating states on adS? 
For a quantum scalar field, as on Minkowski space-time, the rotating and nonrotating vacua coincide~\cite{Kent:2014wda}, irrespective of whether or not there is an SLS.
In this paper we examine what happens for a quantum fermion field on adS.
In particular, we address the following questions:
\begin{enumerate}
    \item Is the rotating fermion vacuum state distinct from the global static fermion vacuum on adS?
    \item Can rigidly-rotating thermal states be defined for fermions on adS? 
    \item What are the properties of these rigidly-rotating states?
\end{enumerate}

These questions are important because a fuller understanding of rotating states on adS may have implications for the physics of strongly-coupled condensed matter systems, due to the connection between these afforded by the adS/CFT (conformal field theory) correspondence~\cite{Aharony:1999ti, Hartnoll:2009sz}.
Furthermore, the maximal symmetry of adS allows us to perform a nonperturbative investigation of the effect of curvature on, e.g., the axial vortical effect, previously considered in, for example, Ref.~\cite{Flachi:2014jra}.

Our purpose in this paper is to provide comprehensive answers to Questions 1--3 for both massless and massive fermions on adS space-time (preliminary answers to these questions were presented in
\cite{Ambrus:2014fka,Ambrus:2019gkt}).
We restrict our attention to the situation where the rate of rotation $\Omega $ is smaller than the inverse radius of curvature $\ell ^{-1}$. This means that there is no SLS, which simplifies the formalism. 
In particular, we are able to exploit the maximal symmetry of the underlying adS space-time, even though this is broken by the rotation. 
After deriving the Kubo-Martin-Schwinger (KMS) relations for two-point functions at finite temperature undergoing rigid rotation, we are able to write the thermal propagator for rotating states as an infinite imaginary-time image sum involving the vacuum propagator. The maximal symmetry allows us to write the vacuum fermion propagator in closed form. 
This greatly facilitates the computation of t.e.v.s, which are the main focus of our work. Our results confirm known results for the vortical effects, namely the expression for the axial vortical conductivity $\sigma^\omega_A$, the vorticity- and acceleration-induced corrections to the energy density and pressure and the circular heat conductivity $\sigma^\tau_\varepsilon$. Furthermore, we uncover curvature corrections to the above quantities which depend on the Ricci scalar $R = -12 \ell^{-2}$. 

As a natural consequence of the chiral vortical effect, a finite axial flux is induced through the equatorial plane of adS. For massless fermions, we show that due to the conservation of the axial current, this flux originates from the adS boundary corresponding to the southern hemisphere and is transported through the northern hemisphere (defined with respect to the orientation of $\bm{\Omega}$). Since massive particles cannot reach the adS boundary in finite time, we show that the axial flux through the adS boundaries is exactly zero when considering quanta of nonvanishing mass $M> 0$. In this case, the axial flux generated through the chiral vortical effect is converted gradually into the pseudoscalar condensate ${\rm PC} = -i {\bar \psi} \gamma^5 \psi$, as required by the divergence equation $\nabla_\mu J_A^\mu = -2M \, {\rm PC}$. Thus, the adS boundary is transparent with respect to the flow of chirality of strictly massless particles, becoming opaque when massive quanta are considered. 

Finally, we discuss the total (volume integral) energy and scalar condensate contained within the adS space and show that they diverge as $(1 - \ell^2 \Omega^2)^{-1}$ as the rotation parameter $\Omega$ approaches the inverse radius of curvature $\ell^{-1}$.

We begin in Section~\ref{sec:adS} with a brief review of the geometry of adS and the formalism for the Dirac equation on this curved space-time background.
We also use relativistic kinetic theory (RKT) to find the stress-energy tensor (SET) of a rigidly-rotating thermal distribution of fermions.
The fermion propagator for rigidly-rotating thermal states is derived in Section~\ref{sec:SF}, using the aforementioned geometric approach. In Sections~\ref{sec:SCPC}--\ref{sec:SET} we study the  quantum t.e.v.s of the scalar (SC) and pseudoscalar (PC) condensates, the vector (VC) and axial (AC) charge currents and the SET. 
Our conclusions and further discussion are presented in Section~\ref{sec:conc}. Appendices~\ref{app:traces}--\ref{app:hyp} compile useful relations concerning spinor traces at finite temperature, large temperature summation formulae and some useful properties of the hypergeometric and Bessel functions, respectively.

Throughout this paper, we use Planck units ($\hbar = c = G = k_B = 1$) and the metric signature $(-,+,+,+)$. Our convention for the Levi-Civita symbol is $\varepsilon^{0123} = 1/\sqrt{-g}$. The analysis in this paper is restricted to the case of vanishing chemical potential $\mu = 0$.

\section{Dirac Fermions on adS}
\label{sec:adS}

In this section we briefly describe the formalism for the Dirac equation on adS space-time, and also derive the SET for rigidly-rotating fermions using an RKT approach.

\subsection{Preliminaries}\label{sec:adS:prelim}

The line element of adS can be written as:
\begin{equation}\tag{1}
 ds^2 = \frac{\ell ^{2}}{\cos^2\wr}
 \left(-d\wt^2 + d\wr^2 +\sin^2 \wr  \, dS_2^2\right),
 \label{eq:ds2}
\end{equation}
where $dS_2^2 = d\theta^2 + \sin^2\theta \, d\varphi^2$ is the line element on 
the two-sphere of unit radius, \mbox{$\wt \in (-\infty, \infty)$} is the time coordinate on the covering space of adS, $0 \le \wr < \pi/2 $ is the radial coordinate, 
and the adS radius of curvature $\ell $ is related to the Ricci scalar 
and cosmological constant $\Lambda$ by
\begin{equation}\tag{2}
 R = 4\Lambda = -\frac{12}{\ell ^{2}}.
 \label{eq:Ricci}
\end{equation}

In Equation~(\ref{eq:ds2}) we have used dimensionless coordinates $\wt$, $\wr$. Dimensionful time and radial coordinates are defined by 
\begin{equation}\tag{3}
t=\ell \wt , \qquad r=\ell \wr .
\label{eq:dimensionful}
\end{equation}

It is convenient to introduce at this point the orthonormal tetrad 
of vectors $e_{\halpha} = e_\halpha^\mu \partial_\mu$ 
($\halpha \in \{\hatt, \hatr, \htheta, \hvarphi\}$),
\begin{align}\tag{4}
 e_\hatt =& \ell ^{-1}\cos\wr \, \partial_\wt, &
 e_\hatr =& \ell ^{-1}\cos \wr \, \partial_\wr, &
 e_\htheta =& \frac{\ell ^{-1}\partial_\theta}{\tan \wr}, &
 e_\hvarphi =& \frac{\ell ^{-1} \partial_\varphi}{\sin \theta \tan \wr},
 \label{eq:tetrad_sph}
\end{align}
which satisfies
$g_{\mu\nu} e^\mu_\halpha e^\nu_\hbeta = \eta_{\halpha\hbeta} \equiv 
{\rm diag}(-1,1,1,1)$, where $\eta_{\halpha\hbeta}$ is the Minkowski 
metric. The following set of one-forms,
\begin{align}\tag{5}
 e^{\hatt} =& \frac{\ell \, d\wt}{\cos\wr},&
 e^\hatr =& \frac{\ell  \, d\wr}{\cos \wr}, &
 e^\htheta =& \ell \, \tan \wr \, d\theta , &
 e^\hvarphi =&  \ell  \, \tan \wr \, \sin \theta \, d\varphi, 
 \label{eq:tetrad_sph_omega}
\end{align}
is dual to the vector tetrad in Equation~\eqref{eq:tetrad_sph} in the sense that
\begin{equation}\tag{6}
 e^\halpha_\mu e^\mu_\hbeta = \delta^\halpha{}_\hbeta, \qquad 
 e^\halpha_\mu e^\nu_\halpha = \delta^\mu{}_\nu, \qquad 
 \eta_{\halpha\hbeta} e^\halpha_\mu e^\hbeta_\nu = g_{\mu\nu}.
 \label{eq:prelim_dual}
\end{equation}

Let us now consider the Cartesian equivalent of the tetrad in Equation~\eqref{eq:tetrad_sph_omega}.
To this end, we introduce the Cartesian-like coordinates $x^\halpha \in \{t,x,y,z\}$ 
by
\begin{equation}\tag{7}
 x = r \sin\theta \cos\varphi, \qquad
 y = r \sin \theta \sin\varphi, \qquad
 z = r \cos\theta,
 \label{eq:geom_xyz_def}
\end{equation}
using (\ref{eq:dimensionful}). Indices on these Cartesian-like coordinates are raised and lowered with the Minkowski rather than adS metric, so that $x^\hatt = -x_\hatt = t$ 
and $x_\hati = x^\hati$.
Starting from the relation between the partial derivatives with respect to the
spherical coordinates and those with respect to the Cartesian coordinates, in Minkowski space-time we have
\begin{align}\tag{8}
 \partial_r =& \frac{x\partial _x+y\partial_y+z\partial _z}{r}, &
 \frac{1}{r} \partial_\theta =& \frac{x \partial_x+y\partial_y+z\partial_z}{r \tan \theta} - \frac{\partial_z}{\sin\theta}, &
 \frac{1}{r \sin \theta} \partial_\varphi =& \frac{-y \partial_x + x \partial_y}{r \sin\theta}.
\end{align}
Returning to adS space-time, we seek the Cartesian gauge tetrad which satisfies
\begin{align}\tag{9}
 e_\hatr =& \frac{x^\hati e_\hati}{r}, &
 e_\htheta =& \frac{x^\hati e_\hati}{r \tan \theta} - \frac{e_\hatz}{\sin\theta}, &
 e_\hvarphi =& \frac{-y e_\hatx + x e_\haty}{r \sin\theta}.
\end{align}
The vectors that we require are~\cite{Cotaescu:2007xv}
\begin{equation}\tag{10}
 e_\hati = \cos \wr \left[ \frac{\wr}{\sin \wr} \left(
 \delta^\hatj{}_\hati - \frac{x^\hatj x_\hati}{r^2}\right) + \frac{x^\hatj x_\hati}{r^2}\right]
 \frac{\partial}{\partial x^\hatj},
 \label{eq:tetrad}
\end{equation}
where $x_\hati = x^\hati$,
while $e_\hatt = \cos \wr\, \partial_\hatt$ is the same as in Equation~\eqref{eq:tetrad_sph}.
The dual one-forms satisfying Equation~\eqref{eq:prelim_dual} are 
\begin{equation}\tag{11}
 e^\hati = \frac{1}{\cos\wr}\left[\frac{\sin \wr}{\wr}
 \left(\delta^\hati{}_\hatj - \frac{x^\hati x_\hatj}{r^2}\right) + \frac{x^\hati x_\hatj}{r^2}\right]dx^\hatj.
 \label{eq:tetrad_omega}
\end{equation}
This Cartesian gauge is useful in establishing analogies with the familiar spinor algebra of Minkowski special relativity, as pointed out in Ref.~\cite{Brill:1957fx}.

We end this section by discussing the connection coefficients $\Gamma^{\hat{\sigma}}{}_{\hgamma\halpha} = 
\eta^{\hsigma\hbeta} \Gamma_{\hbeta\hgamma\halpha}$ required for the covariant 
differentiation of tensors. These can be computed via
\begin{equation}\tag{12}
 \Gamma_{\hbeta\hgamma\halpha} = \frac{1}{2}(c_{\hbeta\hgamma\halpha} + c_{\hbeta\halpha\hgamma} - c_{\hgamma\halpha\hbeta}),
\end{equation}
where the Cartan coefficients $c_{\halpha\hbeta}{}^\hgamma$ are
defined in terms of the commutator $[e_\halpha, e_\hbeta]$ of the tetrad vectors by
\begin{equation}\tag{13}
 c_{\halpha\hbeta}{}^\hgamma = e^\hgamma_\mu [e_\halpha, e_\hbeta]^\mu,
 \qquad 
 [e_\halpha, e_\hbeta]^\mu = e_\halpha^\nu \partial_\nu e_\hbeta^\mu - 
 e_\hbeta^\nu \partial_\nu e_\halpha^\mu.
\end{equation}
For the Cartesian gauge tetrad, the nonvanishing Cartan coefficients are
\begin{align}\tag{14}
 c_{\hatt\hati}{}^\hatt =& (\sin \wr) \frac{x_\hati}{\ell r}, &
 c_{\hati\hatj}{}^{\hatk} =&
 \tan\frac{\wr}{2}
 (\delta^\hatk{}_{\hati} \eta_{\hatj \hatl} - 
 \delta^\hatk{}_{\hatj} \eta_{\hati \hatl}) \frac{x^\hatl}{\ell r},
 \label{eq:cartan}
\end{align}
giving rise to the following nonvanishing connection coefficients:
\begin{align}\tag{15}
 \Gamma^\hatt{}_{\hati\hatt} =&  (\sin \wr) \frac{x_\hati}{\ell r}, &
 \Gamma^\hati{}_{\hatj\hatk} =& 
 - \tan\frac{\wr}{2}
 (\delta^\hati{}_\hatl \eta_{\hatj\hatk} - \delta^\hati{}_\hatk \eta_{\hatj\hatl})
 \frac{x^\hatl}{\ell r}.
 \label{eq:connection}
\end{align}

\subsection{Dirac Equation}\label{sec:adS:dirac}

Following the minimal coupling procedure, the equation for a Dirac field of 
mass $M$ on a curved background is
\begin{equation}\tag{16}
 (i \slashed{D} - M) \psi = 0,
\end{equation}
where $\slashed{D} = \gamma^\halpha D_\halpha$ is the contraction between the 
$\gamma$ matrices $\gamma^\halpha = (\gamma^\hatt, \gamma^\hatx, \gamma^\haty, \gamma^\hatz)$,
which are defined with respect to the Cartesian gauge tetrad in Equation~\eqref{eq:tetrad},
and the spinor covariant derivative $D_\halpha = \nabla_\halpha - \Gamma_\halpha$. 
The spin connection $\Gamma_\halpha$ is computed via
\begin{equation}\tag{17}
 \Gamma_\halpha = -\frac{i}{2} \Gamma_{\hbeta\hgamma \halpha} 
 S^{\hbeta\hgamma},
 \label{eq:Gamma_def}
\end{equation}
where $S^{\hbeta\hgamma} = \frac{i}{4} [\gamma^\hbeta, \gamma^\hgamma]$ 
are the spin part of the generators of Lorentz transformations.

In this paper, we take the $\gamma$ matrices to be in the Dirac representation, as follows:
\begin{equation}\tag{18}
 \gamma^\hatt = 
 \begin{pmatrix}
  1 & 0 \\ 0 & -1 
 \end{pmatrix}, \qquad 
 \gamma^\hati = 
 \begin{pmatrix}
  0 & \sigma^i \\ -\sigma^i  & 0
 \end{pmatrix},\qquad 
 \gamma^5 = 
 \begin{pmatrix}
  0 & 1 \\ 1 & 0
 \end{pmatrix},
\end{equation}
where $\gamma^5 = i \gamma^\hatt \gamma^\hatx \gamma^\haty \gamma^\hatz$ is the chirality matrix and $\sigma^i = (\sigma^x, \sigma^y, \sigma^z)$ are the usual Pauli~matrices:
\begin{equation}\tag{19}
 \sigma^x = 
 \begin{pmatrix}
  0 & 1 \\ 1 & 0
 \end{pmatrix}, \qquad \sigma^y = 
 \begin{pmatrix}
  0 & -i \\ i & 0
 \end{pmatrix}, \qquad \sigma^z = 
 \begin{pmatrix}
  1 & 0 \\ 0 & -1
 \end{pmatrix}.
\end{equation}
In this case, the components $\Gamma_\halpha$ of the spin connection are
\cite{Ambrus:2017cow}:
\begin{align}\tag{20}
 \Gamma_\hatt =& \frac{1}{2\ell } (\sin \wr) \gamma_{\hatt}
 \left(\frac{\vx \cdot \vgamma}{r}\right), &
 \Gamma_\hatk =& \frac{1}{2\ell } \tan\frac{\wr}{2}
 \left[\frac{x_\hatk}{r} + \gamma_\hatk \left(\frac{\vx\cdot\vgamma}{r}\right)\right].
 \label{eq:Gamma}
\end{align}

\subsection{Kinematics of Rigid Motion on adS}
\label{sec:adS:kin}

Let us consider first a fluid at rest. Its four-velocity field is
\begin{equation}\tag{21}
 u_{\rm s} = \ell^{-1} \cos \wr \, \partial_t, \qquad u_{\rm s}^2 = -1.
\end{equation}
The acceleration of this vector field is
\begin{equation}\tag{22}
 a_{\rm s} = \nabla_{u_{\rm s}} u_{\rm s} = \cos^2\wr \, \Gamma^\mu{}_{tt} \, \partial_\mu =
 \ell ^{-2} \sin \wr \cos\wr \, \partial_\wr, \qquad 
 a_{\rm s}^2 = \ell^{-2} \sin^2 \wr,
 \label{eq:static_a}
\end{equation}
where we used the fact that only the following Christoffel symbols are nonvanishing:
\begin{equation}\tag{23}
\begin{array}{ccc}
 \Gamma^\wr{}_{\wt\wt} = \Gamma^\wt{}_{\wr\wt} = \Gamma^\wr{}_{\wr\wr} = \tan \wr, &
 \Gamma^\wr{}_{\theta\theta} = -\tan \wr, & 
 \Gamma^\wr{}_{\varphi\varphi} = -\sin^2\theta \tan \wr, \nonumber\\
 \Gamma^\theta{}_{\wr\theta} = \Gamma^\varphi{}_{\wr\varphi} = \frac{1}{\sin\wr \cos\wr}, &
 \Gamma^\theta{}_{\varphi\varphi} = -\sin\theta \cos\theta, &
 \Gamma^\varphi{}_{\theta\varphi} = \cot \theta.\\
 \label{eq:Christoffel}
\end{array}
\end{equation}

When the rotation is switched on (that is, at finite vorticity), the acceleration $a_{\rm s}$ seen in the static case will receive a centripetal correction.
Since we are interested in global thermodynamic equilibrium, 
the temperature four-vector $\beta^\mu = \beta u^\mu$ (where $\beta = T^{-1}$ 
is the local inverse temperature) must satisfy the Killing equation 
\cite{Cercignani02}:
\begin{equation}\tag{24}
 (\beta u_\halpha)_{;\hbeta} + 
 (\beta u_\hbeta)_{;\halpha} = 0.
\end{equation}
For angular velocity $\Omega $, starting from the Killing vector 
$\beta_0(\partial_t + \Omega \partial_\varphi)$, it can be seen that the four-velocity and temperature are given by~\cite{Ambrus:2016ocv}:
\begin{align}\tag{25}
\begin{array}{lll}
 u =& \Gamma \cos\wr \, (\partial_t + \Omega \partial_\varphi)
 = \Gamma\left(e_\hatt + \wO \wrho e_\hvarphi\right), \nonumber\\
 \beta =& {\displaystyle{\frac{\beta_0}{\Gamma \cos\wr}}}, \qquad
 \Gamma = 
{\displaystyle{\frac{1}{\sqrt{1 - \wrho^2  \wO^2}}}},
\label{eq:RKT_u_coord}\\
\end{array}
\end{align}
where $\beta_0 = T_0^{-1}$ represents the 
inverse temperature at the coordinate origin and we introduced the relative angular velocity $\wO$ and the effective transverse coordinate $\wrho$, as well as an effective vertical coordinate $\wz$ via
\begin{equation}\tag{26}
 \wO = \ell \Omega, \qquad 
 \wrho = \sin\wr \sin\theta, \qquad 
 \wz = \tan\wr \cos \theta.
 \label{eq:wOrhoz}
\end{equation}
We see that the rotation has an effect on the local inverse temperature $\beta $. If $\wO <1$, the Lorentz factor $\Gamma  $ and inverse temperature $\beta $ remain finite for all $\wr\in [0,\pi/2]$, while $\beta^\mu$ remains timelike.
However, if $\wO > 1$, there will be a surface (the speed of light surface, SLS) where 
$\Gamma $ (and hence the local temperature) diverges and $\beta^\mu$ becomes a null vector~\cite{Ambrus:2016ocv}.

\begin{figure}
\begin{center}
 \includegraphics[width=0.45\linewidth]{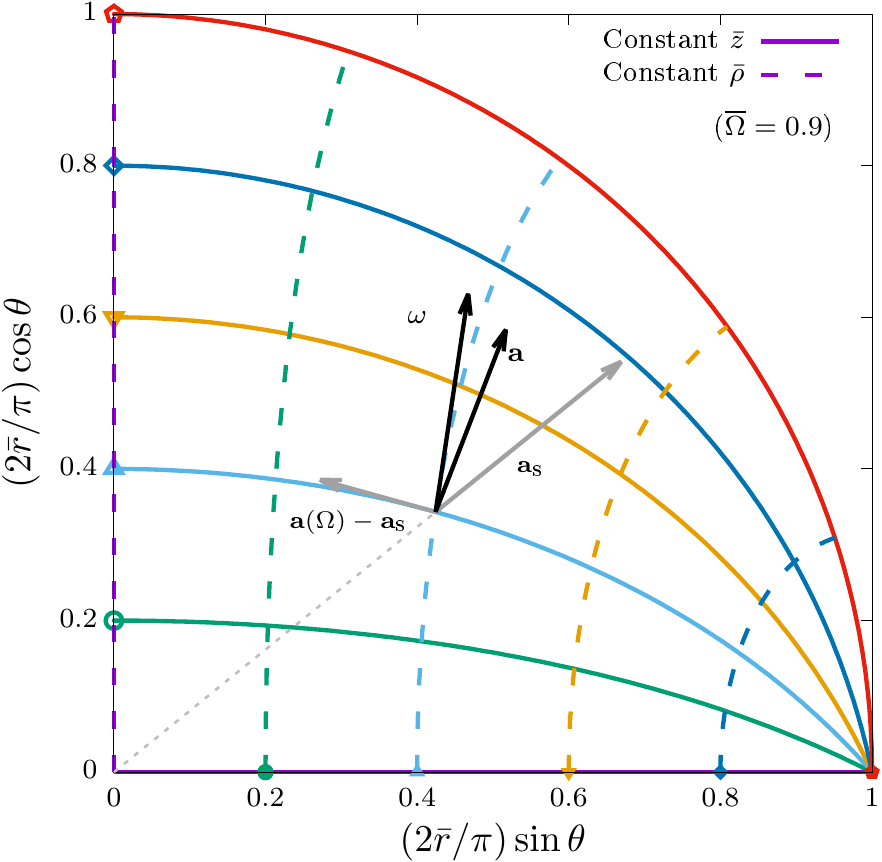}
\end{center}
\caption{Equal $\wz$ and $\wrho$ contours in the $(\wr \sin \theta, \wr \cos \theta)$ plane, shown with solid and dashed lines, respectively. The coordinates are normalised with respect to $\pi / 2$, so that the horizontal and vertical axes have the ranges $[0,1]$. The acceleration $\bm{a}$ and vorticity $\bm{\omega}$ are shown using solid black arrows at the point $({\bar z}, {\bar \rho}) = (\arctan(0.2\pi), \arcsin(0.2\pi))$ for the case when $\wO = 0.9$. The radial and rotational components $\bm{a}_{\rm s}$ and $\bm{a}(\Omega) - \bm{a}_{\rm s}$ of the acceleration are shown with dark gray arrows. 
\label{fig:kin_surf}}

\end{figure}

The inverse transformation corresponding to Equation~\eqref{eq:wOrhoz} is
\begin{align}\tag{27}
\begin{array}{llll}
 \sin\theta =& {\displaystyle {\frac{\wrho}{\sin \wr}}}, & 
 \cos\theta =& {\displaystyle{\frac{\wz}{\tan \wr}}}, \\[0.3cm]
 \sin \wr =& {\displaystyle{\sqrt{\frac{\wz^2 + \wrho^2}{1 + \wz^2}}}}, & 
 \cos \wr =& {\displaystyle{\sqrt{\frac{1 - \wrho^2}{1 + \wz^2}}}},\\
 \end{array}
\end{align}
while the line \hl{element} 
  \eqref{eq:ds2} with respect to $\wrho$ and $\wz$ becomes
\begin{equation}\tag{28}
 \ell^{-2} ds^2 = -\frac{1 + \wz^2}{1 -  \wrho^2} d\wt^2 + \frac{d\wz^2}{1 + \wz^2} + 
 \frac{d\wrho^2(1 +  \wz^2)}{(1 -  \wrho^2)^2} + 
 \frac{\wrho^2(1 + \wz^2)}{1 -  \wrho^2} d\varphi^2,
\end{equation}
with $\sqrt{-g} = \ell^4 \wrho (1 + \wz^2) / (1 -  \wrho^2)^2$.
The surfaces of constant $\wz$ and $\wrho$ are shown in Figure~\ref{fig:kin_surf} using solid and dashed lines, respectively. The acceleration and vorticity vectors $\bm{a}$ and $\bm{\omega}$, shown with black arrows, are discussed below.

The acceleration $a^\mu = u^\nu \nabla_\nu u^\mu = \Gamma^\mu{}_{\lambda \nu} u^\lambda u^\nu$ can be obtained
using the Christoffel symbols in Equation~\eqref{eq:Christoffel}:
\begin{align}
 a =& \ell^{-2} \Gamma^2 \cos \wr  \, \left[
 \sin \wr  \, (1 - \wO^2 \sin^2 \theta) \partial_\wr - 
 \wrho \wO^2 \wz \cot^2\wr \, \partial_\theta\right]\nonumber\\
 =& \ell^{-2} [\Gamma^2 \wrho(1 - \wO^2) \cos^2 \wr \, \partial_\wrho - \wz \partial_\wz]\nonumber\\
 \simeq& -\rho \Gamma^2 \Omega^2 \partial_\rho + O(\ell ^{-1}).
 \label{eq:acc}\tag{29}
\end{align}
The first term on the first line is akin to the static acceleration uncovered 
in \mbox{Equation~\eqref{eq:static_a}.} The last line reproduces the Minkowski expression,
obtained in the limit $\ell ^{-1} \rightarrow 0$ ($\rho = r \sin\theta$). To highlight more clearly the rotational contribution to the acceleration, it is convenient to perform the split $a(\Omega) = a_{\rm s} + [a(\Omega) - a_{\rm s}]$, where the static acceleration is given in Equation~\eqref{eq:static_a} and
\begin{equation}\tag{30}
 a(\Omega) - a_{\rm s} = -\rho \Omega^2 \Gamma^2 (1 - \wrho^2) \cos^2 \wr \, \partial_\rho.
\end{equation}
Thus, it becomes clear that the rotational part $a(\Omega) - a_{\rm s}$ plays a purely centripetal role, being directed along the effective horizontal direction $\wrho = \rho / \ell$ defined in Equation~\eqref{eq:wOrhoz}. Figure~\ref{fig:kin_surf} shows the decomposition of the acceleration for the case when $\wO = 0.9$ into its purely static component $\bm{a}_{\rm s}$, which points in the radial direction from the origin towards the adS boundary; and the purely rotational component $\bm{a}(\Omega) - \bm{a}_{\rm s}$, which is tangent to the line of constant ${\wrho}$.

The vorticity $\omega^\mu = \frac{1}{2} \varepsilon^{\mu\nu\alpha\beta} u_\nu \nabla_\alpha u_\beta$ 
is given by:
\begin{align}
 \omega =& \ell^{-2} \wO \Gamma^2 \cos\wr \cot^2 \wr  \, (\wz  \sin\wr \, \partial_\wr - 
 \wrho \, \partial_\theta)\nonumber\\
 =& \ell^{-2} \wO \Gamma^2 (1- \wrho^2) \partial_\wz \nonumber\\
 \simeq& \Omega \Gamma^2 \partial_z + O(\ell ^{-2}),
 \label{eq:RKT_omega_coord}\tag{31}
\end{align}
where again the Minkowski result was recovered on the last line by setting $\ell ^{-1} \rightarrow 0$. The advantage of using the effective horizontal and vertical coordinates $\wrho$ and $\wz$ is now obvious, since $\omega^\mu$ becomes parallel to the $\partial_\wz$ direction. As seen in Figure~\ref{fig:kin_surf}, the vorticity vector $\bm{\omega}$ is orthogonal to the $\wz = \textrm{const}$ lines. Due to the static component $\bm{a}_{\rm s}$ of the acceleration, the scalar product $\omega \cdot a$ is nonvanishing and is given in Equation~\eqref{eq:aomega} below.

Finally, a fourth vector $\tau ^{\mu }$  which is orthogonal to $u^{\mu }$, $a^{\mu }$ and $\omega ^{\mu }$ simultaneously can 
be introduced via $\tau^\mu = \varepsilon^{\mu\nu\alpha\beta} \omega_\nu a_\alpha u_\beta$.
The following result is obtained:
\begin{align}
 \tau =& \ell^{-3} \Gamma^5 \wO (1 - \wO^2) \cos^3\wr\left(\wrho^2 \wO \, \partial_\wt + \partial_\varphi\right)
 \nonumber\\
 =& -\Omega^3 \Gamma^5 (\rho^2 \Omega \, \partial_t + \partial_\varphi) + O(\ell^{-2}),\tag{32}
\end{align}
where again the Minkowski result is recovered in the limit of small $\ell ^{-1}$.  

We close this subsection with the expressions for the vectors in the kinematic tetrad, expressed with respect to those of the Cartesian tetrad $(e_\hatt, e_\hatx, e_\haty, e_\hatz)$ in Equation~\eqref{eq:tetrad}:
\begin{align}
 u =& \Gamma(e_\hatt + \wrho \wO e_\hvarphi) = \Gamma
 \begin{pmatrix}
  1 \\  -\wrho\wO \sin \varphi \\ \wrho\wO \cos \varphi \\  0
 \end{pmatrix}, \nonumber\\
 a =& \Gamma^2 \ell ^{-1} \sin\wr\, \frac{x^\hati e_\hati}{r} - 
 \wrho \wO^2 \Gamma^2 \ell^{-1} \left[\sin\theta  \, (1 + \cos\wr \cot^2\theta) 
 \frac{x^\hati e_\hati}{r} - \frac{\cos \wr}{\tan \theta} e_\hatz\right] \nonumber\\
 =& \Gamma^2 \ell ^{-1} \sin\wr
 \begin{pmatrix}
  0 \\ \sin\theta\cos\varphi \\ \sin\theta\sin\varphi \\ \cos\theta
 \end{pmatrix} -  \wrho \wO^2 \Gamma^2 \ell^{-1} \sin\theta 
 \begin{pmatrix}
  0 \\ 
  (1 + \cos\wr \cot^2 \theta) \sin\theta \cos\varphi \\
  (1 + \cos\wr \cot^2 \theta) \sin\theta \sin\varphi \\
  (1 - \cos\wr) \cos\theta
 \end{pmatrix},\nonumber\\
 \omega =& \Omega \Gamma^2 \cos\wr  \, \left[\cos\theta(1 - \cos\wr) \frac{x^\hati e_\hati}{r} 
 + \cos\wr  \, e_\hatz\right]
 \nonumber \\
 =&\Omega \Gamma^2 \cos \wr \cos^2\theta
 \begin{pmatrix}
  0 \\ (1 - \cos\wr) \tan\theta \cos\varphi \\ 
  (1 - \cos\wr) \tan\theta \sin\varphi \\ 
  1 + \cos\wr \tan^2\theta
 \end{pmatrix},\nonumber\\
 \tau =& \wrho \wO \Gamma^5 (\ell^{-2} - \Omega^2) \cos^2 \wr \, 
 (\wrho \wO e_\hatt - \sin\varphi \, e_\hatx + \cos\varphi  \, e_\haty) 
 \nonumber \\ =&
 \wrho \wO \Gamma^5 (\ell^{-2} - \Omega^2) \cos^2 \wr
 \begin{pmatrix}
  \wrho \wO \\
  -\sin \varphi\\
  \cos\varphi \\ 0
 \end{pmatrix}.\label{eq:kinematic_cartesian}\tag{33}
\end{align}
With the relations in Equation~\eqref{eq:kinematic_cartesian}, it is not difficult to find the squared norms
\begin{align}
 \omega^2 =& \bm{\omega }^2 = \Omega^2 \Gamma^4 (1 - \wrho^2) \cos^2\wr = \Omega^2 \Gamma^4 + O(\ell^{-2}), \nonumber\\
 a^2 =& \bm{a}^2 = \ell^{-2} [1 - \Gamma^4 (1 - \wrho^2 \wO^4) \cos^2 \wr] = \rho^2 \Omega^4 \Gamma^4 + O(\ell ^{-2}),\nonumber\\
 \tau^2 =& \ell^{-4} \wrho^2 \wO^2 \Gamma^8 (1 - \wO^2)^2 \cos^4 \wr \simeq \rho^2 \Omega^6 \Gamma^8 + O(\ell^{-2}),
 \label{eq:kinematic_sq}\tag{34}
\end{align}
while $u^2 = -1$. It is remarkable that, contrary to the situation on Minkowski space, the acceleration and vorticity are not orthogonal:
\begin{equation}
 a \cdot \omega  = a_\mu \omega^\mu = \ell^{-2} \Gamma^2 \wO \wz \cos^2\wr \simeq O(\ell^{-2}).
 \label{eq:aomega}\tag{35}
\end{equation}

\subsection{Relativistic Kinetic Theory Approach}
\label{sec:adS:RKT}

In order to gain insight into the properties of thermal states undergoing rigid rotation 
on adS space, in this subsection we perform a relativistic kinetic theory analysis. 
In subsequent sections, we will employ a full quantum field theory treatment to reveal 
quantum effects and corrections not captured within this classical framework.
Since rigid rotation leads to a state of general thermodynamic equilibrium, we assume that 
the system is comprised of noninteracting fermions distributed according to the 
Fermi-Dirac~distribution,
\begin{equation}
 f = \frac{{\mathsf {g}}}{(2\pi)^3} 
 \frac{1}{e^{-\beta u \cdot p} + 1},\tag{36}
\end{equation}
where ${\mathsf {g}}$ is the degeneracy factor (${\mathsf {g}} = 2\times 2 = 4$ accounts for 
the degeneracies due to spin and the particle/antiparticle contributions).
For rigidly-rotating thermal states, the local inverse temperature
$\beta = T^{-1}$ and the four-velocity $u^\halpha$ are given in 
Equation~\eqref{eq:RKT_u_coord}. The microscopic four-momentum 
of the fermion gas $p^\halpha$ satisfies $p^2 = -M^2$. For simplicity, only the case of vanishing chemical potential is considered here and henceforth.

It is convenient to parameterize the one-particle momentum space using the spatial 
components $p^\hati = p^\mu \omega_\mu^\hati$ of the particle momentum, expressed
with respect to the tetrad. In this parametrization, the SET
can be obtained via~\cite{Ambrus:2016ocv}
\begin{align}
 T_{\rm RKT}^{\halpha\hbeta} =& \int \frac{d^3\hat{p}}{p^\hatt} f\, p^\halpha p^\hbeta 
 \nonumber\\
 =& (E_{\rm RKT} + P_{\rm RKT}) u^\halpha u^\hbeta + P_{\rm RKT} \eta^{\halpha\hbeta},
 \label{eq:RKT_SET_def}\tag{37}
\end{align}
where the hat over the integration measure was added to indicate that the integration 
is performed with respect to the tetrad components of the momentum vector. Contracting Equation~\eqref{eq:RKT_SET_def} with 
$\eta_{\halpha\hbeta}$ and $u_\halpha u_\hbeta$, the energy density and pressure can be shown to 
satisfy~\cite{Ambrus:2019ayb}:
\begin{equation}
 \begin{pmatrix}
  E_{\rm RKT} \\ E_{\rm RKT} - 3P_{\rm RKT}
 \end{pmatrix}= \frac{2}{\pi^2}
 \int_0^\infty \frac{p^2 dp}{p^\hatt} 
 \begin{pmatrix}
  (p^\hatt)^2 \\ M^2
 \end{pmatrix}
 \frac{1}{e^{p^\hatt/T} + 1},
 \label{eq:RKT_gen}\tag{38}
\end{equation}
where ${\mathsf {g}} = 4$ was used and $T = \beta^{-1} = T_0 \cos\wr / \Gamma$. In order to compute the above integral, the Fermi-Dirac factor can be expanded as follows:
\begin{equation}\tag{39}
 \frac{1}{e^{p^\hatt / T} + 1} = \sum_{j = 1}^{\infty} (-1)^{j + 1} e^{-j p^\hatt / T}.
\end{equation}
It can be seen that both the pressure and energy density have the usual dependence on the local temperature $T$. They will remain regular as long as the local temperature is finite, but diverge if $T\rightarrow \infty $, which can happen if $\Omega \ell >1$.
For this reason we focus our attention in this paper to the situation $\Omega \ell <1$.

We now discuss some large-$T$ asymptotic properties of $E_{\rm RKT}$ and $P_{\rm RKT}$. Using the~relation
\begin{equation}\tag{40}
 \int_1^\infty dX  \, (X^2 - 1)^{\nu - \frac{1}{2}} e^{-XZ} = \frac{2^\nu \Gamma(\frac{1}{2} + Z)}{Z^\nu \sqrt{\pi}} K_\nu(Z),
\end{equation}
where $K_\nu(Z)$ is the modified Bessel function of the third kind and the variables are \mbox{$X = p^\hatt / M$ and $Z = j p^\hatt / T$}, it is possible to write 
\begin{align}
 P_{\rm RKT} =& \frac{2 M^4}{\pi^2} \sum_{j = 1}^\infty \frac{(-1)^{j+1}}{(j M / T)^2} K_2\left(\frac{jM}{T}\right),&
 E_{\rm RKT} - 3P_{\rm RKT} =& \frac{2 M^4}{\pi^2} \sum_{j = 1}^\infty \frac{(-1)^{j+1}}{j M / T} K_1\left(\frac{jM}{T}\right).
 \label{eq:RKT_PE_j}\tag{41}
\end{align}
The above expressions are identical to the vanishing chemical potential limit \linebreak \mbox{($\mu_V = \mu_H = 0$)}  of Equation~\hl{(4.22)} in Ref.~\cite{Ambrus:2019ayb}.
The large temperature limit can be computed using
Equation~\eqref{eq:Kn_smallZ} to expand the modified Bessel functions:
\begin{align}
 \frac{1}{Z} K_1(Z) =& \frac{1}{Z^2} + \frac{1}{2} \ln\left(\frac{Z}{2} e^{{\mathcal {C}} - \frac{1}{2}}\right) + O(Z^2),\nonumber\\
 \frac{1}{Z^2} K_2(Z) =& \frac{1}{2Z^4} - \frac{1}{2Z^2} - \frac{1}{8} \ln \left(\frac{Z}{2}e^{{\mathcal {C}} - \frac{3}{4}}\right) + O(Z^2),\tag{42}
\end{align}
where ${\mathcal {C}}$ is the Euler-Mascheroni \hl{constant} \eqref{eq:digamma}.
Substituting the above into Equation~\eqref{eq:RKT_PE_j} and using Equation~\eqref{eq:sumj_zeta} to perform the summation over $j$,
we can derive the following expansions~\cite{Ambrus:2019ayb}:
\begin{align}
 E_{\rm RKT} + P_{\rm RKT} =& \frac{7\pi^2 T^4}{45} - \frac{M^2 T^2}{6} + \frac{M^4}{8\pi^2} + O(T^{-1}),\nonumber\\
 E_{\rm RKT} - 3P_{\rm RKT} =& \frac{M^2 T^2}{6} - \frac{M^4}{2\pi^2} \ln \frac{\pi T}{M e^{{\mathcal {C}} - \frac{1}{2}}} + O(T^{-1}).
 \label{eq:RKT_highT}\tag{43}
\end{align}

In the limit $\Omega \ell \rightarrow 1$, the Lorentz factor $\Gamma$ \eqref{eq:RKT_u_coord}
diverges in the equatorial plane as $\wr \rightarrow \pi / 2$. This behaviour 
signals the formation of the SLS, however the other macroscopic quantities remain finite.
For example, the local temperature $T = \Gamma \cos\wr / \beta_0$ remains constant
in the equatorial plane, $\lim_{\Omega \ell \rightarrow 1} T(\theta = \pi/2) = \beta_0^{-1}$.
Similarly, $\omega \rightarrow \Omega e_\hatz$ becomes constant and
$a \rightarrow 0$. The circular vector $\tau$ vanishes 
everywhere in the adS space-time, $\lim_{\Omega \ell \rightarrow 1} \tau = 0$, due to the $(\ell^{-2} - \Omega^2)$ 
prefactor in Equation~\eqref{eq:kinematic_cartesian}. These results are summarised for convenience 
below:

\vspace{+12pt}
\end{paracol}
\nointerlineskip
\begin{equation}
 \lim_{\Omega \ell \rightarrow 1} T\left(\theta = \frac{\pi}{2}\right) = \beta_0^{-1}, \quad 
 \lim_{\Omega \ell \rightarrow 1} \omega \left(\theta = \frac{\pi}{2}\right) = \Omega,\quad 
 \lim_{\Omega \ell \rightarrow 1} a\left(\theta = \frac{\pi}{2}\right) = 0, \quad 
 \lim_{\Omega \ell \rightarrow 1} \tau = 0.
 \label{eq:RKT_crit}\tag{44}
\end{equation}
\begin{paracol}{2}
\switchcolumn

We now consider the volume integrals of the quantities in Equation~\eqref{eq:RKT_PE_j}. The volume element $\sqrt{-g}$ can be obtained from Equation~\eqref{eq:ds2},
\begin{equation}\tag{45}
 \sqrt{-g} = \frac{\ell^4 \sin^2 \wr}{\cos^4 \wr} \sin\theta,
\end{equation}
the integration measure is $\ell^{-1} \sqrt{-g} \, d^3x$ and we obtain
\vspace{+12pt}
\end{paracol}
\nointerlineskip
\begin{equation}
 \begin{pmatrix}
  V_{\beta_0,\Omega}^{E_{\rm RKT} - 3P_{\rm RKT}}\\ 
  V_{\beta_0,\Omega}^{P_{\rm RKT}}
 \end{pmatrix} = \frac{4k^3 M}{\pi} \sum_{j = 1}^\infty (-1)^{j + 1} 
 \int_{-1}^1 d\cos\theta \int_0^{\pi / 2} \frac{\sin^2 \wr\, d\wr}{\cos^4 \wr}
 \begin{pmatrix}
  K_1(j M / T) / (j M / T) \\
  K_2(j M / T) / (j M / T)^2
 \end{pmatrix},\tag{46}
\end{equation}
\begin{paracol}{2}
\switchcolumn

\noindent
where 
\begin{equation}\tag{47}
k = \ell M, \qquad T = T_0 \Gamma \cos \wr
\label{eq:kT0}
\end{equation}
and 
$V_{\beta_0,\Omega}^{\mathsf {f}} = \ell^{-1} \int d^3x \sqrt{-g}\, {\mathsf {f}}$ 
is the volume integral of the function ${\mathsf {f}}$ for rotation rate $\Omega $ and inverse temperature at the origin $\beta_0$.
Taking into account the fact that the radial integration covers the whole 
adS space, it is convenient to employ the coordinate 
\mbox{$X = 1 / \Gamma^2 \cos^2\wr$, satisfying}
\begin{equation}\tag{48}
 \sin^2\wr = \frac{X - 1}{X - \wO^2 \sin^2\theta}, \qquad 
 \cos^2\wr = \frac{1 - \wO^2 \sin^2 \theta}{X - \wO^2 \sin^2\theta }, \qquad 
 \frac{dX}{d\wr} = \frac{2(X - 1)}{\sin\wr \cos\wr}. 
 \label{eq:RKT_x}
\end{equation}
Since $X(\wr = 0) = 1$ and $X(\wr = \pi/2) = \infty$ (valid for $|\wO| < 1$), the integration limits with respect to 
$X$ are independent of $\theta$. Additionally, the arguments of the modified Bessel functions do 
not depend on $\theta$, allowing the integration with respect to the angular coordinate 
to be performed first:
\begin{align}
 \begin{pmatrix}
  V_{\beta_0,\Omega}^{E_{\rm RKT} - 3 P_{\rm RKT}} \\ 
  V_{\beta_0,\Omega}^{P_{\rm RKT}}
 \end{pmatrix} & 
 \nonumber \\ & \hspace{-1.5cm} = 
  \frac{2k^3 M}{\pi} \sum_{j = 1}^\infty (-1)^{j + 1} 
 \int_1^\infty dX\, \sqrt{X - 1}
 \begin{pmatrix}
  K_1(j M \sqrt{X} / T_0) / (j M \sqrt{X} / T_0) \\
  K_2(j M \sqrt{X} / T_0) / (j M \sqrt{X} / T_0)^2
 \end{pmatrix}
 \nonumber \\ & \times 
 \int_{-1}^1 \frac{d\cos\theta}{(1 - \wO^2 \sin^2\theta )^{3/2}} \nonumber\\
& \hspace{-1.5cm} = \frac{4k^3 M}{\pi (1 - \wO^2)} \sum_{j = 1}^\infty (-1)^{j + 1} 
 \int_1^\infty dX \, \sqrt{X - 1}
 \begin{pmatrix}
  K_1(j M \sqrt{X} / T_0) / (j M \sqrt{X} / T_0) \\
  K_2(j M \sqrt{X} / T_0) / (j M \sqrt{X} / T_0)^2
 \end{pmatrix}.\label{eq:RKT_vol_aux}\tag{49}
\end{align}
It can be seen that the angular integration (with respect to $\varphi$ and $\theta$) effectively produces a factor $4\pi / (1 - \wO^2)$, showing that the effect of 
rotation on these volume-integrated quantities is essentially given by this
proportionality factor. It is interesting to note that the limit 
$\wO \rightarrow 1$ leads to a divergence of these quantities, which is consistent 
with the divergent behaviour of the Lorentz factor $\Gamma$. Although $E_{\rm RKT}$ 
and $P_{\rm RKT}$, which depend on $T = T_0 \Gamma \cos\wr$, remain finite everywhere, the fact that their value in the equatorial plane is no longer decreasing as $\wr \rightarrow \pi / 2$ (when $T = T_0$ for all $\wr$) leads to infinite contributions due to the infinite volume of adS.

Starting from the following identity~\cite{gradshteyn2014table},
\begin{equation}\tag{50}
 \int_1^\infty dX\, X^{-\frac{\nu}{2}} (X - 1)^{\mu - 1} K_\nu(a \sqrt{X}) = \Gamma(\mu) 2^\mu a^{-\mu} K_{\nu - \mu}(a),
\end{equation}
the integration with respect to $X$ can be performed using the relations 
\begin{equation}\tag{51}
 \int_1^\infty \frac{dX}{\sqrt{X}} \sqrt{X - 1} K_1(\sqrt{aX}) = 
 \int_1^\infty \frac{dX}{X} \sqrt{X - 1} K_2(\sqrt{aX}) = \frac{\pi}{a} e^{-\sqrt{a}},
\end{equation}
leading to 
\begin{align}
 V_{\beta_0,\Omega}^{E_{\rm RKT} - 3 P_{\rm RKT}}
 =& \frac{4k^3 M}{1 - \wO^2} \sum_{j = 1}^\infty (-1)^{j + 1} 
 e^{-j M / T_0} \left(\frac{T_0}{j M}\right)^3 = 
 -\frac{4M \ell^3 T_0^3}{1 - \wO^2} {\rm Li}_3(-e^{-M /T_0}),\nonumber\\
 V_{\beta_0,\Omega}^{P_{\rm RKT}} =& \frac{4k^3 M}{1 - \wO^2} \sum_{j = 1}^\infty (-1)^{j + 1} 
 e^{-j M / T_0} \left(\frac{T_0}{j M}\right)^4
 = -\frac{4 \ell^3 T_0^4}{1 - \wO^2} {\rm Li}_4(-e^{-M /T_0}),\label{eq:RKT_vol}\tag{52}
\end{align}
where ${\rm Li}_n(Z) = \sum_{j = 1}^\infty Z^j / j^n$ is the polylogarithm function 
\cite{NIST:DLMF}. The above relations are exact. It is convenient at this point to 
derive the high-temperature limit of Equation~\eqref{eq:RKT_vol} by expanding the 
polylogarithms:
\begin{align}
 V_{\beta_0,\Omega}^{E_{\rm RKT} - 3 P_{\rm RKT}} =& \frac{\ell^3 M}{1 - \wO^2} \left[3 T_0^3 \zeta(3) - 
 \frac{\pi^2 M T_0^2}{3} + 2 M^2 T_0 \ln 2 - \frac{M^3}{3} + O(T_0^{-1})\right],\nonumber\\
 V_{\beta_0,\Omega}^{E_{\rm RKT}} =& \frac{\ell^3}{1 - \wO^2}\left[
 \frac{7\pi^4 T_0^4}{60} - 6 M T_0^3 \zeta(3) + \frac{\pi^2 M^2 T_0^2}{6} - 
 \frac{M^4}{12} + O(T_0^{-1})\right],
 \label{eq:RKT_vol_highT}\tag{53}
\end{align}
where the Riemann zeta \hl{functio}n \eqref{eq:zeta_def} satisfies $\zeta(3) \simeq 1.202$.
Our focus in the rest of this paper is the computation of quantum corrections to these RKT results.

\section{Feynman Propagator for Rigidly-Rotating Thermal States}
\label{sec:SF}

In the geometric approach employed here, the maximal symmetry of adS is exploited to 
construct the Feynman propagator, which then plays the central role
in computing expectation values with respect to vacuum or thermal 
states. In Section~\ref{sec:SF:vac}, we briefly review the construction
of the vacuum propagator. We discuss the construction of  the propagator for thermal states 
under rigid rotation in Section~\ref{sec:SF:beta}, highlighting that
the approach is valid only for subcritical rotation, when $|\Omega \ell| < 1$.
Finally, in Section~\ref{sec:tev}, we outline how the thermal Feynman propagator is used to construct the t.e.v.s which are the focus of this paper. 

\subsection{Vacuum Feynman Propagator}
\label{sec:SF:vac}

As pointed out by M\"uck~\cite{Muck:1999mh}, the Feynman propagator corresponding to the
global adS vacuum state can be written in the form
\begin{equation}
 iS^F_{\rm vac}(x,x') = \left[\mathcal{A}_F(s) + \mathcal{B}_F(s) \slashed{n}\right] \Lambda(x,x'),
 \label{eq:geom_SF_def}\tag{54}
\end{equation}
where $\mathcal{A}_F$ and $\mathcal{B}_F$ are scalar functions that depend only on 
the geodesic interval $s \equiv s(x,x')$ between the points $x$ and $x'$ 
and satisfy the equations
\begin{align}
 \frac{\partial \mathcal{A}_F}{\partial s} - \frac{3 }{2\ell } \mathcal{A}_F \tan\frac{\ws}{2} 
 + i M \mathcal{B}_F =& 0,\nonumber\\
 \frac{\partial \mathcal{B}_F}{\partial s} + \frac{3 }{2\ell } \mathcal{B}_F \cot\frac{\ws}{2} 
 + i M \mathcal{A}_F =& \frac{1}{\sqrt{-g}} \delta(x,x'),
 \label{eq:geom_SF_AB_eqs}\tag{55}
\end{align}
where we introduced the dimensionless geodesic distance $\ws = s/\ell$.
The solution is~\cite{Ambrus:2015mfa,Ambrus:2017cow}:
\begin{align}
 \mathcal{A}_F =& \frac {\Gamma_k}{16 \pi ^2 \ell ^{3}}
 \cos \left( \frac  {\ws}{2} \right) \left[
 -\sin^{2} \left( \frac {\ws}{2} \right) \right] ^{-2-k}
 {}_{2}F_{1}\left( 1+k, 2+k; 1+2k ; {\rm cosec} ^{2} \left( \frac {\ws}{2} \right) \right),
 \nonumber\\
 \mathcal{B}_F =& \frac {i\Gamma_k}{16 \pi ^2 \ell ^{3}}
 \sin \left( \frac {\ws}{2} \right) \left[
 -\sin^{2} \left( \frac {\ws}{2} \right) \right] ^{-2-k}
 {}_{2}F_{1}\left(k, 2+k; 1+2k ; {\rm cosec} ^{2} \left( \frac {\ws}{2} \right) \right) ,
 \label{eq:SF_AB_gen}\tag{56}
\end{align}
where ${}_{2}F_{1}(a,b;c;z)$ is a hypergeometric function and
$k$ is given in terms of the fermion mass $M$ by (\ref{eq:kT0}), 
while the normalisation constant $\Gamma_k$ is given by
\begin{equation}
 \Gamma_k = \frac{\Gamma \left( 2 + k \right) \Gamma\left(\frac{1}{2}\right)}
 {4^k \Gamma \left( \frac {1}{2} + k \right)}.\label{eq:Gammak}\tag{57}
\end{equation}
In the limit $k \rightarrow 0$, we have
$\Gamma_k \rightarrow 1$ and
Equation~\eqref{eq:SF_AB_gen} reduces to
\begin{align}
 \lim_{k \rightarrow 0} {\mathcal {A}}_F =& 
 \frac{1}{16\pi^2\ell ^{3}} \left(\cos\frac{\ws}{2}\right)^{-3}, &
 \lim_{k \rightarrow 0} {\mathcal {B}}_F =& 
 \frac{i}{16\pi^2 \ell ^{3}} \left(\sin\frac{\ws}{2}\right)^{-3}.
 \label{eq:SF_AB_k0}\tag{58}
\end{align}

The geodesic interval $s(x,x')$, representing the distance between the space-time 
points $x$ and $x'$ along the geodesic connecting them, satisfies
\begin{align}
 \cos \ws =& \frac{\cos\Delta \wt}{\cos\wr \cos\wr'} - 
 \cos \Upsilon \tan \wr \tan \wr',\nonumber\\
 \cos\Upsilon =& \cos\theta \cos \theta' + \sin \theta \sin \theta' \cos \Delta \varphi,
 \label{eq:geom_s_def}\tag{59}
\end{align}
where $\Delta \wt = \wt - \wt'$ and $\Delta \varphi = \varphi - \varphi'$, while 
$\Upsilon $ represents the angle between $\bm{x}$ and $\bm{x}'$. 
The quantity $\slashed{n} = \gamma^\mu n_\mu$ appearing in Equation~\eqref{eq:geom_SF_def}
is written in terms of the tangent at $x$ to the geodesic connecting $x$ and $x'$, 
namely $n_\mu \equiv n_\mu(x,x')= \nabla_\mu s(x,x')$. Its components with respect to 
the tetrad in Equation~\eqref{eq:tetrad} are given by~\cite{Ambrus:2017cow}:
\begin{equation}
 n_\hatt = \frac{\sin \Delta \wt}{\sin \ws\cos \wr}, \qquad 
 n_\hati = -\frac{x_\hati}{r} \frac{\cos\Delta \wt \sin \wr - \cos \Upsilon \sin \wr'
 (1 - \cos\wr)}{\sin \ws \cos\wr\cos \wr'} + 
 \frac{\tan \wr'}{\sin \ws} \frac{x'_\hati}{r'},
 \label{eq:n}\tag{60}
\end{equation}
while the components $n_{\hatt'}$ and $n_{\hati'}$ of the tangent at $x'$ can be 
obtained from the above expressions by performing the change $x^\mu \leftrightarrow x'^\mu$. 
Finally, $\Lambda(x,x')$ represents the bispinor of parallel transport, satisfying 
$\slashed{D} \Lambda(x,x') = 0$. Due to the maximal symmetry of adS, $\Lambda(x, x')$ 
also satisfies~\cite{Muck:1999mh,Ambrus:2017cow}
\begin{align}
 D_{\halpha} \Lambda(x,x') =& \frac{1}{2\ell } \tan\left( \frac{\ws}{2} \right) 
 (n_\halpha + \gamma_\halpha \slashed{n}) \Lambda(x,x'), \nonumber\\
 D_{\halpha'} \Lambda(x,x') =& \frac{1}{2\ell } \tan \left( \frac{\ws}{2} \right) 
 \Lambda(x,x') (n_{\halpha'} + \slashed{n}' \gamma_{\halpha'}),
 \label{eq:D_Lambda}\tag{61}
\end{align}
where $D_{\halpha'} \Lambda(x,x') \equiv e_{\halpha}^\mu(x') \partial_{\mu'} \Lambda(x,x') + 
\Lambda(x,x') \Gamma_\halpha(x')$ denotes the action of the spinor covariant 
derivative on $\Lambda(x,x')$ at $x'$, while $n_{\halpha'} = e_{\halpha}^\mu(x') 
\nabla_{\mu'} s(x,x')$ is the tangent at $x'$ to the geodesic connecting $x$ and $x'$.
A closed form expression for $\Lambda(x,x')$ on adS was found in Ref.~\cite{Ambrus:2017cow}:
\begin{multline}
 \Lambda(x,x') = \frac{\sec\frac{\ws}{2}}{\sqrt{\cos\wr \cos\wr'}} \Bigg[
 \cos\frac{\Delta \wt}{2} \Bigg(
 \cos\frac{\wr}{2} \cos\frac{\wr'}{2}
 + \sin\frac{\wr}{2} \sin\frac{\wr'}{2}
 \frac{\vx \cdot \vgamma}{r} \frac{\vx' \cdot \vgamma}{r'}\Bigg) \\
 + \sin\frac{\Delta \wt}{2} \Bigg(
 \sin\frac{\wr}{2} \cos\frac{\wr'}{2} \frac{\vx \cdot\vgamma}{r} \gamma^\hatt
 +\sin\frac{\wr'}{2} \cos\frac{\wr}{2} \frac{\vx' \cdot \vgamma}{r'} \gamma^\hatt\Bigg)\Bigg],
 \label{eq:Lambda}\tag{62}
\end{multline}
where $\vgamma $ is a vector of Dirac $\gamma $-matrices.
The following expression for $\slashed{n} \Lambda(x,x')$ will prove 
useful in later sections:
\begin{multline}
 \slashed{n} \Lambda(x,x') = \frac{ {\rm cosec} \, \frac{\ws}{2}}{\sqrt{\cos\wr \cos\wr'}} \Bigg[
 \sin\frac{\Delta \wt}{2} \Bigg(
 \cos\frac{\wr}{2} \cos\frac{\wr'}{2}\gamma^\hatt 
 - \sin\frac{\wr}{2} \sin\frac{\wr'}{2}
 \frac{\vx \cdot \vgamma}{r} \frac{\vx' \cdot \vgamma}{r'}\gamma^\hatt \Bigg) \\
 - \cos\frac{\Delta \wt}{2} \Bigg(
 \sin\frac{\wr}{2} \cos\frac{\wr'}{2} \frac{\vx \cdot \vgamma}{r} -
 \cos\frac{\wr}{2} \sin\frac{\wr'}{2} \frac{\vx' \cdot \vgamma}{r'}
 \Bigg)\Bigg].\label{eq:nLambda}\tag{63}
\end{multline}

\subsection{Thermal Two-Point Function for Rigidly-Rotating States}
\label{sec:SF:beta}

The construction of the propagator for rigid rotation on Minkowski space was discussed previously, for example in Refs.~\cite{Vilenkin:1980zv,Chernodub:2020qah,Ayala:2021osy}, based on a mode sum approach. In this paper, we seek to take advantage of the exact expression for the maximally symmetric vacuum propagator, following the geometric method introduced in Refs.~\cite{Brown:1969na,Birrell:1982ix} and applied for static (nonrotating) adS in Ref.~\cite{Ambrus:2017cow}. 
In this approach, the propagator at finite temperature is obtained via a sum over vacuum propagators evaluated on points which are displaced along the thermal contour towards imaginary times. 

In this section, we discuss the extension of the geometric method to the situation of states at finite temperature undergoing rigid rotation. We construct such states as ensemble 
averages with respect to the weight function $\hat{\rho}$~\cite{Vilenkin:1980zv,Kapusta:1989bd,Becattini:2012tc,Panerai:2015xlr} (not to be confused with the effective transverse coordinate $\wrho$ defined in Equation~(\ref{eq:wOrhoz})). As discussed in Refs.~\cite{Becattini:2012tc,Panerai:2015xlr}, $\hat{\rho}$ can be derived in the frame of covariant statistical mechanics by enforcing the maximisation of the von Neumann entropy $-{\rm tr}(\hat{\rho} \ln \hat{\rho})$ under the constraints of fixed, constant mean energy and total angular momentum~\cite{Becattini:2012tc} and has the form
\begin{equation}
 \hat{\rho} = \exp\left[-\beta_0 (\widehat{H} - \Omega \widehat{M}^z)\right],
 \label{eq:rho}\tag{64}
\end{equation}
where $\widehat{H}$ is the Hamiltonian operator and $\widehat{M}^z$ is the 
total angular momentum along the $z$-axis. For simplicity, we consider only the case of vanishing chemical potential, $\mu = 0$. We use the hat 
to denote an operator acting on Fock space.
The operators $\widehat{H}$ and $\widehat{M}^z$ commute with each other
and are associated to the SO(2,3) isometry group of adS. 
As shown in Ref.~\cite{Cotaescu:2000jp}, these operators have the usual form 
(hats are absent from the expressions below because these are the forms of the 
operators before second quantisation, that is, the operators acting on wavefunctions):
\begin{align}\tag{65}
 H =& i \partial_t, &
 M^z =& -i \partial_\varphi + S^z.
\end{align}
For clarity, in this section we work with the dimensionful quantities $t$ and $r$ given in Equation~(\ref{eq:dimensionful}). The spin matrix $S^z$ appearing above is given by
\begin{equation}
 S^z = \frac{i}{2} \gamma^\hatx \gamma^\haty = \frac{1}{2} 
\begin{pmatrix}
 \sigma^z & 0 \\ 0 & \sigma^z
\end{pmatrix}.
 \label{eq:Sz}\tag{66}
\end{equation}
The t.e.v.~of an operator $\widehat{A}$ is 
computed via~\cite{Kapusta:1989bd,Laine:2016hma,Mallik:2016anp}
\begin{equation}
 \braket{\widehat{A}}_{\beta_0,\Omega} = Z_{\beta_0,\Omega}^{-1} {\rm tr}(\hat{\rho} \widehat{A}),
 \label{eq:tev_def}\tag{67}
\end{equation}
where $Z_{\beta_0,\Omega} = {\rm tr}(\hat{\rho})$ is the partition function.

We now consider an expansion of the field operator $\widehat{\Psi}$ with respect 
to a complete set of particle and antiparticle modes, $U_j$ and $V_j = i \gamma^\haty U_j^*$,
\begin{equation}
 \widehat{\Psi}(x) = \sum_j[\hat{b}_j U_j(x) + \hat{d}^\dagger_j V_j(x)],
 \label{eq:Psi_j}\tag{68}
\end{equation}
where the index $j$ is used to distinguish between solutions at the level of the eigenvalues 
of a complete system of commuting operators (CSCO), which contains also $H$ and $M^z$.
In particular, $U_j$ and $V_j$ satisfy the eigenvalue equations
\begin{align}
 H U_j =& E_j U_j, & H V_j =& -E_j V_j,\nonumber\\
 M^z U_j =& m_j U_j, & M^z V_j =& -m_j V_j,
 \label{eq:eigenj}\tag{69}
\end{align}
where the azimuthal quantum number $m_j = \pm \frac{1}{2}, \pm \frac{3}{2}, \dots$ is 
an odd half-integer, while the energy $E_j > 0$ is assumed to be positive for all modes 
in order to preserve the maximal symmetry of the ensuing vacuum state $\ket{0}$. 
These eigenvalue equations are satisfied automatically by the following four-spinors:
\begin{align}
 U_j(x) =& \frac{1}{2\pi} e^{-i E_j t + i m_j \varphi - i S^z \varphi} u_j(r,\theta), \nonumber\\
 V_j(x) =& \frac{1}{2\pi} e^{i E_j t - i m_j \varphi - i S^z \varphi} v_j(r,\theta),\tag{70}
\end{align}
where the four-spinors $u_j$ and $v_j$ do not depend on $t$ or $\varphi$. This 
allows $\widehat{\Psi}$ to be written as
\begin{equation}
 \widehat{\Psi}(x) = \frac{e^{-i S^z \varphi}}{2\pi} 
 \sum_j \left(e^{-iE_j t + i m_j \varphi} \hat{b}_j u_j + e^{iE_j t - im_j \varphi} \hat{d}^\dagger_j v_j\right).
 \label{eq:Psi_Sz}\tag{71}
\end{equation}

The one-particle operators in Equation~\eqref{eq:Psi_j} are assumed to satisfy canonical
anticommutation relations,
\begin{equation}
 \{\hat{b}_j, \hat{b}^\dagger_{j'}\} = \{\hat{d}_j, \hat{d}^\dagger_{j'}\} = \delta(j,j'),
 \label{eq:acommj}\tag{72}
\end{equation}
with all other anticommutators vanishing. The eigenvalue equations in \eqref{eq:eigenj}
imply
\begin{align}
 [\widehat{H}, \hat{b}_j^\dagger]  =& E_j \hat{b}_j^\dagger, & 
 [\widehat{H}, \hat{d}_j^\dagger] =& E_j \hat{d}_j^\dagger, \nonumber\\
 [\widehat{M}^z, \hat{b}_j^\dagger] =& m_j \hat{b}_j^\dagger, & 
 [\widehat{M}^z, \hat{d}_j^\dagger] =& m_j \hat{d}_j^\dagger,
 \label{eq:commj}\tag{73}
\end{align}
so that
\begin{equation}
 \hat{\rho} \hat{b}_j \hat{\rho}^{-1} = e^{\beta_0 \widetilde{E}_j} \hat{b}_j,\qquad
 \hat{\rho} \hat{d}_j^\dagger \hat{\rho}^{-1} = e^{-\beta_0 \widetilde{E}_j} \hat{d}_j^\dagger,
 \label{eq:heisj}\tag{74}
\end{equation}
where the corotating energy is defined via 
\begin{equation}
 \widetilde{E}_j = E_j - \Omega m_j.
 \label{eq:Et}\tag{75}
\end{equation}
Noting that
\begin{align}
 e^{\beta_0 \widetilde{E}_j} U_j(t,\varphi) =& 
 e^{i \beta_0 \partial_t - \beta_0 \Omega(-i \partial_\varphi + S^z)} U_j(t,\varphi) \nonumber\\
 =& e^{-\beta_0 \Omega S^z} U_j(t + i \beta_0 , \varphi + i \beta_0 \Omega),\nonumber\\
 e^{-\beta_0 \widetilde{E}_j} V_j(t,\varphi) =& 
 e^{-\beta_0 \Omega S^z} V_j(t + i \beta_0 , \varphi + i \beta_0 \Omega),
 \label{eq:heis_modesj}\tag{76}
\end{align}
it can be seen that
\begin{equation}
 \hat{\rho} \widehat{\Psi}(t, \varphi) \hat{\rho}^{-1} = 
 e^{-\beta_0 \Omega S^z} \widehat{\Psi}(t + i \beta_0, \varphi + i \beta_0 \Omega),
 \label{eq:heis_Psi}\tag{77}
\end{equation}
where 
\begin{equation}
 e^{-\beta_0 \Omega S^z} = \cosh\left(\frac{\beta_0 \Omega}{2}\right) - 
 2 \sinh\left(\frac{\beta_0 \Omega}{2}\right) S^z.
 \label{eq:expS}\tag{78}
\end{equation}

We now introduce the two-point functions~\cite{Mallik:2016anp}
\begin{align}
 iS^+_{\beta_0,\Omega}(x,x') =& \braket{\widehat{\Psi}(x) \widehat{\overline{\Psi}}(x')}_{\beta_0,\Omega}, &
 iS^-_{\beta_0,\Omega}(x,x') =& -\braket{\widehat{\overline{\Psi}}(x') \widehat{\Psi}(x)}_{\beta_0, \Omega}.\tag{79}
\end{align}
Taking into account Equation~\eqref{eq:heis_Psi}, it is possible to derive the 
KMS relation for thermal states with rotation:
\begin{align}
 S^-_{\beta_0,\Omega}(\tau, \varphi; x') =& 
 i Z_{\beta_0,\Omega}^{-1} {\rm tr}[\hat{\rho} \widehat{\overline{\Psi}}(\tau',\varphi') \widehat{\Psi}(\tau,\varphi)]\nonumber\\
 =& i Z_{\beta_0,\Omega}^{-1} {\rm tr}[\hat{\rho} 
 e^{\beta_0 \Omega S^z} \widehat{\Psi}(\tau - i \beta_0, \varphi - i \beta_0 \Omega)
 \widehat{\overline{\Psi}}(\tau',\varphi')]\nonumber\\
 =& -e^{\beta_0 \Omega S^z} S^+_{\beta_0,\Omega}(\tau - i \beta_0, \varphi - i \beta_0 \Omega; x'),
 \label{eq:KMS}\tag{80}
\end{align}
where the time coordinate was denoted by $\tau$ to indicate that time is an inherently complex 
parameter when finite temperature states are considered. In the above and in what follows, 
the dependence on the coordinates $r$ and $\theta$ is omitted for brevity.
The factor $e^{-i S^z \varphi}$ in Equation~\eqref{eq:Psi_Sz} indicates that 
the two-point functions can be expanded in a double Fourier series with 
respect to $\Delta \tau = \tau - \tau'$ and $\Delta \varphi = \varphi - \varphi'$ as follows:
\begin{equation}
 S^\pm_{\beta_0,\Omega}(\tau, \varphi; \tau', \varphi') = 
 \int_{-\infty}^\infty \frac{dw}{(2\pi)^2} \sum_m 
 e^{-i w\Delta \tau + i m \Delta \varphi}
 e^{-i S^z \varphi} s^\pm_{\beta_0, \Omega; m}(w) e^{i S^z \varphi'},
 \label{eq:spm_def}\tag{81}
\end{equation}
where $m = \pm \frac{1}{2}, \pm \frac{3}{2}, \dots$ and the Fourier coefficients $s^\pm_{\beta_0, \Omega; m}(w)$ are independent of $\Delta \tau$ and $\Delta \varphi$.
By virtue of Equation~\eqref{eq:KMS},
the matrix-valued Fourier coefficients $s^\pm_{\beta_0, \Omega;m}(w)$ can be shown to satisfy~\cite{Birrell:1982ix,Mallik:2016anp}
\begin{equation}
 s^\pm_{\beta_0, \Omega;m}(w) = -e^{\pm \beta_0 {\widetilde {w}}}
 s^\mp_{\beta_0, \Omega;m}(w),
 \label{eq:KSM_s}\tag{82}
\end{equation}
where ${\widetilde {w}} = w - m \Omega$.

We now consider the Schwinger (anticommutator) two-point function,
\begin{equation}
 S(x,x') = S^+_{\beta_0,\Omega}(x,x') - S^-_{\beta_0,\Omega}(x,x'),
 \label{eq:Schwinger_def}\tag{83}
\end{equation}
which is independent of state, since it involves the field anticommutators.
Introducing its Fourier transform, $s_m(w)$, through a relation equivalent 
to that in Equation~\eqref{eq:spm_def}, we find
\begin{align}
 s^+_{\beta_0, \Omega; m}(w) =& [1 - n_{\beta_0}({\widetilde {w}})] s_m(w), &
 s^-_{\beta_0,\Omega; m}(w) =& - n_{\beta_0}({\widetilde {w}}) s_m(w),\label{eq:spm_s}\tag{84}
\end{align}
where the Fermi-Dirac factor $n_{\beta_0}({\widetilde {w}})$ is given by
\begin{equation}
 n_{\beta_0}({\widetilde {w}}) = \frac{1}{e^{\beta_0 {\widetilde {w}}} + 1}.\tag{85}
\end{equation}
At vanishing temperature, the following limits can be obtained:
\begin{equation}
 s_{\infty,\Omega; m}^\pm(w) = \pm
 \Theta(\pm {\widetilde {w}}) s_m(w),\tag{86}
\end{equation}
where $\Theta $ is the usual Heaviside step function. 

We now move onto the thermal Feynman two-point function, defined by
\begin{equation}
 S^F_{\beta_0,\Omega}(x,x') = \Theta_c(\tau - \tau') S^+_{\beta_0,\Omega}(x,x') 
  + \Theta_c(\tau' - \tau) S^-_{\beta_0,\Omega}(x,x'),\tag{87}
\end{equation}
where $\Theta_c(\tau - \tau')$ is the step function on a contour in the complex plane which 
causally descends towards negative values of the imaginary part of $\tau - \tau'$,
such that 
\begin{equation}
 \Theta_c(t - t' - i \varepsilon) = 1,\qquad
 \Theta_c(t - t' + i \varepsilon) = 0, 
 \label{eq:thetac}\tag{88}
\end{equation}
for any real $t-t'$ and $\varepsilon > 0$~\cite{Mallik:2016anp}. 
Replacing the functions $S^\pm_{\beta_0, \Omega}(x,x')$ with their Fourier representation,
given in Equation~\eqref{eq:spm_def}, and using Equation~\eqref{eq:spm_s} to 
replace their Fourier coefficients, we~obtain 
\begin{multline}
 S^F_{\beta_0,\Omega}(x,x') = \int_{-\infty}^\infty \frac{dw}{(2\pi)^2} 
 \sum_m e^{-iw \Delta \tau + i m \Delta \varphi}
 \{\Theta_c(\Delta \tau)[1 - n_{\beta_0}({\widetilde {w}})] 
 - \Theta_c(-\Delta \tau) n_{\beta_0}({\widetilde {w}})\} \\ \times
 e^{-iS^z \varphi} s_m(w) e^{i S^z \varphi'}.
 \label{eq:SF_beta_aux}\tag{89}
\end{multline}
Noting that the Fermi-Dirac factors $n_{\beta_0}({\widetilde {w}})$ and 
$1 - n_{\beta_0}({\widetilde {w}})$ admit the following expansions,
\begin{align}
 n_{\beta_0}({\widetilde {w}}) =& \Theta(-{\widetilde {w}}) - \sum_{j \neq 0}
 (-1)^j e^{-j\beta_0 {\widetilde {w}}} [\Theta(j) \Theta({\widetilde {w}}) - \Theta(-j) \Theta(-{\widetilde {w}})], \nonumber\\
 1 - n_{\beta_0}({\widetilde {w}}) =& \Theta({\widetilde {w}}) + \sum_{j \neq 0}
 (-1)^j e^{-j\beta_0 {\widetilde {w}}} [\Theta(j) \Theta({\widetilde {w}}) - \Theta(-j) \Theta(-{\widetilde {w}})],
 \label{eq:SF_theta_contour}\tag{90}
\end{align}
where $j = \pm 1, \pm 2, \dots$, it can be seen that 
\begin{multline}
 S^F_{\beta_0,\Omega}(x,x') = \int_{-\infty}^\infty \frac{dw}{(2\pi)^2} 
 \sum_m e^{-iw\Delta \tau + i m \Delta \varphi}
 \{\Theta_c(\Delta \tau) \Theta({\widetilde {w}}) - \Theta_c(-\Delta \tau) \Theta(-{\widetilde {w}}) \\
 + \sum_{j \neq 0} (-1)^j e^{-j \beta_0 {\widetilde {w}}} [\Theta(j) \Theta({\widetilde {w}}) - \Theta(-j) \Theta(-{\widetilde {w}})]\}
 e^{-iS^z \varphi} s_m(w) e^{i S^z \varphi'}. 
 \label{eq:SF_beta_aux2}\tag{91}
\end{multline}
The term on the first line can be identified with the 
Feynman propagator at vanishing temperature, which can be read from Equation~\eqref{eq:SF_beta_aux}:
\begin{multline}
 S^F_{\infty,\Omega}(x,x') = \int_{-\infty}^\infty \frac{dw}{(2\pi)^2} 
 \sum_m e^{-iw \Delta \tau + i m \Delta \varphi}
\\ \times [\Theta_c(\Delta \tau)\Theta({\widetilde {w}}) - \Theta_c(-\Delta \tau) \Theta(-{\widetilde {w}})] 
 e^{-iS^z \varphi} s_m(w) e^{i S^z \varphi'}.
 \label{eq:SF_betainf}\tag{92}
\end{multline}
Shifting the time variables $\tau$ and $\varphi$ from the real axis by the imaginary quantities $i j\beta_0$ and $i j \beta_0 \Omega$ (where $j$ can be either negative or positive) and taking into account the relations in Equation~\eqref{eq:SF_theta_contour}, we have
\begin{multline}
 S^F_{\infty,\Omega}(\tau + i j \beta_0, \varphi + i j \beta_0 \Omega;x') =
 e^{i j \beta_0 \Omega S^z} \int_{-\infty}^\infty \frac{dw}{(2\pi)^2} 
 \sum_m e^{-iw\Delta \tau + i m \Delta \varphi}
 \\ \times 
 e^{j \beta_0 {\widetilde {w}}} [\Theta(-j) \Theta({\widetilde {w}}) - \Theta(j) \Theta(-{\widetilde {w}})]
 e^{-iS^z \varphi} s_m(w) e^{i S^z \varphi'}.\tag{93}
\end{multline}
Performing the flip $j \rightarrow -j$ in Equation~\eqref{eq:SF_beta_aux2}, the following equality can be derived:
\begin{equation}
 S^F_{\beta_0,\Omega}(x,x') = \sum_{j = -\infty}^\infty (-1)^j e^{-j \beta_0 \Omega S^z}
 S_{\infty,\Omega}^F(\tau +ij \beta_0, \varphi + i j \beta_0 \Omega; x').
 \label{eq:SF_beta_gen}\tag{94}
\end{equation}
In Equation~\eqref{eq:SF_beta_gen}, it is understood that $\tau$ and $\tau'$ are 
always taken on the real axis, while the step functions appearing in Equation~\eqref{eq:SF_beta_aux} are evaluated according to Equation~\eqref{eq:thetac}.

We now discuss the connection with the vacuum Feynman propagator, $S^F_{\rm vac}(x,x')$,
introduced in Equation~\eqref{eq:geom_SF_def}. It can be seen that $S^F_{\rm vac}(x,x')$
admits a representation similar to that in Equation~\eqref{eq:SF_betainf}, with 
$\Omega$ set equal to zero:
\begin{multline}
 S^F_{\rm vac}(x,x') = \int_{-\infty}^\infty \frac{dw}{(2\pi)^2} 
 \sum_m e^{-iw \Delta \tau + i m \Delta \varphi} 
\\ \times [\Theta_c(\Delta \tau)\Theta(w) - \Theta_c(-\Delta \tau) \Theta(-w)] 
 e^{-iS^z \varphi} s_m(w) e^{i S^z \varphi'}.
 \label{eq:SF_vac}\tag{95}
\end{multline}

The difference between Equations~\eqref{eq:SF_betainf} and \eqref{eq:SF_vac} 
can appear only if the maximally symmetric vacuum, employed to construct 
$S^F_{\rm vac}(x,x')$, differs from the ``rotating'' vacuum. 
As discussed by Iyer in the context of rotating 
states on Minkowski space~\cite{Iyer:1982ah}, such a difference is due to the presence
of quantum particle modes (defined with respect to the Minkowski vacuum) 
for which the corotating energy $\widetilde{E} = E - \Omega m$, defined in 
terms of the particle (Minkowski) energy $E$ and the projection 
$m$ of the particle total angular momentum along the rotation vector, is negative.
The existence of such modes is quite generally intimately related to the presence 
of an SLS. In Ref.~\cite{Nicolaevici:2001yy}, it was shown that
the Unruh-de Witt detector in rigid rotation cannot be excited by a neutral scalar 
field in any space-time, as long as no SLS forms in the (noninertial) reference 
frame of the detector. While such a general result is not available for the 
Dirac field, our analysis of rigidly-rotating states bounded by a cylinder 
indicates that the modes with $\widetilde{E} < 0$ are not permitted, as long as 
the SLS is excluded from within the space contained inside the boundary~\cite{Ambrus:2016ocv}.
This is also confirmed for the bounded rigidly-rotating states of the Klein-Gordon 
field~\cite{Duffy:2002ss,Ambrus:2019vkx}. For the specific case of adS, we expect 
that $\widetilde{E} > 0$ as long as $|\wO| \le 1$, due to the quantisation of energy on adS~\cite{Cotaescu:1998ts}, which reads
\begin{equation}\tag{96}
 E_{n} = M + \ell^{-1}\left(n + \frac{3}{2}\right), \qquad n = 0, 1, 2 \dots.
\end{equation}
In the above, the main quantum number $n = 2n_r + l$ can be written as the sum of the radial quantum number $n_r = 0, 1, 2, \dots$ and the auxiliary quantum number $l$, taking even or odd integer values when $n$ is even or odd. The total angular momentum quantum number $j = l \pm \frac{1}{2}$ can be obtained from $l$, while the magnetic quantum number $m$ satisfies $-j \le m \le j$. Thus, the co-rotating energy is
\begin{equation}\tag{97}
 \widetilde{E}_{n_r, j} = M + \ell^{-1} \left(2n_r + \frac{3}{2} \pm \frac{1}{2}\right) + \ell^{-1}(j - \wO m).
\end{equation}
Since $j \ge m$, it is clear that $\widetilde{E}_{n_r, j} > 0$ as long as $|\wO| \le 1$.

To summarise, the above discussion shows that, when $|\wO| \le 1$, 
the Feynman propagator at vanishing temperature and finite rotation,
$S^F_{\infty,\Omega}(x,x')$, can be replaced by the maximally symmetric
Feynman propagator, $S^F_{\rm vac}(x,x')$, as follows:
\begin{equation}\tag{98}
 S^F_{\beta_0,\Omega}(x,x') = \sum_{j = -\infty}^\infty (-1)^j e^{-j \beta_0 \Omega S^z}
 S_{\rm vac}^F(\tau +ij \beta_0, \varphi + i j \beta_0 \Omega; x').
 \label{eq:SF_beta}
\end{equation}

\subsection{Thermal Expectation Values}
\label{sec:tev}

In the rest of this paper, we consider the t.e.v.s of the scalar condensate (SC), the pseudoscalar condensate
(PC), the vector ($J^\mu_V$) (VC) and axial ($J^\mu_A$)  (AC) charge currents, 
as well as the stress-energy tensor ($T^{\mu\nu}$) (SET)~\cite{Groves:2002mh}:
\begin{align}
 \begin{pmatrix}
  SC \\
  PC \\
  J_V^\halpha \\
  J_A^\halpha \\
  T_{\halpha\hsigma}
 \end{pmatrix}  & \equiv 
 \begin{pmatrix}
  \frac{1}{2}\braket{:[\widehat{\overline{\Psi}}, \widehat{\Psi}]:}_{\beta_0,\Omega} \\
  \frac{1}{2}\braket{:[\widehat{\overline{\Psi}}, -i\gamma^5 \widehat{\Psi}]:}_{\beta_0,\Omega} \\
  \braket{:\widehat{J}_V^\halpha:}_{\beta_0,\Omega} \\
  \braket{:\widehat{J}_A^\halpha:}_{\beta_0,\Omega} \\
  \braket{:\widehat{T}_{\halpha\hsigma}:}_{\beta_0,\Omega}
 \end{pmatrix}
 \nonumber \\ & = 
 -\lim_{x' \rightarrow x} {\rm tr} \left[
 \begin{pmatrix}
  1 \\ -i\gamma^5 \\ \gamma^\halpha \\ \gamma^\halpha \gamma^5 \\ 
  -\frac{i}{2}[\gamma_{(\halpha} (D_{\hsigma)} - g^{\hsigma'}{}_\hsigma D_{\hsigma')})]
 \end{pmatrix}
 i\Delta S^F_{\beta_0, \Omega}(x,x') \Lambda(x',x)\right],
 \label{eq:tevs_gen}\tag{99}
\end{align}
where $\Delta S^F_{\beta_0,\Omega}(x,x') \equiv S^F_{\beta_0, \Omega}(x,x') - S^F_{\rm vac}(x,x')$ is the 
thermal propagator with the vacuum part subtracted. 
The bispinor of parallel transport $\Lambda(x',x)$ 
has the role of parallel transporting the spinor structure at point $x'$ towards $x$ as the coincidence
limit is taken~\cite{Groves:2002mh} and is not acted upon by the derivative operators. Furthermore, 
since $\Lambda(x,x)$ is just the identity matrix, this term makes only trivial contributions to the 
t.e.v.s considered here and can thus be ignored in what follows.
Upon replacing the thermal propagator using Equation~\eqref{eq:SF_beta}, it can be seen that the t.e.v.s 
appearing in Equation~\eqref{eq:tevs_gen} involve taking traces and derivatives of the vacuum propagator,
when the point $x$ is detached from the real axis along the thermal contour. Let us 
denote by $x_j^\halpha = (t_j, x_j, y_j, z_j)$ the Cartesian-like coordinates corresponding 
to the $j$'th term in Equation~\eqref{eq:SF_beta}. While 
$t_j = t + i j \beta_0$ is obtained via a translation, the spatial coordinates are 
obtained via a rotation:
\begin{equation}
 x_j^\hati = R_z(i \Omega j \beta_0)^\hati{}_\hatl x^\hatl, \qquad 
 R_z(i \Omega j \beta_0)^\halpha{}_\hsigma = 
 \begin{pmatrix}
  1 & 0 & 0 & 0\\
  0 & \cosh \Omega j \beta_0 & -i \sinh \Omega j \beta_0 & 0 \\
  0 & i \sinh \Omega j \beta_0 & \cosh \Omega j \beta_0 & 0 \\
  0 & 0 & 0 & 1
 \end{pmatrix},\tag{100}
\end{equation}
where the temporal line and column were added for future convenience.
It is easy to see that the derivatives with respect to $x^\halpha$ can be 
transformed into derivatives $\partial_{\hsigma;j}$ with respect to the rotated coordinates,
$x^\halpha_j$, as follows:
\begin{equation}\tag{101}
 \partial_\halpha = R_z(i \Omega j \beta_0)^\hsigma{}_\halpha \partial_{\hsigma;j}.
\end{equation}
A similar transformation law can be found for the $\gamma^\halpha$ matrices:
\begin{equation}\tag{102}
 e^{\Omega j \beta_0 S^\hatz} \gamma_\halpha e^{-\Omega j \beta_0 S^\hatz} = 
 \gamma_\hsigma R_z(i \Omega j \beta_0)^\hsigma{}_\halpha.
\end{equation}
Further noting that 
\begin{equation}\tag{103}
 e^{j\beta_0 \Omega S^\hatz} \left(\frac{\bm{x} \cdot \bm{\gamma}}{r}\right) 
 e^{-j\beta_0 \Omega S^\hatz} = \frac{\bm{x}_j \cdot \bm{\gamma}}{r},
\end{equation}
it can be seen that the covariant derivative 
$D_\halpha = e_\halpha^\hbeta(x) (\partial / \partial x^\hbeta) - \Gamma_\halpha(x)$ 
also transforms according to
\begin{equation}\tag{104}
 e^{\Omega j \beta_0 S^\hatz} D_\halpha e^{-\Omega j \beta_0 S^\hatz} = 
 D_{\hsigma;j} R_z(i \Omega j \beta_0)^\hsigma{}_\halpha,
 \label{eq:D_transf}
\end{equation}
where $D_{\hsigma;j} = e_\hsigma^\hbeta(x_j) (\partial / \partial x_j^\hbeta) - \Gamma_\hsigma(x_j)$ 
is the covariant derivative acting at the point $x^\halpha_j$ on the thermal contour.

Due to the relation between the thermal Feynman propagator and the vacuum one given in Equation~\eqref{eq:SF_beta}, 
it is possible to write the t.e.v. of a normal-ordered operator $:\widehat{A}:$ as $A = \sum_{j \neq 0} A_j$. 
Leaving the case of the SET for later, the following terms are uncovered 
for the SC, PC and charge currents:
\begin{align}
 SC_j =& (-1)^{j+1} \mathcal{A}_{F;j} {\rm tr}(e^{-j \beta_0 \Omega S^z}\Lambda_j), \nonumber\\
 PC_j =& -i (-1)^{j+1} \mathcal{A}_{F;j} {\rm tr}(e^{-j \beta_0 \Omega S^z} \gamma^5 \Lambda_j), \nonumber\\
 J^\halpha_{V;j} =& (-1)^{j+1} R_z(i \Omega j \beta_0)_\hsigma{}^\halpha 
 \mathcal{B}_{F;j} {\rm tr}(e^{-j \beta_0 \Omega S^z} \gamma^\hsigma \slashed{n}_j \Lambda_j), \nonumber\\
 J^\halpha_{A;j} =& (-1)^{j+1} R_z(i \Omega j \beta_0)_\hsigma{}^\halpha 
 \mathcal{B}_{F;j} {\rm tr}(e^{-j \beta_0 \Omega S^z} \gamma^\hsigma \gamma^5 \slashed{n}_j \Lambda_j), 
 \label{eq:jterms}\tag{105}
\end{align}
where $\mathcal{A}_{F;j} \equiv \mathcal{A}_F(s_j)$ and 
$\mathcal{B}_{F;j} \equiv \mathcal{B}_F(s_j)$ depend on the geodesic distance $s_j \equiv s(t + i j \beta_0, r, \varphi + i j \beta_0 \Omega, \theta ; t, r, \varphi , \theta )$ along the imaginary 
contour,  which satisfies:
\begin{align}
 \cos \frac{s_j}{\ell } =& 1 + \frac{2}{\Gamma_j^2 \cos^2\wr} \sinh^2 \frac{j \beta_0}{2\ell },\tag{106}
\end{align}
where $\wrho $ was introduced in Equation~\eqref{eq:wOrhoz} and we introduce here the notations 
$\Gamma_j$ and $\wO_j$ by
\begin{equation}
 \Gamma_j = \frac{1}{\sqrt{1 - \wrho^2 \wO_j^2}}, \qquad 
 \wO_j = \frac{\sinh(\Omega j \beta_0 / 2)}{\sinh(j \beta_0 / 2\ell)}.
 \label{eq:Oj_def}\tag{107}
\end{equation}
We have also defined the quantities
$\slashed{n}_j$ and $\Lambda_j$,  which are, respectively, the tangent at the point $x_{j}=(t+ij\beta _{0},r,\varphi + ij\beta _{0}\Omega ,\theta )$ and the bispinor of parallel transport between the points $x_{j}$ and $x'=(t,r, \varphi , \theta )$. 

\section{Scalar and Pseudoscalar Condensates}
\label{sec:SCPC}

We begin our detailed study of t.e.v.s by considering first the simplest ones, namely the SC and PC. 
The $j$th terms appearing in Equation~(\ref{eq:jterms}) for the SC and PC are
\begin{align}
 SC_j =& (-1)^{j+1} \mathcal{A}_{F;j} 
 \left[ 
 \cosh \frac{\Omega j \beta_0}{2} {\rm tr}(\Lambda_j) - 
 \sinh\frac{\Omega j \beta_0}{2} {\rm tr}(\gamma^5 \gamma^\hatt \gamma^\hatz \Lambda_j)\right],\nonumber\\
 PC_j =& -i(-1)^{j+1} \mathcal{A}_{F;j} 
 \left[ 
 \cosh \frac{\Omega j \beta_0}{2} {\rm tr}(\gamma^5 \Lambda_j) - 
 \sinh\frac{\Omega j \beta_0}{2} {\rm tr}(\gamma^\hatt \gamma^\hatz \Lambda_j)\right],
 \label{eq:SCPCj_aux}\tag{108}
\end{align}
where the relations $e^{-\Omega j \beta_0 S^\hatz} = \cosh \frac{\Omega j \beta_0}{2} - 
2 S^\hatz \sinh\frac{\Omega j \beta_0}{2}$ and 
$2S^\hatz = \gamma^5 \gamma^\hatt \gamma^\hatz$ were employed.
Specialising the results in Equation~\eqref{eq:tr_FC} for the traces appearing above
to $(x,x') \rightarrow (x_j, x')$, Equation~\eqref{eq:SCPCj_aux} simplifies to
\begin{align}
 SC_j =& \frac{4 (-1)^{j+1} \mathcal{A}_{F;j}}{\cos(\ws_j / 2)} \cosh \frac{j \beta_0}{2\ell } \cosh \frac{\Omega j \beta_0}{2}, \nonumber\\
 PC_j =& \frac{4 \wz (-1)^{j+1} \mathcal{A}_{F;j}}{\cos(\ws_j / 2)}
 \sinh \frac{j \beta_0}{2\ell } \sinh \frac{\Omega j \beta_0}{2},
 \label{eq:SCPC}\tag{109}
\end{align}
where the effective vertical coordinate $\wz = \tan\wr \cos\theta$ was introduced 
in Equation~\eqref{eq:wOrhoz}. In the above, $j \neq 0$ takes both positive and negative values.
It can be seen that the SC persists in the absence of rotation (as also remarked in Ref.~\cite{Ambrus:2017cow}),
while the PC only forms at nonvanishing $\Omega$. We now introduce the following notation:
\begin{equation}
 \zeta_j = -\frac{1}{\sin^2 \frac{\ws_j}{2}} = \frac{\Gamma_j^2 \cos^2 \wr}{\sinh^2(j \beta_0 / 2\ell)},
 \label{eq:zetaj_def}\tag{110}
\end{equation}
where $\Gamma_j$ was defined in Equation~\eqref{eq:Oj_def}. At large temperature, $\zeta_j$ can be expanded as shown in Equation~\eqref{eq:zetaj_largeT}.
Using Equation~\eqref{eq:SF_AB_gen} to replace $\mathcal{A}_{F;j}$ in Equation~\eqref{eq:SCPC}, we find
\begin{align}
 SC =& \frac{\Gamma_k}{2\pi^2\ell ^{3}} 
 \sum_{j = 1}^\infty (-1)^{j+1} \zeta_j^{2+k} \cosh \frac{j \beta_0}{2\ell } \cosh \frac{\Omega j \beta_0}{2}
 {}_2F_1(1 + k, 2 + k; 1 + 2k; -\zeta_j),\nonumber\\
 PC =& \frac{\wz\, \Gamma_k}{2\pi^2\ell ^{3}}
 \sum_{j = 1}^\infty (-1)^{j+1} \zeta_j^{2+k} \sinh \frac{ j \beta_0}{2\ell } \sinh \frac{\Omega j \beta_0}{2}
 {}_2F_1(1 + k, 2 + k; 1 + 2k; -\zeta_j).
 \label{eq:SCPC_k}\tag{111}
\end{align}

In the limit of critical rotation, when $\wO \rightarrow 1$, 
the quantity $\zeta_j$ introduced in \mbox{Equation~\eqref{eq:zetaj_def}} becomes
\begin{equation}\tag{112}
 \lim_{\wO \rightarrow 1} \zeta_j = \frac{\Gamma^2 \cos^2\wr}{\sinh^2 \frac{j \beta_0}{2\ell }}.
 \label{eq:zetaj_crit}
\end{equation}
In the equatorial plane, $\zeta_j$ becomes coordinate-independent, since $\Gamma = 1 / \cos\wr$ 
when $\theta = \pi / 2$. While the PC trivially vanishes in the equatorial plane due to the 
$\wz$ prefactor, the SC attains a constant value:
\vspace{+12pt}
\end{paracol}
\nointerlineskip
\begin{equation}
 \lim_{\wO \rightarrow 1} SC\left(\theta = \frac{\pi}{2}\right) = 
 \frac{\Gamma_k}{2\pi^2\ell ^{3}} \sum_{j = 1}^\infty 
 \frac{(-1)^{j+1} \cosh^2\frac{j \beta_0}{2\ell }}{(\sinh \frac{ j \beta_0}{2\ell })^{4+2k}}
 {}_2 F_1\left(1 + k, 2+k; 1+2k; -\sinh^{-2}\frac{j \beta_0}{2\ell }\right).
 \label{eq:SC_crit}\tag{113}
\end{equation}
\begin{paracol}{2}
\switchcolumn

We now discuss the massless limit, in which Equation~\eqref{eq:SCPC_k} reduces to:
\begin{align}
 \lim_{k \rightarrow 0} SC =& \frac{\cos^4\wr}{2\pi^2\ell ^{3}} \sum_{j = 1}^\infty 
 \frac{(-1)^{j+1} \cosh \frac{ j \beta_0}{2\ell } \cosh \frac{\Omega j \beta_0}{2}}
 {(\sinh^2 \frac{j \beta_0}{2\ell } + \cos^2 \wr - 
 \wrho^2 \sinh^2\frac{\Omega j \beta_0}{2})^2},\nonumber\\
 \lim_{k \rightarrow 0} PC =& \frac{\cos^4\wr}{2\pi^2\ell ^{3}}
 \wz \sum_{j = 1}^\infty
 \frac{(-1)^{j+1} \sinh \frac{j \beta_0}{2\ell } \sinh \frac{\Omega j \beta_0}{2}}
 {(\sinh^2 \frac{ j \beta_0}{2\ell } + \cos^2 \wr - 
 \wrho^2 \sinh^2\frac{\Omega j \beta_0}{2})^2}.\tag{114}
\end{align}
In the high temperature limit, it can be seen that 
\begin{equation}
 \lim_{k \rightarrow 0} SC = \frac{1}{4\pi^2\ell ^{3}}, \qquad 
 \lim_{k \rightarrow 0} PC = 0.
 \label{eq:SC_highT_m0}\tag{115}
\end{equation}
The above result reveals a nonvanishing value of the SC in the large temperature limit, which is present even for massless fermions. It is interesting to note that the result in Equation~\eqref{eq:SC_highT_m0} is exactly cancelled by the renormalised vacuum expectation value (v.e.v.) of the SC~\cite{Ambrus:2015mfa} (note that
the result in Ref.~\cite{Ambrus:2015mfa} must be multiplied by $-1$):
\begin{equation}\tag{116}
 \lim_{k \rightarrow 0} \braket{SC}_{\rm vac} = -\frac{1}{4\pi^2\ell ^{3}}.
\end{equation}
At finite mass, the SC can be expected to receive extra temperature-dependent contributions. According to the analysis on Minkowski space~\cite{Ambrus:2014uqa,Ambrus:2019ayb,Ambrus:2019cvr}, the t.e.v. of the SC for small masses $M = k\ell ^{-1}$ is given by
\begin{equation}\tag{117}
 SC_{\rm Mink} = M \left[\frac{T^2}{6} + \frac{3\bm{\omega}^2 - \bm{a}^2}{24\pi^2} - \frac{M^2}{2\pi^2} \ln \frac{\pi T}{M e^{{\mathcal {C}} - \frac{1}{2}}} + O(T^{-1})\right],
 \label{eq:SC_Mink}
\end{equation}
where the logarithmic term is based on the classical result for 
$(E_{\rm RKT} - 3P_{\rm RKT})/M$ in \mbox{Equation~\eqref{eq:RKT_highT}.}
We now seek to obtain this limit starting from \mbox{Equation~\eqref{eq:SCPC_k}}.
Using the Formula \eqref{eq:hyp_largez}, the hypergeometric function in Equation~\eqref{eq:SCPC_k} can be expanded around $\zeta_j^{-1} = 0$ as follows
\begin{multline}
 {}_2F_1(1 + k, 2 + k; 1+ 2k; -\zeta_j) = \frac{ \zeta_j^{-1-k}}{\Gamma_k}\Bigg\{k + 
 \zeta_j^{-1} \Bigg[-1 - k + k^2 - k^3 \\
 - 2k(1 - k^2)\left(\ln \zeta_j^{-1/2} + \psi(k) + {\mathcal {C}} \right)\Bigg] + O(\zeta_j^{-2})\Bigg\},\tag{118}
\end{multline}
where the normalisation constant $\Gamma_k$ introduced in Equation~\eqref{eq:Gammak} emerges after using the properties in Equation~\eqref{eq:digamma}.
Substituting the above result into Equation~\eqref{eq:SCPC_k}, the sum over $j$ can be performed by first considering an expansion at large temperatures, as shown in Equation~\eqref{eq:zetaj_largeT} for $\zeta_j$. Using the summation formulae in Equation~\eqref{eq:SC_summation},
the large temperature expansion of the SC can be obtained as follows:
\begin{multline}
 SC = \frac{M T^2}{6} - \frac{M}{2\pi^2} \left(M^2 + \frac{R}{12}\right) \ln \pi T \ell  + \frac{M}{24\pi^2}(3\bm{\omega}^2 - \bm{a}^2)\\ 
 + \frac{1}{4\pi^2\ell ^{3}} \left\{1 - \frac{5k}{6} - k^2 - k^3 - 2k(1 - k^2) [\psi(1+k) + {\mathcal {C}}]\right\}  + O(T^{-1}).
 \label{eq:SC_highT}\tag{119}
\end{multline}
A comparison with Equation~\eqref{eq:RKT_highT} confirms that the leading order term $MT^2 /6$ is consistent with that from the classical result $(E_{\rm RKT} - 3P_{\rm RKT})/M$. The logarithmic term receives a quantum correction due to the Ricci scalar term, $R / 12$, which seems to be consistent with the replacement $M^2 \rightarrow M^2 + \frac{R}{12}$ suggested in Equation~(\hl{7}) of Ref.~\cite{Flachi:2014jra}.

The result in Equation~\eqref{eq:SC_highT} is validated against a numerical computation based on Equation~\eqref{eq:SCPC_k} in Figure~\ref{fig:SC_vs_r}, where the profiles of the SC in the equatorial plane are shown at various values of the parameters $\Omega$, $k$ and $T_0$. Panel (a) confirms the high temperature limit correpsonding to massless quanta derived in Equation~\eqref{eq:SC_highT_m0} at $T_0 = 2\ell^{-1}$. When $\wO = 1$, the SC stays independent of $\wr$ and agrees with the prediction in Equation~\eqref{eq:SC_highT_m0}. For smaller values of $\wO$, deviations can be observed as $\wr \rightarrow \pi/2$, at larger distances from the boundary when $\wO$ is smaller. The $T_0 = 0.5\ell^{-1}$ curves shown in panel (a) indicate that the large temperature limit in Equation~\eqref{eq:SC_highT_m0} loses validity when $T_0 \lesssim \ell^{-1}$. In panel (b), the high temperature limit for arbitrary masses, derived above in Equation~\eqref{eq:SC_highT}, is validated in the high temperature regime ($T_0 = 2\ell^{-1}$) for various values of $\wO$. The agreement at $\wO = 1$ is excellent for both $k =0$ and $k =2$, while deviations can be seen in the vicinity of the boundary when $\wO$ is decreased. Finally, panel (c) shows the SC as a function of the inverse of the distance to the boundary, $(1 - 2 \wr / \pi)^{-1}$, at $k = 0$ and $|\wO| < 1$ ($0$ and $0.9$ are considered), represented using a logarithmic scale. It can be seen that as the temperature is increased, the validity of the asymptotic value $1 / 4\pi^2$ is extended to smaller distances from the boundary.

\begin{figure}[H]

\begin{tabular}{cc}
 \includegraphics[width=0.45\linewidth]{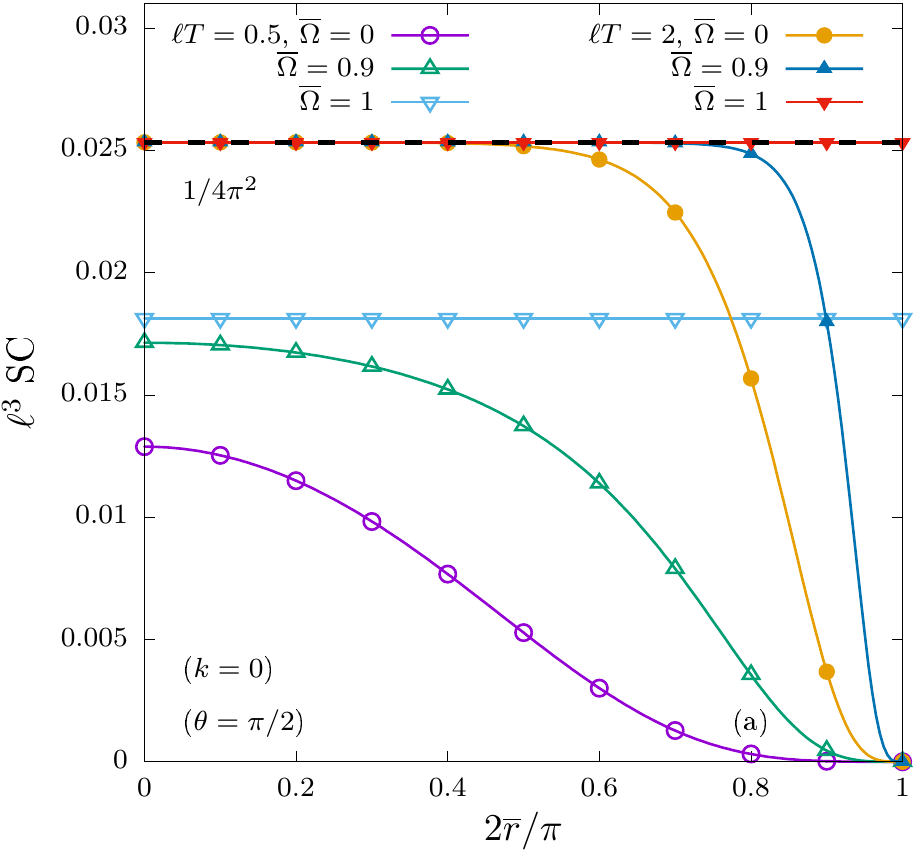} & 
 \includegraphics[width=0.45\linewidth]{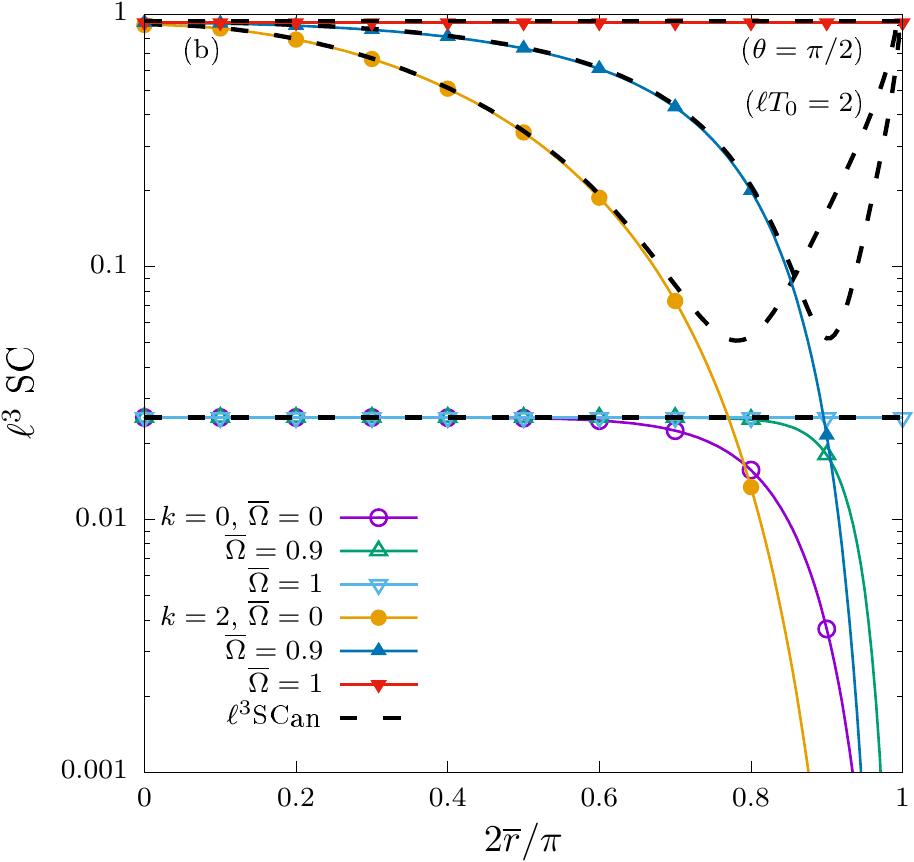}
\end{tabular}
\begin{center}
\includegraphics[width=0.45\linewidth]{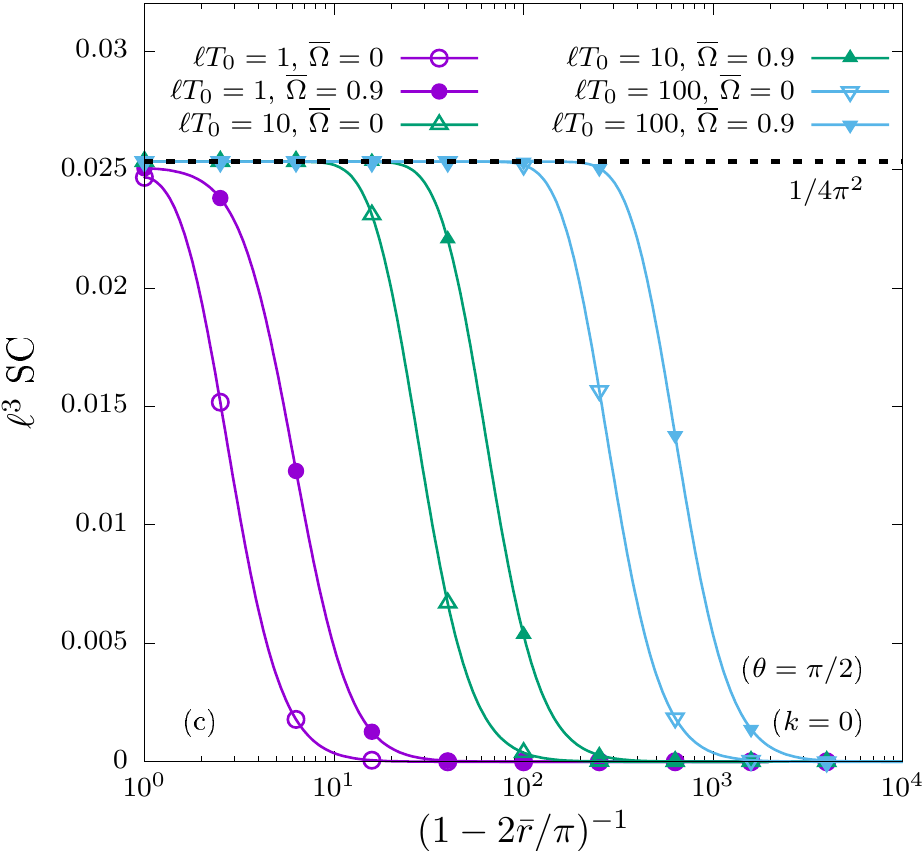}
\end{center}
\caption{Profiles of the scalar condensate  (SC) in the equatorial plane ($\theta = \pi/2$). (\textbf{a}) Massless ($k = 0$) quanta at low ($T_0 = 0.5 \ell^{-1}$) and high ($T_0 = 2 \ell^{-1}$) temperatures, for various values of  $\overline{\Omega}$. (\textbf{b}) High-temperature results for $k = 0$ and $k = 2$ at various $\overline{\Omega}$. (\textbf{c}) Validity of the vanishing mass high-temperature result in Equation~\eqref{eq:SC_highT_m0} as a function of  the radial coordinate, as $T_0$ and $\overline{\Omega}$ are increased. The black dotted lines represent the high-temperature result in Equation~\eqref{eq:SC_highT}, which reduces to Equation~\eqref{eq:SC_highT_m0} when $k = 0$.
\label{fig:SC_vs_r}}

\end{figure}

We further validate the result in Equation~\eqref{eq:SC_highT} in the limit of critical rotation ($\wO = 1$), when the SC is constant in the equatorial plane. Figure~\ref{fig:SC_vs_T} shows the SC as a function of temperature for various values of $k$. In panel (a), it can be seen that the numerical results approach the asymptotic form in Equation~\eqref{eq:SC_highT} around $\ell T_0 \simeq 1$. Panel (b) displays the difference ${\rm SC} - {\rm SC}_{\rm an}$, where ${\rm SC}_{\rm an}$ refers to the asymptotic expression in Equation~\eqref{eq:SC_highT}. At $\ell T_0 > 1$, all curves give essentially zero, thus validating the contributions of order $O(T_0^0)$. At $\ell T_0 < 1$, the asymptotic expression loses its validity due to the terms which are logarithmic in or independent of $T_0$. Since  the SC vanishes as $T_0 \rightarrow 0$, these terms will dominate ${\rm SC} - {\rm SC}_{\rm an}$, as confirmed by the dotted black lines.

\begin{figure}[H]

\begin{tabular}{cc}
 \includegraphics[width=0.45\linewidth]{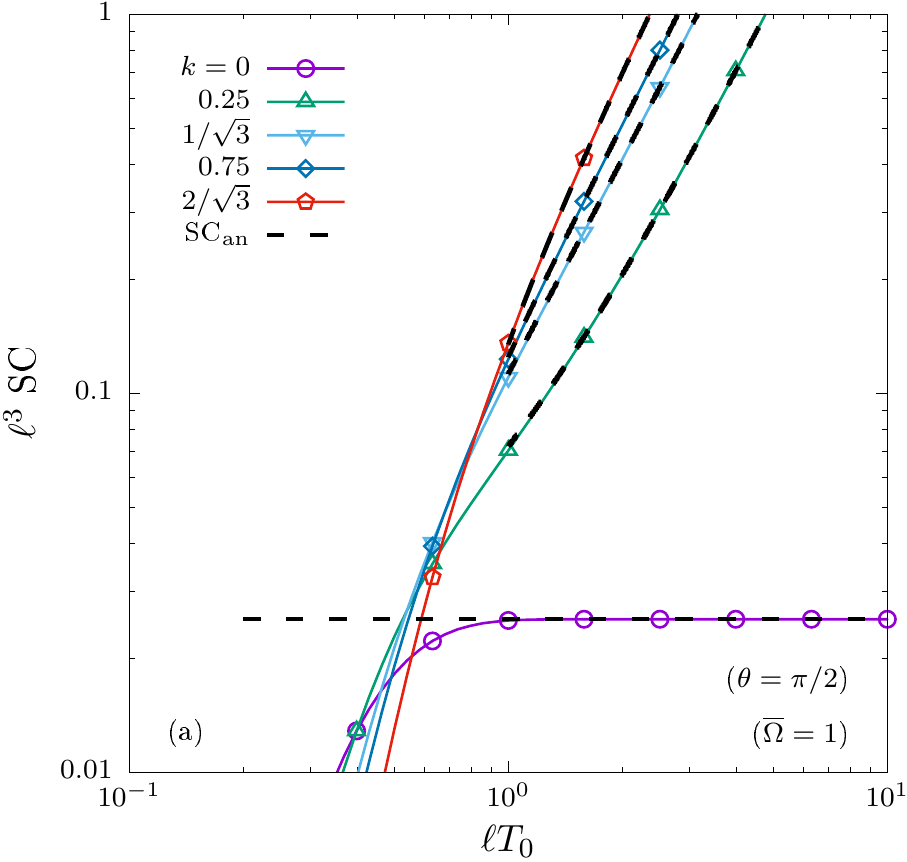} & 
 \includegraphics[width=0.45\linewidth]{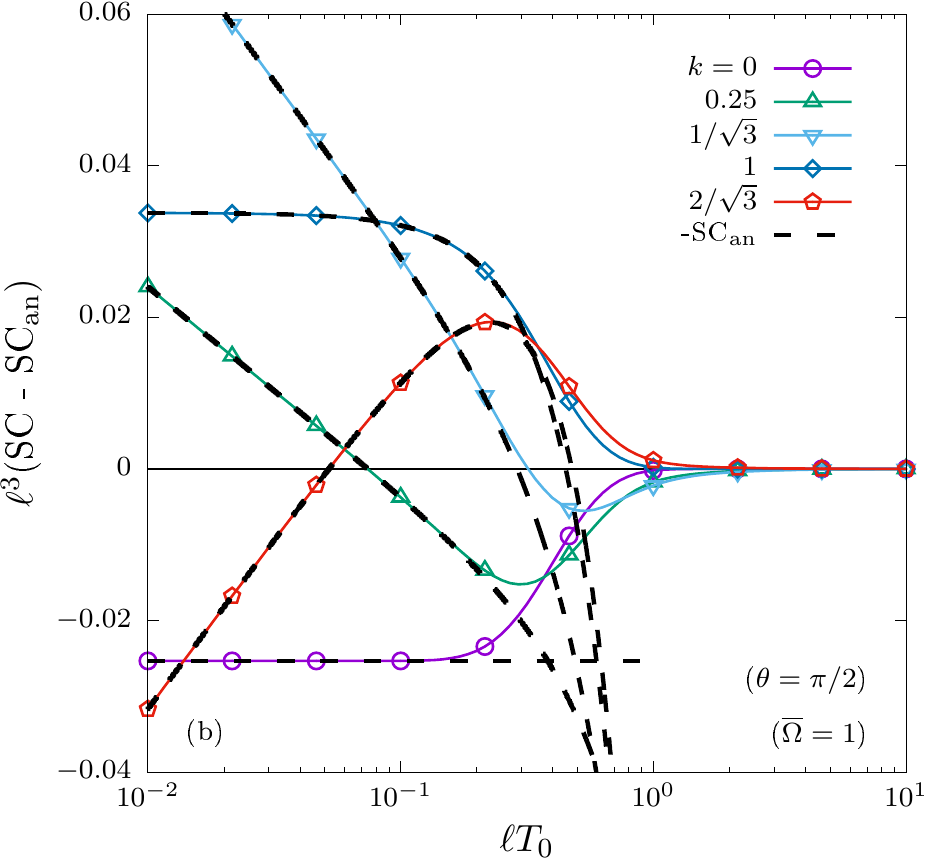}
\end{tabular}
\caption{(\textbf{a}) Log-log plot of the SC in the equatorial plane ($\theta = \pi/2$) as a function of $T_{0}$ for $\overline{\Omega} = 1$ and various values of $k$. (\textbf{b}) Linear-log plot of the difference ${\rm SC} - {\rm SC}_{\rm an}$ between the numerical result and the high temperature analytical expression in Equation~(\ref{eq:SC_highT}).
\label{fig:SC_vs_T}}

\end{figure}

The $O(T^0)$ term appearing in Equation~\eqref{eq:SC_highT} could originate from
the fact that we 
have only taken into account the thermal part of the expectation value of the SC. We now consider 
its full expectation value by adding the vacuum contribution $SC^{\rm Had}_{\rm vac}$ to the result in Equation~\eqref{eq:SC_highT}. According to Ref.~\cite{Ambrus:2015mfa}, Hadamard regularisation gives
\begin{align}
 SC^{\rm Had}_{\rm vac} =& \frac{1}{4\pi^2\ell ^{3}} \left\{1 - \frac{k}{6} - k^2 + k^3 + 2k(1 - k^2) \left[\psi(k) + {\mathcal {C}} - \ln (\nu_{\rm Had}\ell \sqrt{2}) \right]\right\} \nonumber\\
 =& -\frac{1}{4\pi^2\ell ^{3}} \left\{1 + \frac{k}{6} - k^2 - k^3 - 2k(1 - k^2) \left[\psi(1 + k) + {\mathcal {C}} - \ln(\nu_{\rm Had}\ell \sqrt{2}) \right]\right\},\tag{120}
\end{align}
where $\nu_{\rm Had}$ is an arbitrary (unspecified) regularisation mass scale and the \hl{relation} \eqref{eq:digamma}
was employed to go from the first to the second line. 
Adding together the thermal and the vacuum contributions gives
\begin{multline}
 SC^{\rm Had}_{\rm total} \equiv SC + SC_{\rm vac}^{\rm Had} = 
 \frac{M T^2}{6} - \frac{M}{2\pi^2} \left(M^2 + \frac{R}{12}\right) \ln \frac{\pi T}{\nu_{\rm Had} \sqrt{2}} 
 \\ + \frac{M}{24\pi^2}\left(3\bm{\omega}^2 - \bm{a}^2 + \frac{R}{2}\right) 
 + O(T^{-1}).\label{eq:SC_highT_total}\tag{121}
\end{multline}
Imposing agreement with the Minkowski value \eqref{eq:SC_Mink} in the limit $\ell ^{-1} \rightarrow 0$, we obtain
\begin{equation}
 \nu^{SC}_{\rm Had} = \frac{M}{\sqrt{2}} e^{{\mathcal {C}} - \frac{1}{2}},\label{eq:SC_nuHad}\tag{122}
\end{equation}
where $M = k\ell ^{-1}$ is the mass of the field quanta and ${\mathcal {C}}$ is the Euler-Mascheroni constant.

We now discuss the volumetric properties of the SC and PC. 
Due to the $\wz$ prefactor in (\ref{eq:SCPC}), the PC is antisymmetric with respect to the equatorial plane and thus 
its volume integral vanishes identically. This property will become important in understanding the flow of axial charge, which will be discussed in the following section. 
On the other hand, the total SC  contained in adS space 
can be obtained as follows:
\begin{equation}
 V^{\rm SC}_{\beta_0, \Omega} = \int d^3x \sqrt{-g} \, SC = 2\pi \ell ^{3} \int_{-1}^1 d\cos\theta \int_0^{\pi / 2} \frac{  d\wr\, \sin^2\wr}{\cos^4\wr} SC.
 \label{eq:SC_V}\tag{123}
\end{equation}
Due to the coordinate dependence of the volume element $\sqrt{-g}\, d^3x$, it can be seen that the contributions from high values of $\wr$ have a higher weight than those within the adS bulk. For this reason, it is convenient to change the argument of the hypergeometric function appearing in Equation~\eqref{eq:SCPC_k} using Equation~\eqref{eq:hyp_smallz}, leading to
\begin{multline}
 V^{\rm SC}_{\beta_0,\Omega} = \frac{\Gamma_k}{2\pi^2\ell ^{3}} \sum_{j =1}^\infty (-1)^{j+1} \cosh\frac{j \beta_0}{2\ell } \cosh\frac{\Omega j \beta_0}{2} 
 \\ \times \int d^3x\, \sqrt{-g} \left(\frac{\zeta_j}{1 + \zeta_j}\right)^{2+k} {}_2F_1\left(k, 2 + k; 1 + 2k; \frac{\zeta_j}{1 + \zeta_j}\right).\label{eq:SC_V_aux}\tag{124}
\end{multline}
Now using Equation~\eqref{eq:hyp_ser} to express the hypergeometric function as a series, 
the integral can be performed in the following two steps:
\begin{multline}
 \int d^3x\, \sqrt{-g} \left(\frac{\zeta_j}{1 + \zeta_j}\right)^{2+\nu}
 \\ = \frac{\pi^{3/2}\ell ^{3} \Gamma(\frac{1}{2} + \nu)}{2  \Gamma(2 + \nu)} \left(\cosh \frac {j \beta_0}{2\ell }\right)^{-1-2\nu} \int_{-1}^1 \frac{d\cos\theta}{(\sinh^2 \frac{j \beta_0}{2\ell } - \sinh^2\frac{\Omega j \beta_0}{2} \sin^2\theta)^{3/2}} \\ 
 = \frac{\pi^{3/2}\ell ^{3} \Gamma(\frac{1}{2} + \nu)}{ \Gamma(2 + \nu)} \frac{(\cosh \frac{j \beta_0}{2\ell })^{-1-2\nu}}{\sinh\frac{j \beta_0}{2\ell } (\sinh^2 \frac{j \beta_0}{2\ell } - \sinh^2 \frac{\Omega j \beta_0}{2})},
 \label{eq:SC_V_aux2}\tag{125}
\end{multline}
where $\nu = k + n$. The above result shows that all terms in the hypergeometric function will make contributions which diverge as $\wO = \Omega \ell  \rightarrow 1$. Substituting the result in \mbox{Equation~\eqref{eq:SC_V_aux2}} together with the expansion \eqref{eq:hyp_ser} into Equation~\eqref{eq:SC_V_aux}, the sum over $n$ can be \mbox{performed, yielding:}
\begin{equation}
 V^{\rm SC}_{\beta_0, \Omega} = \frac{2}{4^{k+1}} \sum_{j = 1}^{\infty} \frac{(-1)^{j+1} \cosh\frac{\Omega j \beta_0}{2} (\cosh \frac{j \beta_0}{2\ell })^{-2k}}{\sinh\frac{j \beta_0}{2\ell }(\sinh^2 \frac{j \beta_0}{2\ell } - \sinh^2 \frac{\Omega j \beta_0}{2})} \,
 {}_2 F_1\left(k, \frac{1}{2} + k; 1 + 2k;{\rm sech}^2\frac{ j \beta_0}{2\ell }\right).
 \label{eq:SC_V_aux3}\tag{126}
\end{equation}
Using Equation~\eqref{eq:hyp_an}, 
the hypergeometric function appearing in Equation~\eqref{eq:SC_V_aux3} has a simple closed-form expression:
\begin{equation}
 {}_2 F_1\left(k, \frac{1}{2} + k; 1 + 2k;{\rm sech}^2\frac{j \beta_0}{2\ell }\right) = \left(4 e^{-j \beta_0/\ell } \cosh^2 \frac{j \beta_0}{2\ell }\right)^k,\tag{127}
\end{equation}
which allows $V^{\rm SC}_{\beta_0, \Omega}$ to be simplified to
\begin{align}
 V^{\rm SC}_{\beta_0, \Omega} &= 
 \sum_{j = 1}^\infty \frac{(-1)^{j+1}}{2 \sinh\frac{j \beta_0}{2\ell }} 
 \frac{e^{-j M \beta_0} \cosh \frac{\Omega j \beta_0}{2}}{\sinh^2\frac{j \beta_0}{2\ell } - \sinh^2\frac{\Omega j \beta_0}{2}}\nonumber\\
 &= -\frac{4T_0^3\ell ^{3}}{1 -\wO^2}{\rm Li}_3(-e^{-\beta_0 M}) - 
 \frac{(3  - \wO^2) T_0\ell }{6(1 - \wO^2)} \ln (1 + e^{-\beta_0 M}) + O(T_0^{-1}).
 \label{eq:SC_V_highT_aux}\tag{128}
\end{align}
The result on the second line of (\ref{eq:SC_V_highT_aux}) gives the closed form coefficients of the terms  cubic and linear in the temperature, though these coefficients are temperature-dependent due to the exponential $e^{-\beta_{0} M}$. Expanding these coefficients for small $\beta_0$, we obtain
\begin{multline}
 V^{\rm SC}_{\beta_0, \Omega} = \frac{\ell^3}{1 - \wO^2} \Bigg[ 
 3\zeta(3) T_0^3 - \frac{\pi^2 M T_0^2}{3} + \frac{T_0}{6}\left(12M^2 + \Omega^2 +\frac{R}{4}\right) \ln 2 \\
 - \frac{M}{12} \left(4M^2 + \Omega^2 + \frac{R}{4}\right) + O(T_0^{-1})\Bigg].
 \label{eq:SC_V_highT}\tag{129}
\end{multline}
It is remarkable that, while the large temperature limit of the SC at vanishing mass given in Equation~\eqref{eq:SC_highT_m0} is temperature-independent, the leading order $T_0^3$ contribution to $V^{\rm SC}_{\beta_0,\Omega}$ is mass-independent. A comparison with the classical (nonquantum) result for $(E_{\rm RKT} - 3P_{\rm RKT}) / M$ obtained in Equation~\eqref{eq:RKT_vol_highT} shows that quantum corrections appear at the next-to-next-to-leading order, in the form of the extra term $\Omega^2 + R / 4$.

\begin{figure}[H]

\begin{tabular}{cc}
 \includegraphics[width=0.45\linewidth]{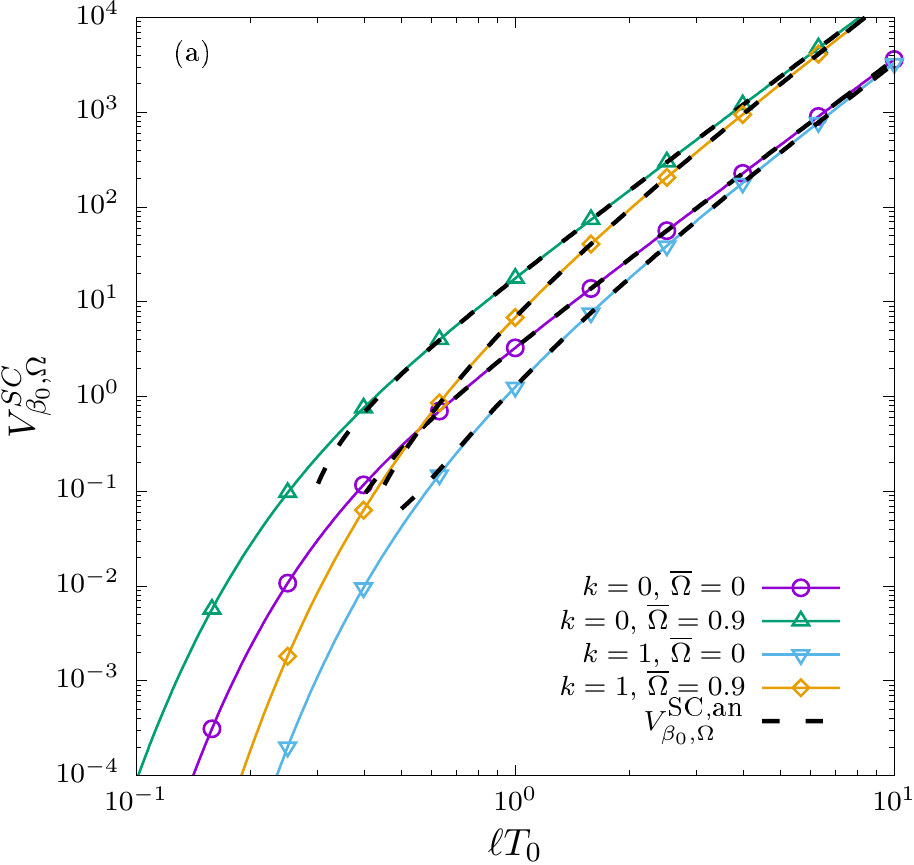} &
 \includegraphics[width=0.45\linewidth]{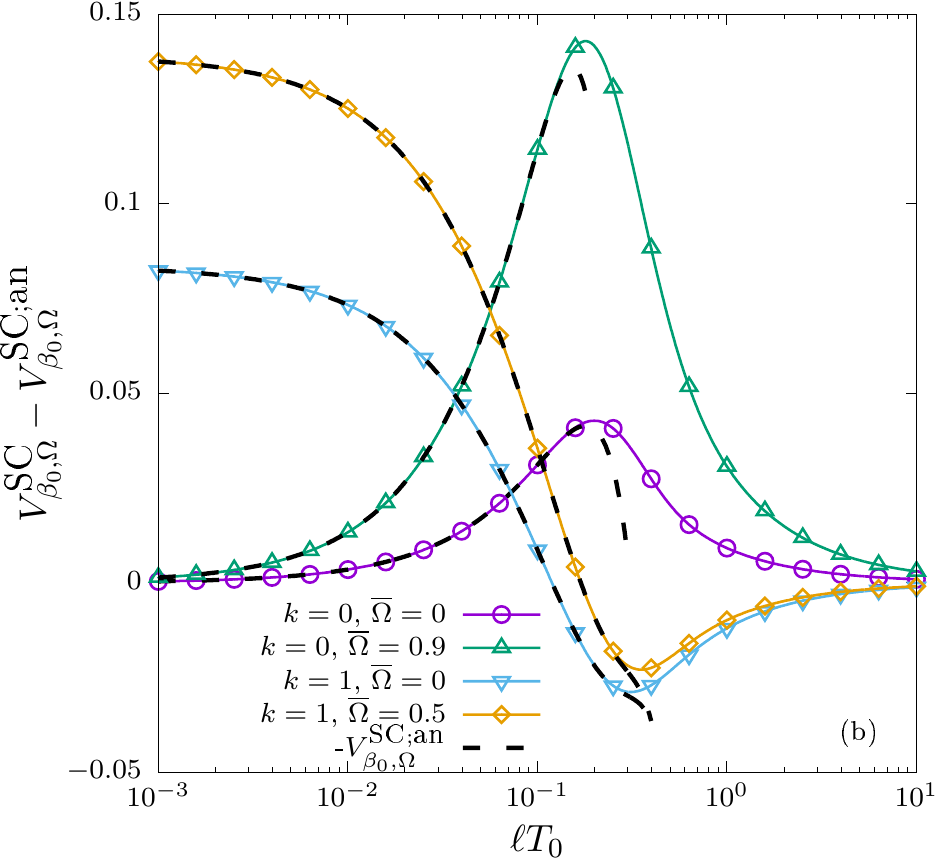}
\end{tabular}
\begin{center}
 \includegraphics[width=0.45\linewidth]{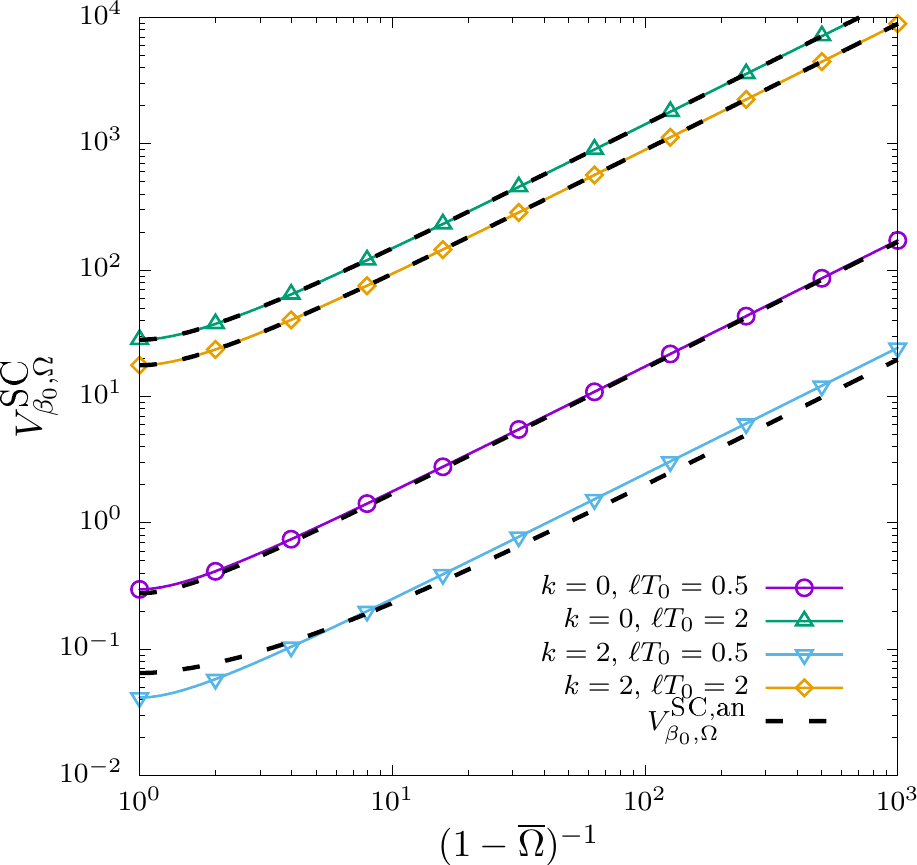}
\end{center}
\caption{(\textbf{a}) Log-log plot of $V^{\rm SC}_{\beta_0,\Omega}$ \eqref{eq:SC_V} as a function of $\ell T_0$. (\textbf{b}) Difference $V^{\rm SC}_{\beta_0,\Omega} - V^{{\rm SC}; {\rm an}}_{\beta_0,\Omega}$ as a function of $\ell T_0$. (\textbf{c}) $V_{\beta_0,\Omega}^{\rm SC}$ as a function of $(1 - \overline{\Omega})^{-1}$ for various values of $k$ and $\ell T_0$. The dashed lines represent the high temperature limit given in Equation~(\ref{eq:SC_V_highT}).
\label{fig:VSC}}

\end{figure}

The asymptotic expression $V_{\beta_0, \Omega}^{\rm SC; an}$ in Equation~\eqref{eq:SC_V_highT} is compared to the numerical result obtained by performing the summation on the first line of Equation~\eqref{eq:SC_V_aux2} in Figure~\ref{fig:VSC}. Panel (a) confirms that the asymptotic expression becomes valid when $\ell T_0 \gtrsim 1$. In panel (b), the difference $V_{\beta_0,\Omega}^{\rm SC} - V_{\beta_0,\Omega}^{\rm SC;an}$ between the numerical and analytical results is shown for various values of $k$ and $\wO$. It can be seen that the curves tend to zero as $\ell T_0 \rightarrow \infty$, confirming the validity of all the terms in Equation~\eqref{eq:SC_V_highT}, including the constant. Since $V_{\beta_0,\Omega}^{\rm SC} \rightarrow 0$ when $T_0 \rightarrow 0$, this latter term becomes dominant at small $T_0$ and its value is confirmed by the dotted black lines. Finally, panel (c) shows $V_{\beta_0,\Omega}^{\rm SC}$ as a function of $(1- \wO)^{-1}$, thus confirming its divergent behaviour as $\wO \rightarrow 1$. The agreement with the analytical solution in Equation~\eqref{eq:SC_V_highT} is excellent at high $T_0$ and small $k$. Small discrepancies can be seen in the case when $k = 2$ and $\ell T_0 = 0.5$.

\section{Charge Currents}
\label{sec:CC}

In the previous section, we studied in detail the simplest t.e.v.s, namely those of the SC and PC. 
Using Equation~\eqref{eq:tr_CC}, the components of the VC
can be conveniently written in matrix form:
\begin{align}
 J^\halpha_{V;j} =& 
 \frac{4i(-1)^{j+1} \mathcal{B}_{F;j}}{\sin\frac{s_j}{2\ell } \cos \wr} 
 R_z(i \Omega j \beta_0)_\hsigma{}^\halpha
 \begin{pmatrix}
  \sinh \frac{j \beta_0}{2\ell } \cosh\frac{\Omega j \beta_0}{2}\\
  - \wrho \cosh \frac{j \beta_0}{2\ell } \sinh\frac{\Omega j \beta_0}{2}
   \sin(\varphi + i \Omega j \beta_0) \\
  \wrho \cosh \frac{j \beta_0}{2\ell } \sinh\frac{\Omega j \beta_0}{2}
   \cos(\varphi + i \Omega j \beta_0) \\
  0
 \end{pmatrix}^\hsigma \nonumber\\
 =& \frac{4i(-1)^{j+1} \mathcal{B}_{F;j}}{\sin\frac{s_j}{2\ell } \cos \wr} 
 \begin{pmatrix}
  \sinh \frac{j \beta_0}{2\ell } \cosh\frac{\Omega j \beta_0}{2}\\
  -\wrho \cosh \frac{j \beta_0}{2\ell } \sinh\frac{\Omega j \beta_0}{2}
   \sin \varphi \\
  \wrho \cosh \frac{j \beta_0}{2\ell } \sinh\frac{\Omega j \beta_0}{2}
   \cos\varphi \\
  0
 \end{pmatrix}^\halpha.
 \label{eq:VCC}\tag{130}
\end{align}
All components are odd with respect to the change $j \rightarrow -j$, and therefore 
the t.e.v. of the VC vanishes identically:
\begin{equation}\tag{131}
 J^\halpha_V = \sum_{j \neq 0} J^\halpha_{V;j} = 0.
\end{equation}

Moving on to the AC, the following components can be obtained:
\begin{align}
 J^\halpha_{A;j}  =& 
 \frac{4i(-1)^j \mathcal{B}_{F;j}}{\sin\frac{s_j}{2\ell } \cos \wr} 
 \sinh \frac{j \beta_0}{2\ell } \sinh\frac{\Omega j \beta_0}{2}
 \nonumber \\ & \hspace{1cm} \times 
 R_z(i \Omega j \beta_0)_\hsigma{}^\halpha
 \begin{pmatrix}
  0\\
  (1-\cos\wr) \sin\theta \cos \theta \cos(\varphi + i \Omega j \beta_0)\\
  (1-\cos\wr) \sin\theta \cos \theta \sin(\varphi + i \Omega j \beta_0)\\
  \cos^2\theta +  \cos\wr \sin^2 \theta 
 \end{pmatrix}^\hsigma \nonumber\\
 =& \frac{4i (-1)^j \mathcal{B}_{F;j}}{\sin\frac{s_j}{2\ell } \cos\wr} 
 \sinh \frac{j \beta_0}{2\ell } \sinh\frac{\Omega j \beta_0}{2} \cos^2\theta
 \begin{pmatrix} 
  0 \\
  (1 - \cos\wr) \tan\theta \cos\varphi \\ 
  (1 - \cos\wr) \tan\theta \sin \varphi \\
  1 + \cos\wr \tan^2 \theta
 \end{pmatrix}^\halpha.
 \label{eq:ACC}\tag{132}
\end{align}
Comparing with the expression for the kinematic vorticity $\omega^\halpha$ in Equation~\eqref{eq:kinematic_cartesian}, it can be seen that
\begin{equation}
 J_{A;j}^\halpha = \sigma^\omega_{A;j} \omega^\halpha,
 \label{eq:sA_highT_def}\tag{133}
\end{equation}
where the contribution $\sigma^\omega _{A;j}$ ($j \neq 0$) to the vortical conductivity $\sigma^\omega_A$ is given by
\begin{equation}
 \sigma^\omega_{A;j} = \frac{4i(-1)^j \mathcal{B}_{F;j}}{\Omega \Gamma^2 \sin\frac{s_j}{2\ell } \cos^2\wr}
 \sinh\frac{ j \beta_0}{2\ell } \sinh\frac{\Omega j \beta_0}{2}.
 \label{eq:sA_highT_aux}\tag{134}
\end{equation}
Substituting the expression for $\mathcal{B}_F$ given in Equation~\eqref{eq:SF_AB_gen} into 
Equation~\eqref{eq:sA_highT_aux} and summing over $j$ gives
\begin{align}
 \sigma^\omega _A 
 & = \sum_{j\neq 0}\sigma^\omega_{A;j} \nonumber
 \\ & = \frac{\Gamma_k}{2 \pi^2 \ell ^{3} \Omega \Gamma^2 \cos^2\wr}
 \sum_{j = 1}^\infty (-1)^{j+1} \zeta_j^{2+k} \sinh\frac{j \beta_0}{2\ell } \sinh\frac{\Omega j \beta_0}{2}
 {}_2F_1(k, 2+k; 1 + 2k; -\zeta_j).
 \label{eq:sA_highTk}\tag{135}
\end{align}
In the case of critical rotation ($\Omega \ell  = 1$), the quantity $\zeta_j$  takes the form in 
Equation~\eqref{eq:zetaj_crit} and the axial vortical conductivity attains 
a constant value in the equatorial plane:
\begin{equation}
 \lim_{\Omega \ell \rightarrow 1} \sigma^\omega _A \left(\theta = \frac{\pi}{2}\right) =
 \frac{\Gamma_k}{2\pi^2 \ell ^{2}} \sum_{j = 1}^\infty 
 \frac{(-1)^{j+1}}{(\sinh \frac{j \beta_0}{2\ell })^{2+2k}}
 {}_2 F_1\left(k, 2+k; 1+2k; -{\rm {cosech }} ^{2}\frac{j \beta_0}{2\ell }\right).\tag{136}
\end{equation}

In order to investigate the large temperature limit of $\sigma^\omega_A$, we employ Equation~\eqref{eq:hyp_largez} to find
\begin{equation}
 {}_2F_1(k,2+k; 1+2k;-\zeta_j) = \frac{\zeta_j^{-k}}{\Gamma_k} \left[1 - 
 \frac{k^2}{\zeta_j} + O(\zeta_j^{-2})\right].\tag{137}
\end{equation}
Using the summation formulae in Equation~\eqref{eq:sA_summation},
we obtain
\begin{equation}
 \sigma^\omega _{A; \textrm{an}} = \frac{T^2}{6} + \frac{1}{24\pi^2} \left(\bm{\omega }^2 + 3\bm{a}^2 + \frac{R}{4} - 6M^2\right) + O(T^{-1}),
 \label{eq:sA_highT}\tag{138}
\end{equation}
where $R$ is the Ricci scalar (\ref{eq:Ricci}) and $M = k \ell^{-1}$ is the mass of the field quanta.
The leading order term is consistent with the vanishing chemical potential limit ($\mu = \mu_5 = 0$) of the result reported in Section~6 of Ref.~\cite{buzzegoli18}. The terms involving $\bm{\omega}^2$ and $\bm{a}^2$ are in agreement with the results derived using the equilibrium density operator formalism in Ref.~\cite{Prokhorov:2018bql}, while the mass term agrees with the correction found in Equations (\hl{7.10}) and (\hl{40}) of Refs.~\cite{Ambrus:2019ayb} and~\cite{Lin:2018aon}, respectively, as well as with the high temperature expansion of the results derived in Ref.~\cite{buzzegoli17} (see last entry for $W^A$ in Table 2). Curvature corrections were also reported in Equations~(2) and (15) of Ref.~\cite{Flachi:2017vlp}, however the coefficient of the Ricci scalar differs by a factor of $-2$ to our result. The results in Ref.~\cite{Flachi:2017vlp} must be multiplied by two in order to account for both $L$ and $R$ chiral states.

On Minkowski space, the second term in Equation~\eqref{eq:sA_highT}, which
is independent of temperature, 
is generated when computing the expectation value of the axial current 
with respect to the static (Minkowski) vacuum (as opposed to the rotating vacuum).
In adS, this term is generated without recourse to the nonrotating vacuum, but only in the large temperature limit. The validity of Equation~\eqref{eq:sA_highT} is probed in a variety of regimes in \mbox{Figure~\ref{fig:sA}}. In panel (a), the massless case $k = 0$ is studied at various values of the temperature $\ell T_0$ and rotation parameter $\wO = \ell \Omega$. The agreement is excellent even at $\ell T_0= 0.5$. In panel (b), the high-temperature case ($\ell T_0 = 2$) is tested for massless and massive quanta and discrepancies can be seen at $k = 2$, when the mass and the temperature are of the same order of magnitude. Panel (c) shows the $T_0$ dependence of $\sigma^\omega_A$ for the case of critical rotation ($\wO = 1$), in the equatorial plane ($\theta = \pi / 2$), where $\sigma^\omega_A$ no longer depends on $\wr$. The asymptotic behaviour can be seen to be independent of $k$, as expected from the structure of the leading order term $T^2 / 6$ in Equation~\eqref{eq:sA_highT}. The asymptotic form is established at larger temperature when $k = 2 / \sqrt{3}$ than for $k =0$. Finally, panel (d) shows the difference $\sigma^\omega_A - \sigma^\omega_{A;\textrm{an}}$, which drops to zero when $\ell T_0 \gtrsim 1$. It should be noted that the $O(T^0)$ term is necessary to obtain the convergence to zero, however this term loses its validity at small temperature (the $T_0 \rightarrow 0$ asymptotics are governed by the temperature-independent contribution to $-\sigma_{A;\textrm{an}}^\omega$).

\begin{figure}[H]

\begin{tabular}{cc}
 \includegraphics[width=0.45\linewidth]{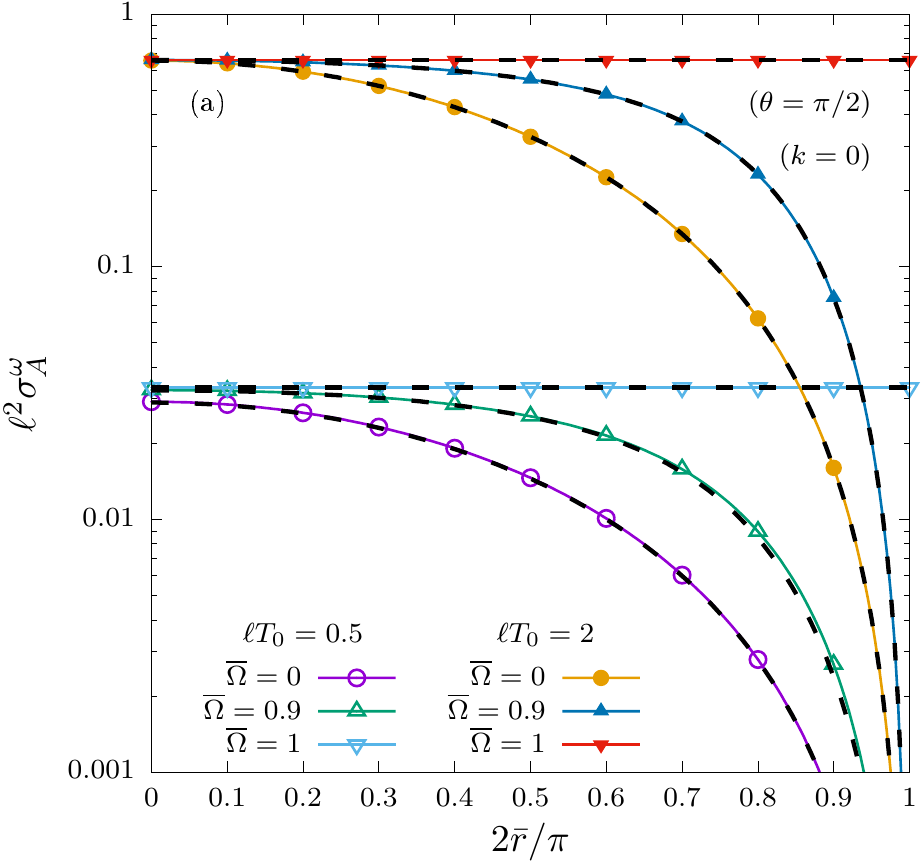} & 
 \includegraphics[width=0.45\linewidth]{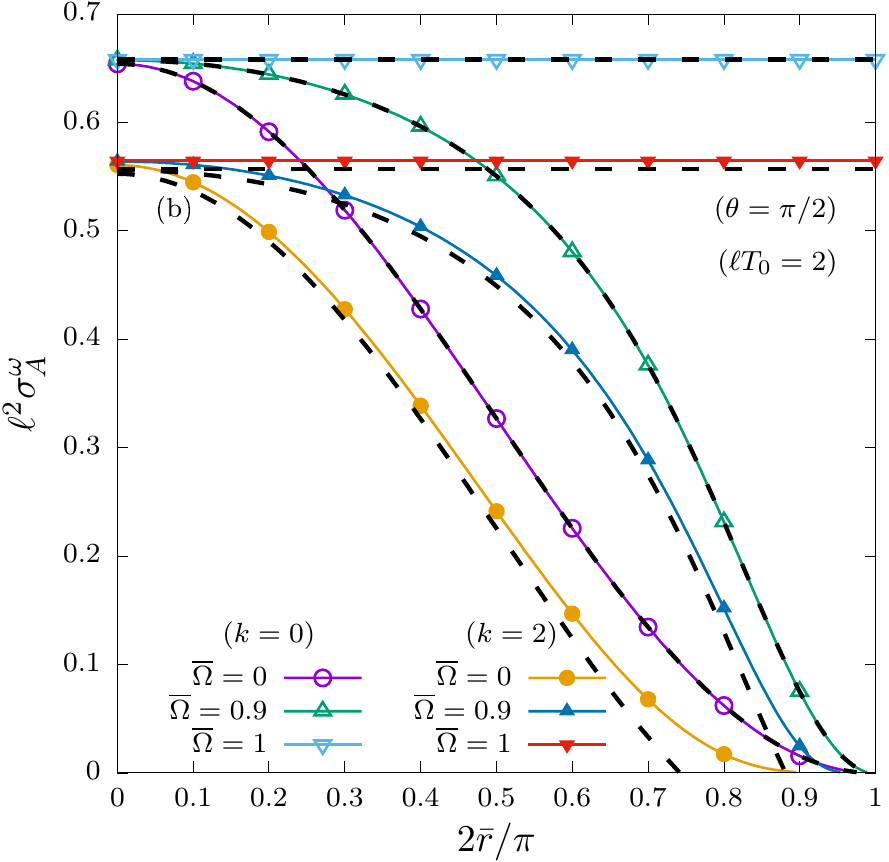} \\
 \includegraphics[width=0.45\linewidth]{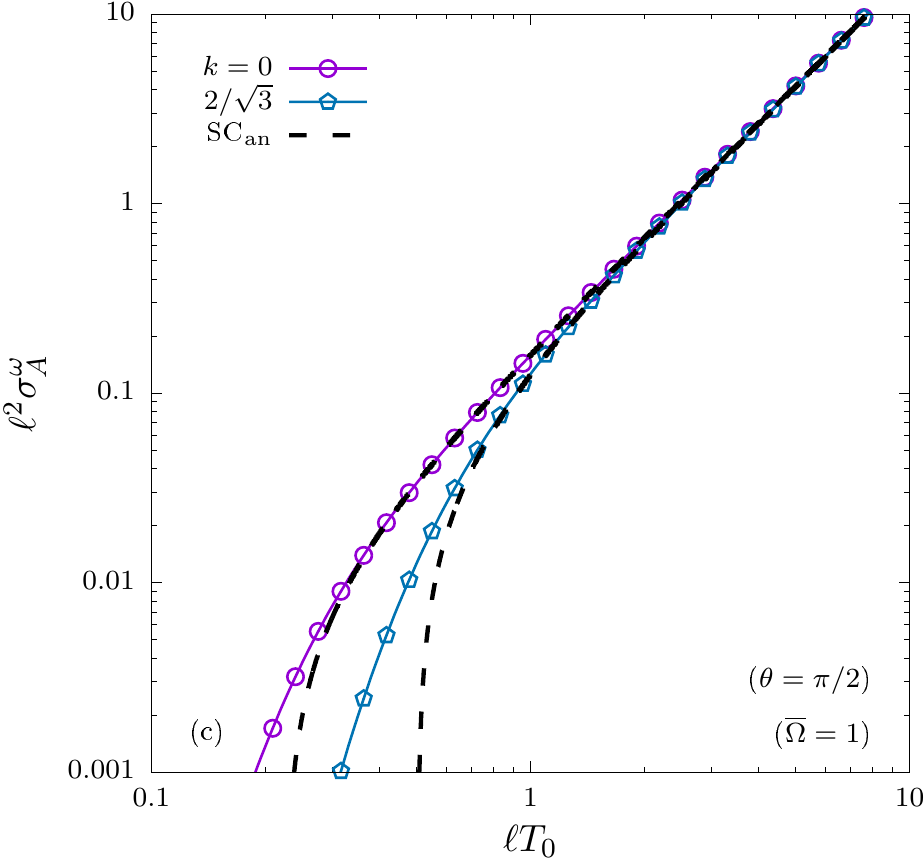} & 
 \includegraphics[width=0.45\linewidth]{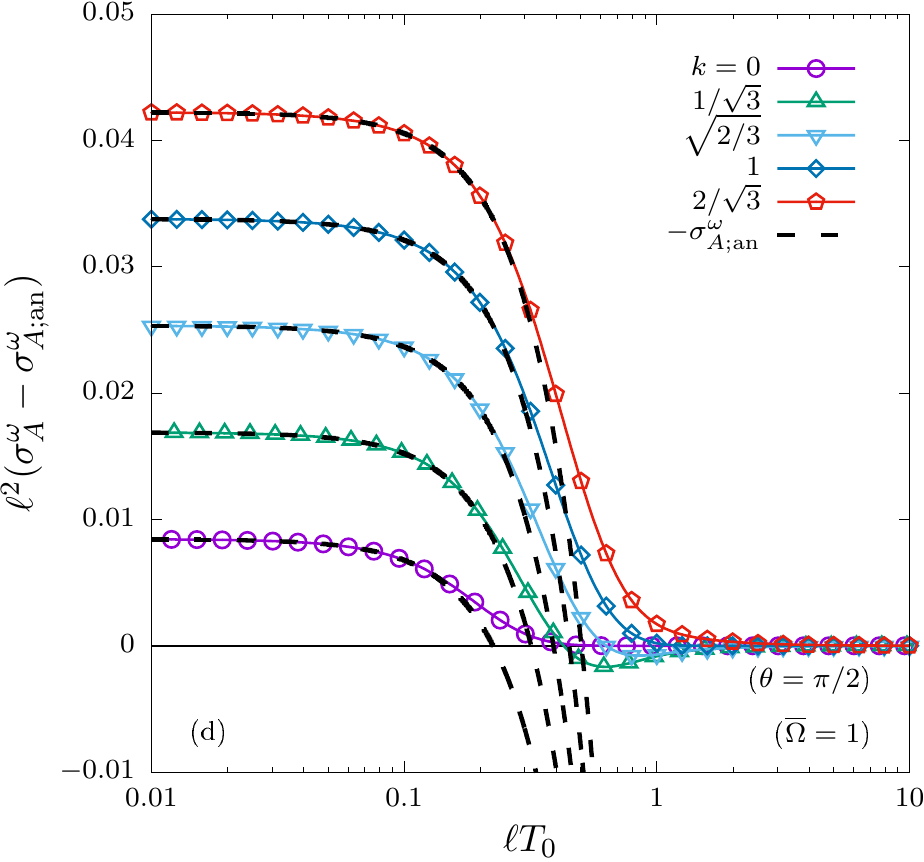}\\
\end{tabular}
\caption{(\textbf{a},\textbf{b}) Radial profiles of the axial vortical conductivity $\sigma^\omega_A$ for various values of $\overline{\Omega}$, evaluated in the equatorial plane ($\theta = \pi/2$) (\textbf{a}) for massless ($k = 0$) quanta at low ($T_0 = 0.5 \ell^{-1}$) and high ($T_0 = 2 \ell^{-1}$) temperatures; (\textbf{b}) at high temperature ($\ell T_0 = 2$) for $k = 0$ and $k = 2$. Temperature dependence (\textbf{c}) of $\sigma^\omega_A$; and (\textbf{d}) of $\sigma^\omega_A - \sigma^\omega_{A;{\textrm {an}}}$, in the case of critical rotation $\wO = 1$ ($\theta = \pi / 2$). The analytical result $\sigma^\omega_{A; \textrm{an}}$, represented using the black dotted lines, is given in Equation~\eqref{eq:sA_highT}.
\label{fig:sA}}

\end{figure}

We end this section with a discussion of the flux of the AC through adS space time. The divergence of the AC becomes
\begin{equation}
 \nabla_\mu J^\mu_A = \frac{1}{\sqrt{-g}} \frac{\partial}{\partial x^\mu} (\sqrt{-g} J^\mu_A) = 
 \frac{1}{\sqrt{-g}} \frac{\partial(J^\wz_A \sqrt{-g})}{\partial \wz} = -2k \ell ^{-1} PC,
 \label{eq:JA_div}\tag{139}
\end{equation}
where PC is the pseudoscalar condensate.
Integrating the above equation over the volume $V$ bounded by the two-surfaces given by constant $\wz$ values $\wz_0$ and $\wz_1$, we find
\begin{equation}
 \int_0^1 d\wrho \int_0^{2\pi} d\varphi\, \int_{\wz_0}^{\wz_1} d\wz\, \sqrt{-g} 
 \nabla_\mu J^\mu_A = \ell[F_A(\wz_1) - F_A(\wz_0)] = -2k\ell ^{-1} \int_V d^3x \sqrt{-g} \, PC,
 \label{eq:FA_cons}\tag{140}
\end{equation}
showing that the flux $F_A$ is independent of $\wz $ for massless fermions, when 
$k = 0$. Its explicit expression for arbitrary $k$ is
\begin{equation}
 F_A(\wz) = \ell^{-1} \int_0^1 d\wrho \int_0^{2\pi} d\varphi\, \sqrt{-g} J^\wz_A(\wz).\tag{141}
\end{equation}
Substituting the expressions \eqref{eq:RKT_omega_coord} and \eqref{eq:sA_highTk} into Equation~\eqref{eq:sA_highT_def} allows $F_A$ to be written~as
\begin{multline}
 F_A = \frac{\Gamma_k}{\pi \ell} \sum_{j = 1}^\infty (-1)^{j + 1}
 \sinh \frac{j \beta_0}{2\ell } \sinh\frac{\Omega j \beta_0}{2} 
\\  \times \int_0^1 d\wrho\, \wrho \left(\frac{1 - \wrho^2}{1 + \wz^2}\right)^k 
 \frac{{}_2 F_1(k, 2+k; 1+2k; -\zeta_j)}{
 (\sinh^2\frac{j \beta_0}{2\ell } - \wrho^2 \sinh^2\frac{\Omega j \beta_0}{2})^{2 + k}},
 \label{eq:ACC:FA_k_aux}\tag{142}
\end{multline}
where the quantity $\zeta_j$ given in Equation~\eqref{eq:zetaj_def} reduces to
\begin{equation}
 \zeta_j = \frac{1 - \wrho^2}{1 + \wz^2} \frac{1}{\sinh^2\frac{j \beta_0}{2\ell } - \wrho^2 \sinh^2\frac{\Omega j \beta_0}{2}}.\label{eq:zetaj_zrho}\tag{143}
\end{equation}
We now perform the $\wrho$ integration in Equation~\eqref{eq:ACC:FA_k_aux}.
Using Equation~\eqref{eq:hyp_ser} to replace the hypergeometric series, $F_A$ becomes

\end{paracol}
\nointerlineskip
\begin{multline}
 F_A = \frac{\Gamma_k}{2 \pi \ell} \sum_{j = 1}^\infty (-1)^{j + 1} 
 \sinh \frac{j \beta_0}{2\ell } \sinh\frac{\Omega j \beta_0}{2} \sum_{n = 0}^\infty \frac{(-1)^n \Gamma(k + n) \Gamma(2 + k + n) \Gamma(1 + 2k)}{n! \Gamma(k) \Gamma(2+k) \Gamma(1 + 2k + n) (1 + \wz^2)^{k+n}}\\\times
 \int_0^1 \frac{Y^{k + n} dY}{
 [\sinh^2\frac{j \beta_0}{2\ell } - (1 - Y) \sinh^2\frac{\Omega j \beta_0}{2}]^{2 + k + n}},
 \label{eq:ACC:FA_k_aux2}\tag{144}
\end{multline}
\begin{paracol}{2}
\switchcolumn

\noindent
where $Y = 1 - \wrho^2$. The integration with respect to $Y$ can be performed trivially,
\begin{equation}
 \int_0^1 \frac{Y^\nu  \, dY}{[a - b(1 - Y)]^{2+\nu}} = 
 \int_0^{1/a}  \frac{X^\nu \, dX}{a - b} = 
 \frac{a^{-1-\nu}}{(a - b)(1 + \nu)},\tag{145}
\end{equation}
where $X = Y / (a - b + bY)$. This allows the summation over $n$ in (\ref{eq:ACC:FA_k_aux2}) to be 
performed in terms of another hypergeometric series, yielding
\begin{multline}
 F_A = \frac{\Gamma_k}{2 \pi \ell (1 + k)} \sum_{j = 1}^\infty \frac{(-1)^{j + 1} \sinh\frac{\Omega j \beta_0}{2} (\sinh\frac{j \beta_0}{2\ell})^{-1-2k}}{(\sinh^2\frac{j \beta_0}{2\ell } - \sinh^2\frac{\Omega j \beta_0}{2})(1 + \wz^2)^k}
\\ \times {}_2F_1\left(k,1+k; 1+2k; -\frac{(1+\wz^2)^{-1}}{\sinh^2 \frac{j \beta_0}{2\ell}}\right).
 \label{eq:ACC:FA_k}\tag{146}
\end{multline}
The argument of the hypergeometric function shows that the limit of high temperature ($\beta_0 \rightarrow 0$) cannot be taken simultaneously with the limit of high $\wz$. In the absence of a uniformly asymptotic expansion, we now proceed to derive separately the large $\wz$ and the large $T_0$ behaviours.

In the case of large $\wz$, we can treat the argument of the hypergeometric function as a small number, which permits the series representation \eqref{eq:hyp_ser} to be employed:
\begin{multline}
 {}_2F_1\left(k,1+k; 1+2k; -\frac{1}{(1+\wz^2) \sinh^2 \frac{j \beta_0}{2\ell}}\right) = 
 1 - \frac{k(1+k)}{(1+2k)(1 + \wz^2) \sinh^2 \frac{j \beta_0}{2\ell}} \\ + O[(1+\wz^2)^{-2}]. \tag{147}
\end{multline}
An expansion with respect to small $\beta_0$ must be performed in order to extract the large temperature behaviour of $F_A(\wz)$. Exchanging the summation over $j$ with the above series expansions is not strictly valid and will lead to a nonuniform asymptotic expansion. Taking into account the $(\sinh\frac{j \beta_0}{2\ell})^{-1-2k}$ term appearing in Equation~\eqref{eq:ACC:FA_k}, the large temperature expansion involves both negative and positive powers of $j$. Restricting only to terms with vanishing or negative powers, the summation over $j$ can be performed using Equation~\eqref{eq:sumj_zeta}, leading to
\begin{multline}
 F_A(\wz) = \frac{\wO}{1 - \wO^2}
 \frac{\Gamma(1 + k)}{2 \ell \sqrt{\pi}\, \Gamma(\frac{1}{2} + k)} 
 \left(\frac{\ell T_0}{\sqrt{1 + \wz^2}}\right)^{2k} \Bigg\{
 \Bigg[\zeta(2 + 2k) \left(1 - \frac{1}{2^{1+2k}}\right)(2\ell T_0)^2 \\
 - \frac{\zeta(2k)}{6} \left(1 - \frac{1}{2^{2k-1}}\right)(3 + 2k + \wO^2)
 + O(T_0^{-2})\Bigg] 
\\ - \frac{4k(1+k)}{1+2k} \left(\frac{\ell T_0}{\sqrt{1 + \wz^2}}\right)^2 \Bigg[ 
 \zeta(4 + 2k) \left(1 - \frac{1}{2^{3 + 2k}}\right)(2 \ell T_0)^2
  \\ - \frac{\zeta(2 + 2k)}{6}\left(1 - \frac{1}{2^{1 + 2k}}\right) (5 + 2k + \wO^2) + 
 O(T_0^{-2})\Bigg]\Bigg\}.
 \label{eq:FA_large_z}\tag{148}
\end{multline}
The same divergent factor $(1 - \wO^2)^{-1}$ that was seen for the SC in Equation~\eqref{eq:SC_V_highT} appears in the total axial flux. The validity of Equation~\eqref{eq:FA_large_z} is restricted to large $\wz$, but also to large $T_0$ compared to the mass $k\ell^{-1}$. 
The dependence of $F_{A}$ on $\wz$ is shown in Figure~\ref{fig:FA}a. At $k = 0$, the coordinate-independence of $F_A$ is confirmed and the agreement with the result in Equation~\eqref{eq:FA_large_z} is excellent even for small temperature, $\ell T_0 = 0.5$. At finite mass ($k = 1$), the flux $F_A$ decreases like $\wz^{-k}$. The leading-order coefficient is temperature-dependent and the result in Equation~\eqref{eq:FA_large_z} becomes inaccurate at small temperatures ($\ell T_0 = 0.5 < k = 1$).

\begin{figure}[H]

\begin{center}
\includegraphics[width=0.45\linewidth]{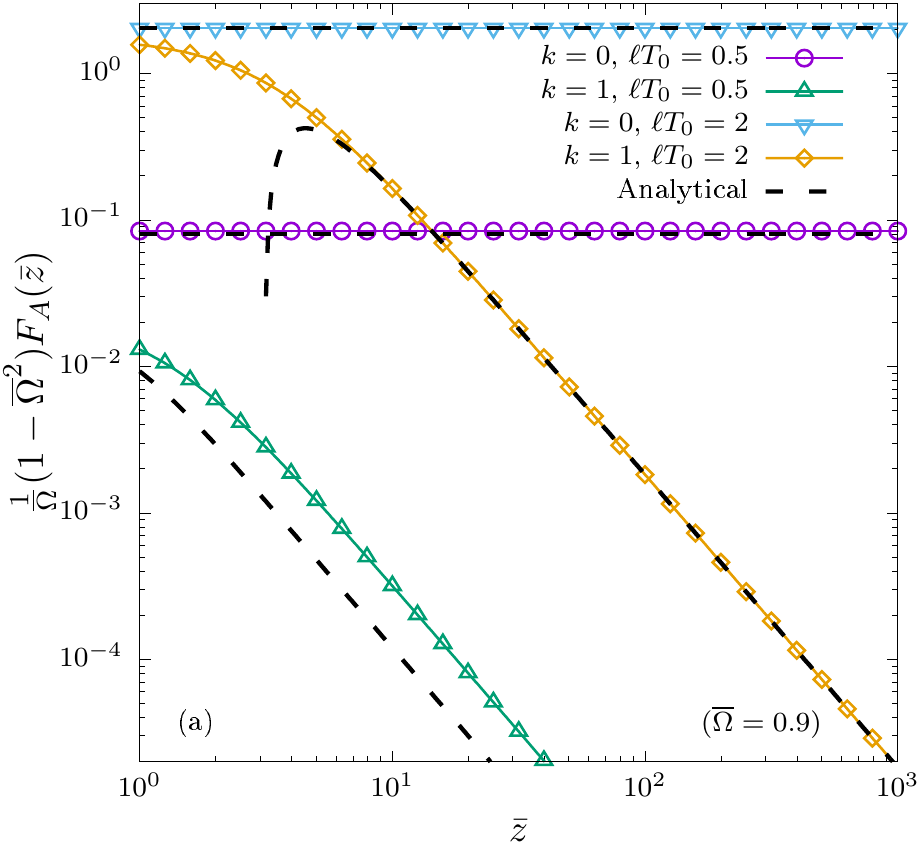}
\end{center}
\vspace{-5pt}
\begin{tabular}{cc}
 \includegraphics[width=0.45\linewidth]{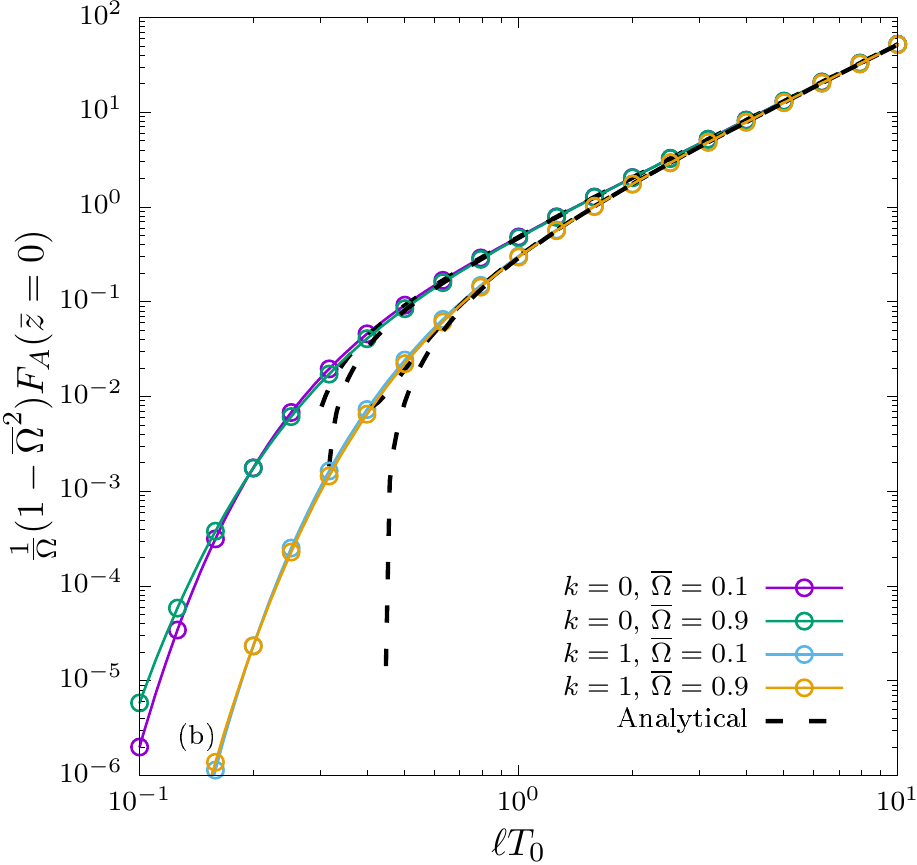} &
 \includegraphics[width=0.45\linewidth]{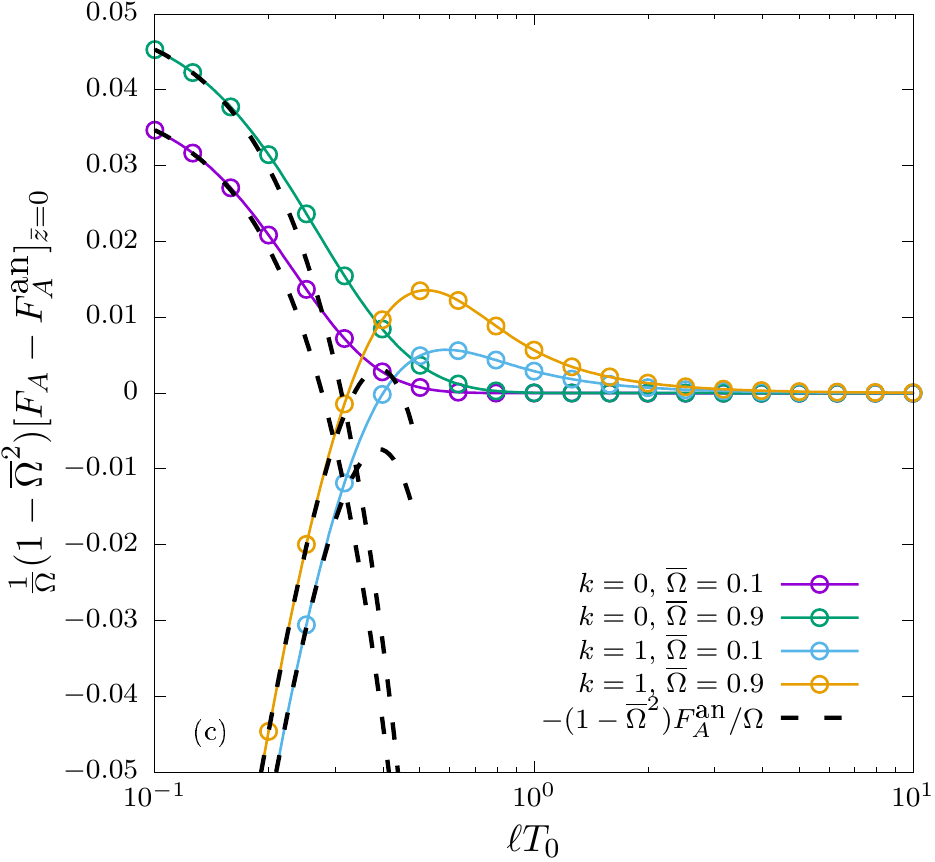}\\
\end{tabular}
\caption{(\textbf{}a) Dependence of $F_A$ on $\wz$ for massless ($k = 0$) and massive ($k = 1$) quanta at small ($\ell T_0 = 0.5$) and large ($\ell T_0 = 2$) temperatures. Temperature dependence of (\textbf{b}) $F_A$; and (\textbf{c}) $F_A - F_A^{\textrm{an}}$ for $\wz = 0$, normalised with respect to the prefactor $\Omega / (1 - \wO^2)$. The analytical results $F_A^{\textrm{an}}$ at large $\wz$ (panel \textbf{a}) and large $T_0$ (panels (\textbf{b},\textbf{c})) are given in Equations~\eqref{eq:FA_large_z} and \eqref{eq:FA_large_T}, respectively.
\label{fig:FA}}

\end{figure}

At high temperature, Equation~\eqref{eq:hyp_largez} can be employed to expand the hypergeometric function appearing in Equation~\eqref{eq:ACC:FA_k}:
\begin{multline}
 {}_2F_1\left(k,1+k; 1+2k; -\frac{(1+\wz^2)^{-1}}{\sinh^2 \frac{j \beta_0}{2\ell}}\right)
 = \frac{\Gamma(1 + 2k)}{[\Gamma(1 + k)]^2} (1 +\wz^2)^k\left(\sinh\frac{j \beta_0}{2\ell}\right)^{2k}
 \\ \times \Bigg\{1  + k^2(1 + \wz^2) \sinh^2 \frac{j \beta_0}{2\ell} \\\times
 \left[\ln(1+\wz^2) + \ln\left(\sinh^2 \frac{j\beta_0}{2\ell}\right) - \psi(1) - \psi(2) + \psi(k) + \psi(1 + k)\right] + O(\beta_0^4)\Bigg\}.\tag{149}
\end{multline}
Noting the properties of the digamma function in Equation~\eqref{eq:digamma},  we can write
\mbox{$F_A(\wz)$ as} 
\begin{multline}
 F_A(\wz) = \frac{1}{2 \pi \ell} \sum_{j = 1}^\infty \frac{(-1)^{j + 1} \sinh\frac{\Omega j \beta_0}{2}}{(\sinh^2\frac{j \beta_0}{2\ell } - \sinh^2\frac{\Omega j \beta_0}{2})\sinh\frac{j \beta_0}{2\ell}}\Bigg\{1 - k(1 + \wz^2) \sinh^2 \frac{j \beta_0}{2\ell} \\\times
 \left[1 + k - 2 k \mathcal{C} - 2k \psi(1 + k) - k \ln(1+\wz^2) - k\ln\left(\sinh^2 \frac{j\beta_0}{2\ell}\right)\right] + O(\beta_0^4)\Bigg\}.\tag{150}
\end{multline}
Using the summation Formula \eqref{eq:sumj_FA}
it can be shown that $F_A$ takes the \mbox{approximate form}
\begin{multline}
 F_A(\wz) = \frac{\Omega}{1 - \wO^2}\Bigg\{\frac{\pi \ell^2 T_0^2}{6} - 
 \frac{3 + \wO^2}{24\pi} - \frac{k^2(1 + \wz^2)}{2\pi} \ln \frac{\pi \ell T_0}{\sqrt{1 + \wz^2}}\\
 - \frac{k(1 + \wz^2)}{4\pi} [1 + k - 2k {\mathcal {C}} - 2k \psi(1 + k)] + O(T_0^{-1})\Bigg\}.
 \label{eq:FA_large_T}\tag{151}
\end{multline}

We now summarise the results derived above. Equation~\eqref{eq:FA_cons} indicates that the axial flux $F_A$ integrated over surfaces of constant $\wz$ is independent of $\wz$ for massless particles. The surfaces corresponding to the limits $\wz \rightarrow \infty$ and $-\infty$ represent the boundaries of the top and bottom halves of adS and thus a net axial charge flux is transferred from the bottom hemisphere through to the top hemisphere. This situation is reminiscent of the fate of null geodesics, which are able to travel to the adS boundary and to be reflected back into the adS bulk in finite time~\cite{Hawking:1973uf}. The boundaries of adS are thus forced open to unleash the axial current induced by the vorticity via the axial vortical effect.

The situation changes completely when the fermions have finite mass. The $(1 + \wz^2)^{-k}$ factor appearing in Equation~\eqref{eq:ACC:FA_k} ensures that the axial flux exhibits a power law decay of the form $\wz^{-2k}$ as $\wz \rightarrow \infty$. The net axial current flux through the equator must be compensated by the buildup of a negative PC in the lower hemisphere and by a PC with opposite sign in the upper hemisphere. Thus, the emergence of a nonvanishing expectation value of the PC for massive free fermions is required by observing that adS is completely opaque to massive particles, which require infinite time to reach its boundary. For this reason, no axial flux can reach the adS boundary, being instead converted into the PC via Equation~\eqref{eq:JA_div}, which then builds up inside the bulk of adS.

We now discuss the validity of the asymptotic expression in Equation~\eqref{eq:FA_large_T}. Panel (b) of Figure~\ref{fig:FA} shows $\frac{1}{\Omega}(1 - \wO^2) F_A$ evaluated on the equatorial plane ($\wz = 0$) as a function of $\ell T_0$ at various values of $\wO$. The analytical approximation can be seen to become valid when $\ell T_0 \gtrsim 0.5$. Panel (c) shows $\frac{1}{\Omega}(1 - \wO^2)[F_A - F_A^{\rm an}]$ evaluated on the equatorial plane ($\wz = 0$) as a function of $\ell T_0$, for various values of $\wO$ and $k$. All curves tend to zero as $\ell T_0$ is increased, confirming all terms in Equation~\eqref{eq:FA_large_T}. Since $F_A \rightarrow 0$ as $T_0 \rightarrow 0$, the logarithmic and temperature-independent terms become dominant at low temperatures, as confirmed by the black dotted lines.

\section{Stress-Energy Tensor}
\label{sec:SET}

The previous two sections have contained detailed analysis of the SC, PC and AC. These have revealed divergences in the volumetric integrals of t.e.v.s in the limit of critical rotation, $\Omega \ell \rightarrow 1$. We have also examined the high-temperature limit, and the role of quantum- and curvature-corrections. We now turn to the remaining (and most complex) physical quantity---the SET.  

For the computation of the components of the SET given in Equation~(\ref{eq:tevs_gen}), it is convenient to perform the covariant derivatives
$D_\hsigma$ and $D_{\hsigma'}$  using the properties of the vacuum Feynman propagator. 
The difficulty encountered in this case is due to the fact that $S^F_{\rm vac}$ in Equation~(\ref{eq:SF_beta}) is evaluated on the 
thermal contour, while the covariant derivatives are taken on the real contour. This problem is 
solved by noting that:
\begin{equation}
 D_\halpha(x) e^{-j \beta_0 \Omega S^\hatz} = 
 R_z(i \Omega j \beta_0)^{\hsigma}{}_\halpha e^{-j \beta_0 \Omega S^\hatz} 
 D_{\hsigma;j},
 \label{eq:DR_RD}\tag{152}
\end{equation}
where $D_\halpha(x) = e_\halpha^\mu(x) \partial_\mu - \Gamma_\halpha(x)$ is the usual spinor covariant 
derivative, while 
$D_{\hsigma; j}$ acts on the shifted coordinate $x_j$, as indicated in Equation~\eqref{eq:D_transf}. On the other hand, the covariant derivative $D_{\hsigma';j}$ acts on the coordinate $x'$ from the right and thus does not require rotation. Writing $T_{\halpha\hsigma} = \sum_{j \neq 0} T_{\halpha\hsigma; j}$, we find
\begin{equation}
 T_{\halpha\hsigma; j} = \frac{i}{2} (-1)^j {\rm tr} \left[e^{-j \beta_0 \Omega S^\hatz} \gamma_\hlambda R^\hlambda{}_{(\halpha} (R^\hbeta{}_{\hsigma)} D_{\hbeta; j} - D_{\hsigma')}) i S_{{\rm {vac;}} j}\right],
 \label{eq:SET_gen_aux}\tag{153}
\end{equation}
where we took into account that the bivector $g^{\hsigma'}{}_\hsigma$ and bispinor $\Lambda(x,x')$ of parallel transport are equal to the identity when $x' \rightarrow x$ (only the vacuum propagator is evaluated on the thermal contour).

The covariant derivatives appearing 
in Equation~\eqref{eq:SET_gen_aux} can be written using the decomposition 
\eqref{eq:geom_SF_def} of the vacuum Feynman  propagator, as follows:
\begin{align}
 D_\hsigma iS^F_{\rm vac}(x,x') =& 
 \left[\frac{1}{2\ell }  \gamma_\hsigma 
 \gamma_\hzeta   \mathcal{A}_F(s)   \tan\frac{\ws}{2} + 
 \eta_{\hsigma\hzeta} \frac{\partial}{\partial s} 
 \left(\frac{\mathcal{A}_F}{\cos \frac{\ws}{2}}\right) \cos\frac{\ws}{2}
 \right] n^\hzeta \Lambda(x,x') \nonumber\\
 & - \left[ \frac{1}{2\ell } \eta_{\hsigma\hzeta} \mathcal{B}_F(s)  \cot\frac{\ws}{2} 
  - n_\hsigma n_\hzeta  \frac{\partial}{\partial s} \left(
 \frac{\mathcal{B}_F}{\sin\frac{\ws}{2}}\right) \sin\frac{\ws}{2}  \right] 
 \gamma^\hzeta \Lambda(x,x'),\nonumber\\
 D_{\hsigma'} iS^F_{\rm vac}(x,x') =& -g^\hsigma{}_{\hsigma'}
 \left[ \frac{1}{2\ell }  \gamma_\hzeta
 \gamma_\hsigma \mathcal{A}_F(s) \tan\frac{\ws}{2}  + 
  \eta_{\hsigma\hzeta} \frac{\partial}{\partial s} 
 \left(\frac{\mathcal{A}_F}{\cos \frac{\ws}{2}}\right) \cos\frac{\ws}{2}
 \right] n^\hzeta \Lambda(x,x') \nonumber\\
 & + g^\hsigma{}_{\hsigma'} \left[
 \frac{1}{2\ell }  \eta_{\hsigma\hzeta} \mathcal{B}_F(s) \cot\frac{\ws}{2} 
  - 
n_\hsigma n_\hzeta  \frac{\partial}{\partial s} \left(
 \frac{\mathcal{B}_F}{\sin\frac{\ws}{2}}\right)  \sin\frac{\ws}{2} 
 \right] \gamma^\hzeta \Lambda(x,x'),
 \label{eq:SF_D}\tag{154}
\end{align}
where Equation~\eqref{eq:D_Lambda} and the properties $D_\halpha n_\hsigma = -\ell^{-1} \cot \ws\, (\eta_{\halpha\hsigma} + n_\halpha n_\hsigma)$, \linebreak $D_{\halpha'} n_\hsigma = \ell^{-1} {\rm cosec} \, \ws \, (g_{\halpha' \hsigma} - n_{\halpha'} n_\hsigma)$ and $n_{\hsigma'} = -g_{\hsigma'\halpha} n^\halpha$ were employed to eliminate the derivatives acting on $\Lambda$ and $\slashed{n}$ (see also Equations~(6.16) and (6.17) in Ref.~\cite{Ambrus:2017cow}). The terms involving $\mathcal{A}_F(s)$ vanish when taking the trace over the spinor indices in Equation~\eqref{eq:SET_gen_aux}.
Taking into account Equation~\eqref{eq:geom_SF_AB_eqs}, the 
derivative acting on $\mathcal{B}_F$ can be eliminated as follows:
\begin{align}
  \frac{\partial}{\partial s} \left(
 \frac{\mathcal{B}_F}{\sin\frac{\ws}{2}}\right)  \sin\frac{\ws}{2} =& 
 -\frac{2}{\ell }  \mathcal{B}_F \cot\frac{\ws}{2}  - i M \mathcal{A}_F - \frac{1}{\sqrt{-g}}\delta(x,x')
 \nonumber\\
 =& -\frac{2+k}{\ell } \mathcal{C}_F \cot\frac{\ws}{2} - \frac{1}{\sqrt{-g}} \delta(x,x'),\tag{155}
\end{align}
where the notation $\mathcal{C}_F$ is introduced below:
\begin{equation}
 \mathcal{C}_F = \frac{i \Gamma_k}{16\pi^2\ell ^{3}} \sin\left(\frac{\ws}{2}\right) \left(-\sin^2 \frac{\ws}{2 }\right)^{-2-k}
 {}_2F_1\left(k,3 + k; 1 + 2k; {\rm cosec}^2\left(\frac{\ws}{2 }\right)\right).
 \label{eq:CF_def}\tag{156}
\end{equation}
After a little algebra, the t.e.v. of the SET can be written
using Equations~\eqref{eq:DR_RD} and \eqref{eq:SF_D} as follows:
\begin{multline}
 T_{\halpha\hsigma} = i \sum_{j \neq 0} (-1)^j \frac{1}{2\ell } \cot\frac{s_j}{2\ell } 
 \left[\mathcal{B}_{F;j} \eta_{\halpha\hsigma} {\rm tr}(e^{-\Omega j \beta_0 S^\hatz} \Lambda_j) \right.\\
 \left. - (2 + k) \mathcal{C}_{F;j} 
 R_z(i \Omega j \beta_0)^\hlambda{}_{(\halpha}[R^\hbeta{}_{\hsigma)} n_{\hbeta;j} - 
 n_{\hsigma');j}] {\rm tr}(e^{-\Omega j \beta_0 S^\hatz} \gamma_\hlambda \slashed{n}_j \Lambda_j)\right] . 
 \label{eq:SET_aux}\tag{157}
\end{multline}
The traces appearing above are summarised in Equation~\eqref{eq:tr_SET}. 
The components of the tangent to the geodesic when $x^\mu$ is on the thermal contour 
can be evaluated from Equation~\eqref{eq:n}. The relevant term appearing in 
Equation~\eqref{eq:SET_aux} is
\begin{equation}
 R_z(i\Omega j \beta_0)^\hbeta{}_{\hsigma} n_{\hbeta;j} - n_{\hsigma';j} = 
 \frac{2i}{\sin \frac{s_j}{\ell } \cos\wr} 
 \begin{pmatrix}
  \sinh \frac{j \beta_0}{\ell } \\ 
  \wrho \sinh\Omega j \beta_0 \sin\varphi \\
  -\wrho  \sinh\Omega j \beta_0 \cos\varphi \\
  0
 \end{pmatrix}_\hsigma. \tag{158}
\end{equation}

\subsection{Thermometer Frame Decomposition}
\label{sec:thermometer}

We now consider the decomposition of the SET with respect to the $\beta$
(thermometer) frame,
defined by setting the fluid four-velocity $u^{\mu }$  equal to that corresponding to rigid 
rotation, given in Equation~\eqref{eq:RKT_u_coord}~\cite{van12,van13,landsteiner13lnp,becattini15epjc}:
\begin{equation}
 T_{\halpha\hsigma} = (E + P) u_\halpha u_\hsigma + 
 P \eta_{\halpha\hsigma} + 
 u_\halpha W_\hsigma + W_\halpha u_\hsigma + \Pi_{\halpha\hsigma},
 \label{eq:SET_dec}\tag{159}
\end{equation}
where $E$ and $P$ are the energy density and the isotropic pressure, respectively.
The dynamic pressure, which is proportional to
the projector $\Delta_{\halpha\hsigma} = u_\halpha u_\hsigma + \eta_{\halpha\hsigma}$,
is not included above, since the expansion scalar $\nabla_\mu u^\mu$ vanishes for rigidly rotating flows.
The heat flux in the fluid rest frame, $W_\halpha$, and 
the anisotropic stress $\Pi_{\halpha\hsigma}$, represent quantum deviations 
from the perfect fluid form, giving rise to anomalous transport~\cite{landsteiner13lnp}. The above quantities can be obtained by inverting 
the decomposition \eqref{eq:SET_dec}:
\begin{gather}
 E = T_{\halpha\hsigma} u^\halpha u^\hsigma, \qquad 
 P = \frac{1}{3} \Delta^{\halpha\hsigma} T_{\halpha\hsigma}, \qquad 
 W^\hlambda = -u^\halpha \Delta^{\hlambda\hsigma} T_{\halpha\hsigma}, \nonumber\\
 \Pi^{\hlambda\hbeta} = \left(\Delta^{\hlambda\halpha} \Delta^{\hbeta\hsigma} - 
 \frac{1}{3} \Delta^{\hlambda\hbeta} \Delta^{\halpha\hsigma}\right) 
 T_{\halpha\hsigma}.\tag{160}
\end{gather}

From the structure of Equation~\eqref{eq:SET_aux}, it can be shown that 
the components $T^{\hatr\hatr}$ and $T^{\htheta\htheta}$, expressed with respect to the 
tetrad in Equation~\eqref{eq:tetrad}, are equal. To see this, we start with the expressions
\begin{multline}
 \begin{pmatrix}
  T^{\hatr\hatr} \\
  T^{\htheta\htheta}
 \end{pmatrix} = 
 \begin{pmatrix}
  \sin^2\theta \\
  \cos^2\theta
 \end{pmatrix}(T^{\hatx\hatx} \cos^2\varphi  + 2 T^{\hatx\haty} \sin\varphi \cos\varphi  + 
 T^{\haty\haty} \sin^2\varphi )  +
 \begin{pmatrix}
  \cos^2\theta \\
  \sin^2\theta
 \end{pmatrix} T^{\hatz\hatz} \\ 
 + 2 
 \begin{pmatrix}
  1 \\ 1
 \end{pmatrix}
 \sin\theta \cos\theta(T^{\hatx\hatz} \cos\varphi  +  T^{\haty\hatz} \sin\varphi ).\tag{161}
\end{multline}
The second term in Equation~\eqref{eq:SET_aux} does not contribute to the 
above expressions, and hence the values of $T^{\hatr\hatr}$ and $T^{\htheta\htheta}$ are exactly given by the term proportional to 
$\eta_{\halpha\hsigma}$. Similar arguments can be used to show that 
$T^{\hatt\hatr} = T^{\hatt\htheta} = T^{\hatr\htheta} = T^{\hatr\hvarphi} = T^{\htheta\hvarphi} = 0$.

Due to the fact that the heat flux $W^\halpha$ is, by construction, orthogonal to $u^\halpha$, it can be 
expressed as a linear combination of the vectors $a$, $\omega$ and $\tau$ of the kinematic tetrad (\ref{eq:kinematic_cartesian}). Since $T^{\hatt\hatr} = T^{\hatt\htheta} = 0$, it is clear 
that $W$ must be proportional to $\tau$, because $a$ and $\omega$ have nonvanishing 
components only along $e_\hatr$ and $e_\htheta$. Thus, we find
\begin{equation}
 W^\halpha = \sigma^\tau_\varepsilon \tau^\halpha,
 \label{eq:Walpha_def}\tag{162}
\end{equation}
where $\sigma^\tau_\varepsilon$ is the circular heat conductivity.
Similarly, $\Pi^{\halpha\hsigma}$ must be orthogonal to $u^\halpha$, symmetric 
and traceless. Since $T^{\hatr\hatr} = P + \Pi^{\hatr\hatr}$ is equal to $T^{\htheta\htheta} = P + \Pi^{\htheta\htheta}$, only one degree of freedom 
is required to characterise $\Pi^{\halpha\hsigma}$. Its most general form satisfying
the above restrictions is 
\begin{equation}
\Pi^{\halpha\hsigma} = \Pi_1 \tau^\halpha \tau^\hsigma + 
A a^\halpha a^\hsigma + B \omega^\halpha \omega^\hsigma + C(a^\halpha \omega^\hsigma + 
\omega^\halpha a^\hsigma).\tag{163}
\end{equation} 
The coefficients $A$, $B$ and $C$ must be taken such that 
$\Pi^{\halpha\hsigma}$ remains traceless, $\Pi^{\hatr\hatr} = \Pi^{\htheta\htheta}$ 
and $\Pi^{\hatr\htheta} =0$:
\begin{equation}
 A = -\frac{1}{2} \Pi_1 \omega^2, \qquad 
 B = -\frac{1}{2} \Pi_1 a^2, \qquad 
 C = \frac{1}{2} \Pi_1 (a \cdot \omega).\tag{164}
\end{equation}
The final form can be written in terms of a single quantity $\Pi _{1}$ as follows
 \begin{equation}
 \Pi^{\halpha\hsigma} = \ell^{-4} \Pi_1  \Gamma^6 (\Gamma^2 - 1) (1 - \wO^2)^2 \cos^4 \wr
 \begin{pmatrix}
  \wrho^2 \wO^2 \Gamma^2 & 0 & 0 & \wrho \wO \Gamma^2 \\
  0 & -\frac{1}{2} & 0 & 0 \\
  0 & 0 & -\frac{1}{2} & 0 \\
  \wrho \wO \Gamma^2 & 0 & 0 & \Gamma^2
 \end{pmatrix},\tag{165}
\end{equation}
where the order of the coordinates is $(\hatt, \hatr, \htheta, \hvarphi)$.

It is convenient to compute the energy $E$ and pressure $P$ from $E - 3P = k \ell ^{-1} (SC)$, where 
the SC is given in Equation~\eqref{eq:SCPC_k}, and the combination $E + P$, given below:
\begin{multline}
 E + P = \frac{\Gamma_k}{6\pi^2\ell ^{4}}(2+ k) \sum_{j = 1}^\infty 
 (-1)^{j+1} \zeta_j^{2 + k} \cosh\frac{j \beta_0}{2\ell } \cosh\frac{\Omega j \beta_0}{2} 
 \, {}_2 F_1(k, 3 + k; 1 + 2k; -\zeta_j)\\
 \times \left[\frac{2 \Gamma^2 \zeta_j}{\cos^2 \wr} 
 \left(\sinh \frac{j \beta_0}{\ell } - \wrho^2 \wO \sinh\Omega j \beta_0 \right) 
 \left(\tanh\frac{j \beta_0}{2\ell } - \wrho^2 \wO \tanh\frac{\Omega j \beta_0}{2}\right) - 1\right].\label{eq:EpP}\tag{166}
\end{multline}
The circular heat conductivity is
\begin{multline}
 \sigma^\tau_\varepsilon = \frac{(2+k) \Gamma_k}{4\pi^2 \wO \Gamma^2\ell ^{2}(1 - \wO^2) \cos^4\wr} 
 \sum_{j = 1}^\infty (-1)^{j+1} \zeta_j^{3+k} {}_2F_1(k, 3+k;1+2k;-\zeta_j) \\\times \Bigg[(1 + \wrho^2 \wO^2)\sinh\frac{j \beta_0}{2\ell } \sinh\frac{\Omega j \beta_0}{2} 
 \left(\cosh^2 \frac{j \beta_0}{2\ell } + \cosh^2 \frac{\Omega j \beta_0}{2}\right) \\
 -  2\wO \cosh\frac{j \beta_0}{2\ell } \cosh\frac{\Omega j \beta_0}{2} \left(\sinh^2\frac{j \beta_0}{2\ell } 
 + \wrho^2 \sinh^2\frac{\Omega j \beta_0}{2}\right)\Bigg],
 \label{eq:SET_se}\tag{167}
\end{multline}
while the coefficient $\Pi_1$ can be computed via
\begin{multline}
 \Pi_1 = -\frac{(2+k) \Gamma_k}{3\pi^2 \wO^2 \Gamma^6(1 - \wO^2)^2 \cos^6\wr} 
 \sum_{j = 1}^\infty (-1)^{j+1} \zeta_j^{3+k} {}_2F_1(k, 3+k;1+2k;-\zeta_j) \\\times \Bigg[
 \wO \sinh\frac{j \beta_0}{2\ell } \sinh\frac{\Omega j \beta_0}{2}\left( 
 \cosh^2\frac{j \beta_0}{2\ell } + \cosh^2\frac{\Omega j \beta_0}{2}\right) \\
 - \cosh\frac{j \beta_0}{2\ell } \cosh\frac{\Omega j \beta_0}{2}\left( 
 \wO^2 \sinh^2\frac{j \beta_0}{2\ell } + 
 \sinh^2\frac{\Omega j \beta_0}{2}\right)\Bigg].
 \label{eq:SET_Pi1}\tag{168}
\end{multline}
In the limit of critical rotation, $\wO = \Omega \ell  \rightarrow 1$, the quantities $E + P$, 
$\sigma^\omega_\varepsilon$ and $\Pi_1$ take the following constant values on the equatorial plane:
\clearpage
\end{paracol}
\nointerlineskip
\begin{align}
 \lim_{\Omega \ell \rightarrow 1} (E + P)\rfloor_{\theta = \frac{\pi}{2}} =& 
 \frac{(2 + k) \Gamma_k}{2\pi^2 \ell ^{4}} \sum_{j = 1}^\infty 
 \frac{(-1)^{j+1} \cosh^2 \frac{j \beta_0}{2\ell }}{(\sinh \frac{ j \beta_0}{2\ell })^{4+2k}}
 {}_2 F_1\left(k, 3+k; 1+2k; -{\rm {cosech}}^{2}\frac{j \beta_0}{2\ell }\right),\nonumber\\
 \lim_{\Omega \ell  \rightarrow 1} \sigma_\varepsilon^\tau\rfloor_{\theta = \frac{\pi}{2}} =& 
 \frac{(2 + k) \Gamma_k}{4\pi^2 \ell ^{4}} \sum_{j = 1}^\infty 
 \frac{(-1)^{j+1} \cosh^2 \frac{j \beta_0}{2\ell }}{(\sinh \frac{ j \beta_0}{2\ell })^{5+2k}}
 {}_2 F_1\left(k, 3+k; 1+2k; -{\rm {cosech}}^{2}\frac{j \beta_0}{2\ell }\right)
 \nonumber\\
 & \times \left(\frac{j \beta_0}{2\ell } \cosh\frac{j \beta_0}{2\ell } - \sinh\frac{j \beta_0}{2\ell }\right), \nonumber\\
 \lim_{\Omega \ell \rightarrow 1} \Pi_1\rfloor_{\theta = \frac{\pi}{2}} =& \frac{(2+k) \Gamma_k}{12\pi^2} 
 \sum_{j = 1}^\infty \frac{(-1)^{j+1}}{(\sinh \frac{ j \beta_0}{2\ell })^{6+2k}} {}_2F_1\left(k, 3+k;1+2k; -{\rm {cosech}}^{2}\frac{j \beta_0}{2\ell }\right)\nonumber\\
 & \times \Bigg[\left(\frac{j \beta_0}{2\ell} \right)^2 \cosh\frac{j \beta_0}{\ell} - \frac{j \beta_0}{2\ell} \sinh \frac{j \beta_0}{2\ell} \cosh^2 \frac{j \beta_0}{2\ell} + \frac{1}{4} \sinh^2 \frac{j \beta_0}{\ell}\Bigg].\tag{169}
\end{align}
\begin{paracol}{2}
\switchcolumn

We now turn to  the large temperature behaviour, when the hypergeometric function can be expanded using Equation~\eqref{eq:hyp_largez}:
\begin{equation}
 {}_2F_1(k, 3+k; 1+2k; -\zeta_j) = \frac{\zeta_j^{-k}}{\Gamma_k(2 + k)} \left[2 - \frac{k^2}{\zeta_j} - \frac{k^2(1 - k^2)}{2 \zeta_j^2} + O(\zeta_j^{-3})\right].\tag{170}
\end{equation}
In this case, we find

\end{paracol}
\nointerlineskip
\begin{align}
 E + P =& \frac{7\pi^2 T^4}{45} +  \frac{T^2}{18}\left(3\bm{\omega}^2 + \bm{a}^2 + \frac{R}{12} - 3M^2\right) + \frac{M^2}{24\pi^2}\left(\bm{a}^2 - 3 \bm{\omega}^2 + \frac{R}{3} + 3M^2\right) \nonumber\\ &+ 
 \frac{1}{2160\pi^2} \Bigg[45 \bm{\omega }^4 + 46 \bm{\omega}^2\left(\bm{a}^2 + \frac{R}{12}\right) - 
 51\left(\bm{a}^2 + \frac{R}{12}\right)^2 + 44 (\bm{\omega} \cdot \bm{a})^2
 + 44 \bm{\omega}^2 \frac{R}{12}\Bigg] \nonumber\\
 &+ O(T^{-1}), \nonumber\\
 \sigma_\varepsilon^\tau =& -\frac{T^2}{18} -\frac{1}{360\pi^2} \left[39 \bm{\omega}^2 + 31\left(\bm{a}^2 + \frac{R}{12}\right) - 15M^2\right] + O(T^{-1}), \nonumber\\
 \Pi_1 =& -\frac{2}{27 \pi^2} + O(T^{-1}).\label{eq:SET_highT}\tag{171}
\end{align}
\begin{paracol}{2}
\switchcolumn

\noindent
The validity of the formulae in Equation~(\ref{eq:SET_highT}) for $\sigma^\tau_\varepsilon$ and $\Pi_1$ derived above is investigated by comparison with the exact numerical evaluation of the expressions in \mbox{Equations~\eqref{eq:SET_se}} and \eqref{eq:SET_Pi1} in Figures~\ref{fig:se} and \ref{fig:Pi1}. The energy density $E$ is discussed further below and the results are investigated in Figure~\ref{fig:E}.

Panels (a) and (b) of Figure~\ref{fig:se} show the radial profiles of the circular heat conductivity $\sigma^\tau_\varepsilon$ taken in the equatorial plane for various values of $T_0$, $k$ and $\wO$. The numerical results are in excellent agreement with the analytical result obtained in Equation~\eqref{eq:SET_highT} at high $T_0$ and $k$. Small discrepancies can be seen in the vicinity of the boundary (for $\wO < 1$) and at $T_0 \lesssim 0.5 \ell^{-1}$. The agreement between the numerical results and the analytical expression is maintained also at $k = 2$. Panels (c) and (d) show $- \ell^2 \sigma^\tau_\varepsilon$ and $\ell^2(\sigma^\tau_\varepsilon - \sigma^\tau_{\varepsilon; {\rm an}})$ as functions of the temperature, confirming the validity of all the terms in Equation~\eqref{eq:SET_highT}.

We now discuss the properties of $\Pi_1$, presented graphically in Figure~\ref{fig:Pi1}. Panels (a--c) show profiles of $\Pi_1$ in the equatorial plane. In panel (a), where massless quanta $k = 0$ are considered, a peculiarity of $\Pi_1$ is revealed, namely that it is coordinate-independent when $\wO = 0$. This is due to the fact that the $\cos^6\wr$ term in the denominator of the prefactor in Equation~\eqref{eq:SET_Pi1} is cancelled by the coordinate-dependent part of $\zeta_j^3 = \cos^6\wr / \sinh^6(j \beta_0 / 2\ell)$ (the hypergeometric function reduces to unity when $k = 0$). At finite $\wO$, we see that $\Pi_1$ becomes point-dependent, more strikingly for smaller temperature (no point dependence can be distinguished on the scale of the plot when $\ell T_0 = 2$). The value of $-\Pi_1$ at the origin exhibits a monotonic increase with $\wO$. Panel (b) presents results for $\wO = 0$ and various values of $k$, showing that $-\Pi_1$ becomes point-dependent when $k > 0$, decreasing in the vicinity of the boundary as $k$ is increased. Panel (c) shows results at high temperature ($\ell T_0 = 2$) for vanishing and large ($k = 3$) masses. In the vanishing mass case (also for small masses), $\Pi_1$ is very well approximated by its high temperature limit in Equation~\eqref{eq:SET_highT}. Finally, panel (d) shows $-\Pi_1$ computed at the origin for various values of $k$ and $\wO$ as a function of the temperature $\ell T_0$. It can be seen that the large temperature limit $2 / 27\pi^2$ is achieved in all cases as $T_0$ is increased.

\begin{figure}[H]

\begin{tabular}{cc}
 \includegraphics[width=0.45\linewidth]{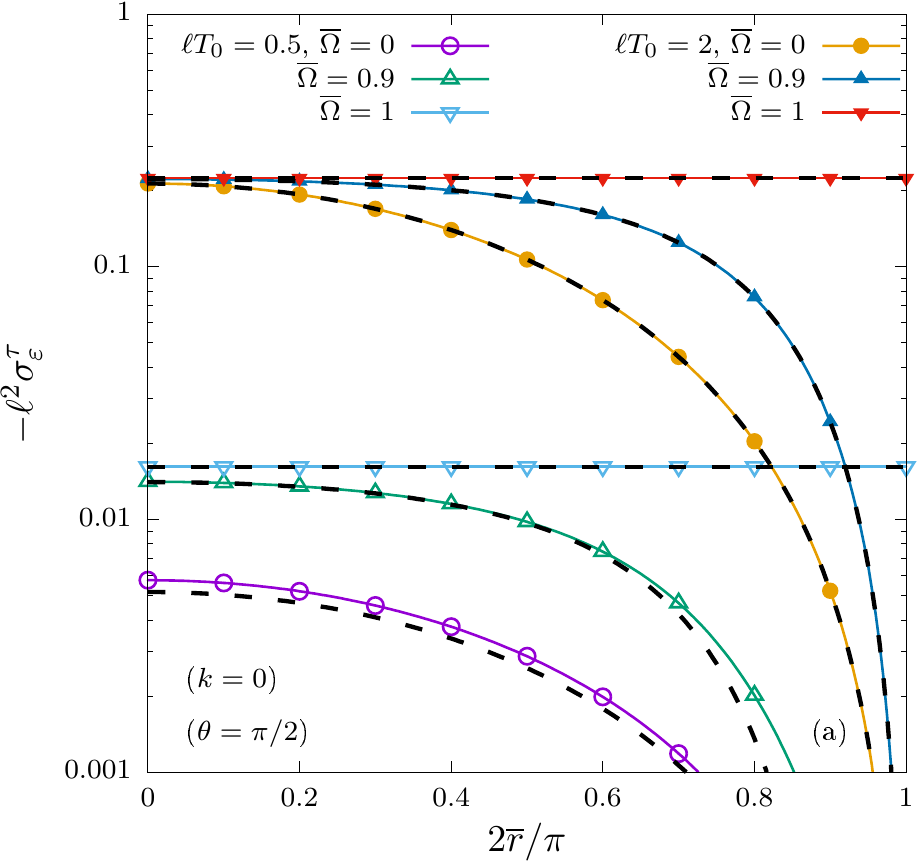} & 
 \includegraphics[width=0.45\linewidth]{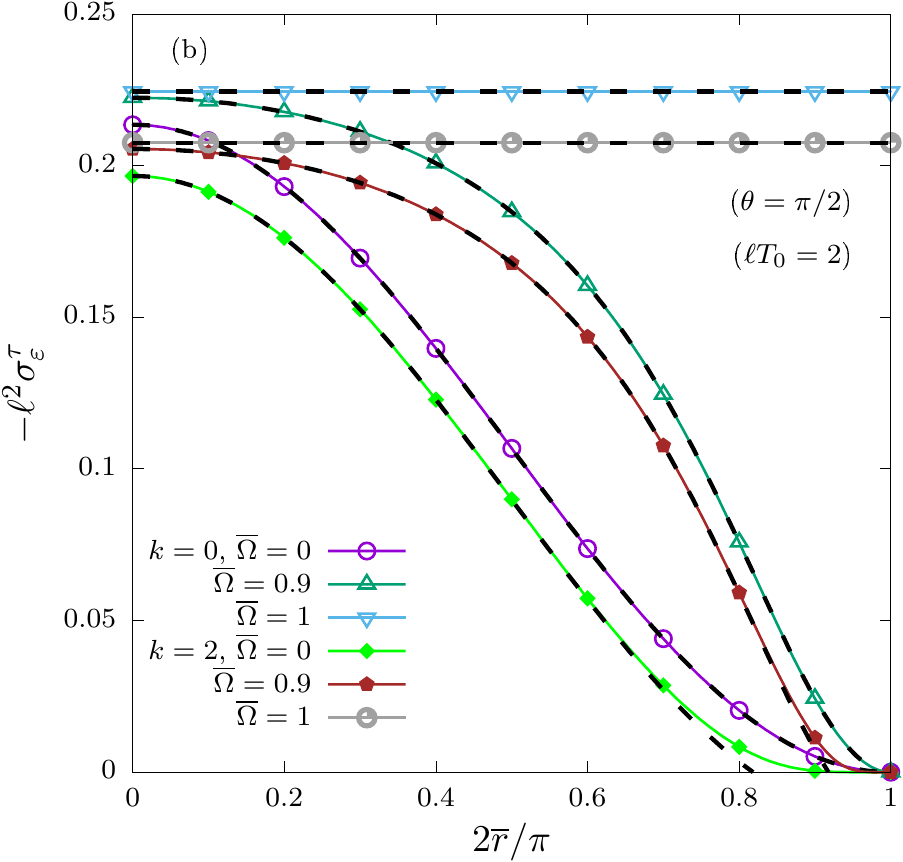}\\
 \includegraphics[width=0.45\linewidth]{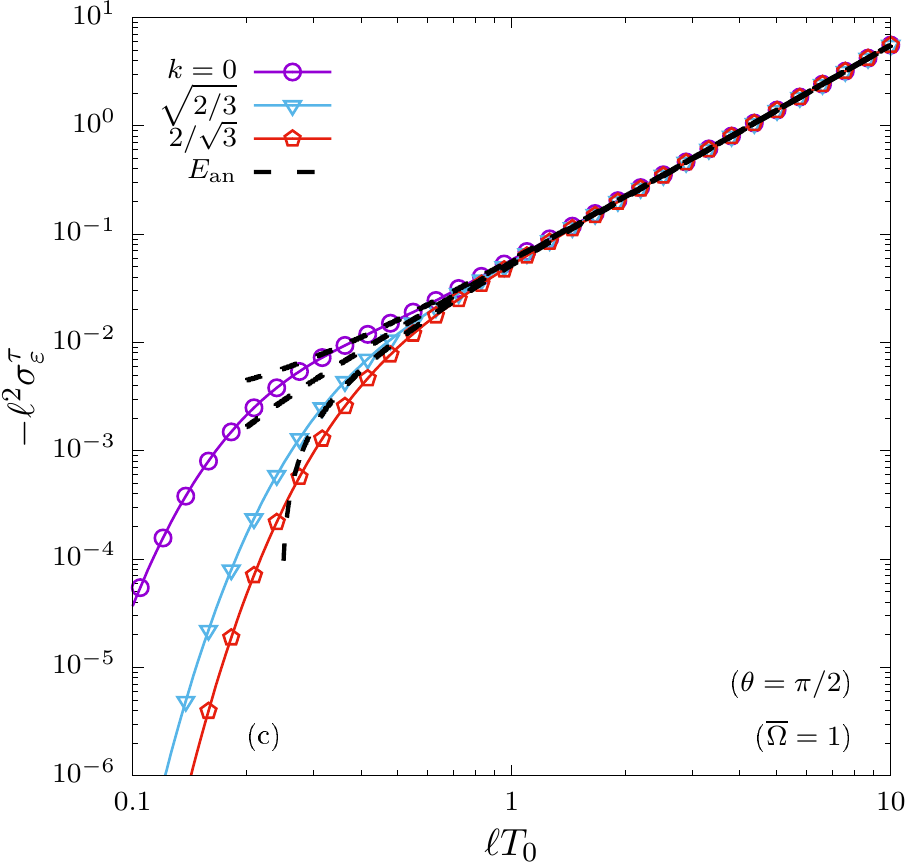} & 
 \includegraphics[width=0.45\linewidth]{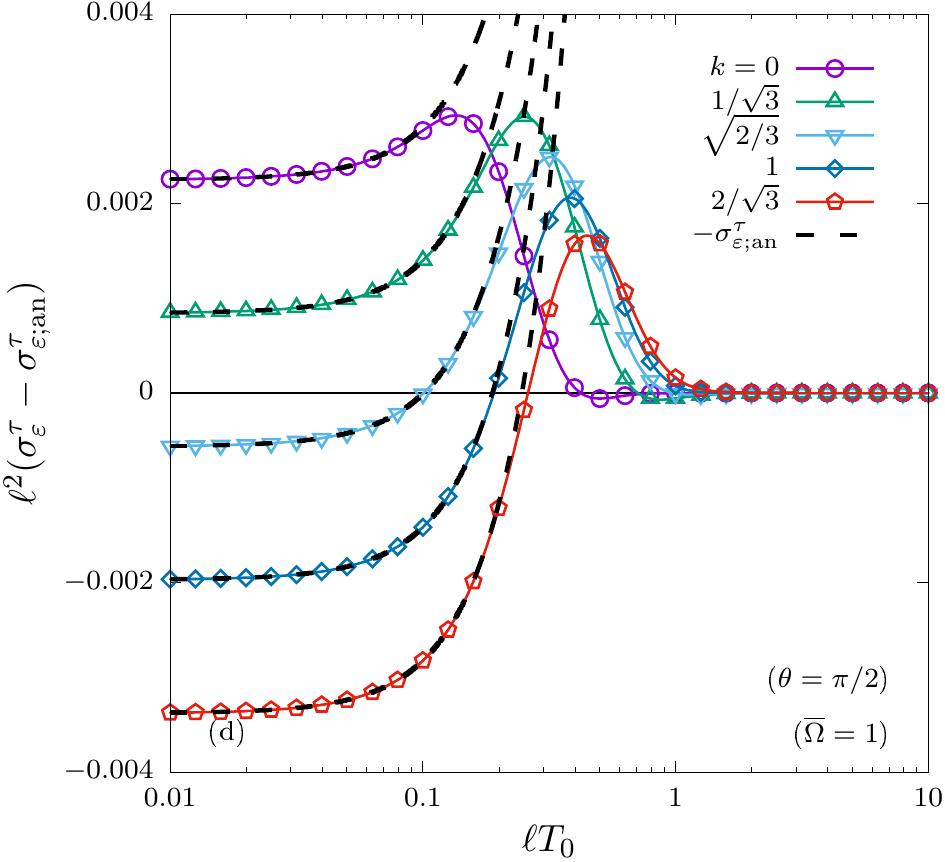} 
\end{tabular}
\caption{(\textbf{a},\textbf{b}) Profiles of the circular heat conductivity $-\sigma^\tau_\varepsilon$ in the equatorial plane ($\theta = \pi/2$). (\textbf{a}) Massless ($k = 0$) quanta at low ($T_0 = 0.5 \ell^{-1}$) and high ($T_0 = 2 \ell^{-1}$) temperatures, for various values of  $\overline{\Omega}$. (\textbf{b}) High-temperature results for $k = 0$ and $k = 2$ at various $\overline{\Omega}$. (\textbf{c}) 
Log-log plot of $-\sigma^\tau_\varepsilon$ at critical rotation ($\overline{\Omega} = 1$) in the equatorial plane ($\theta = \pi/2$) as a function of $T$ for various values of $k$. (\textbf{d}) Linear-log plot of the difference $\sigma^\tau_\varepsilon - \sigma^\tau_{\varepsilon; {\rm an}}$ between the numerical result and the high temperature analytical expression in Equation~(\ref{eq:SET_highT}) at critical rotation $\wO = 1$ and $\theta = \pi/2$. The black dotted lines represent the high-temperature result in Equation~\eqref{eq:SET_highT}.
\label{fig:se}}

\end{figure}

\begin{figure}[H]

\begin{tabular}{cc}
 \includegraphics[width=0.45\linewidth]{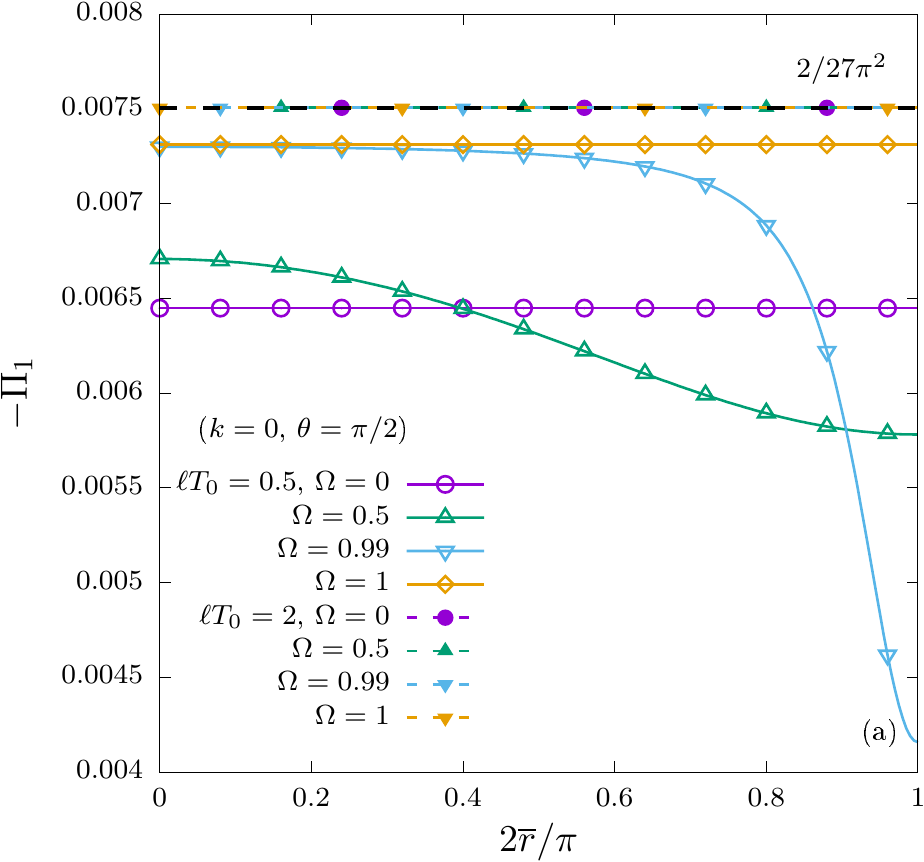} & 
  \includegraphics[width=0.45\linewidth]{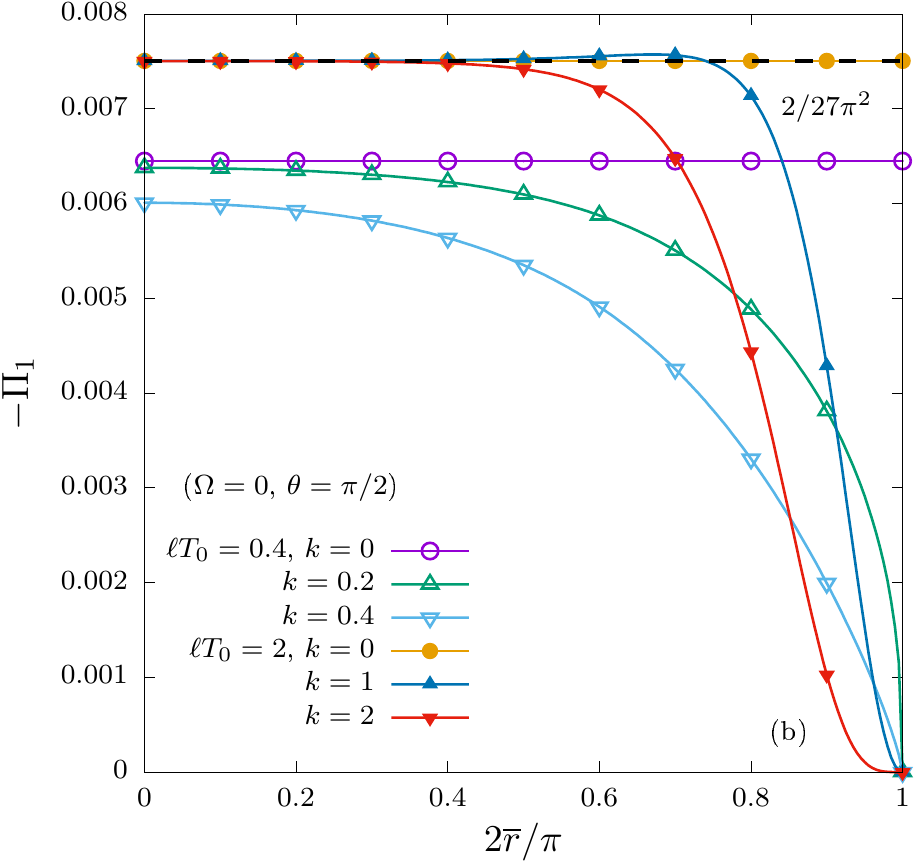} \\
 \includegraphics[width=0.45\linewidth]{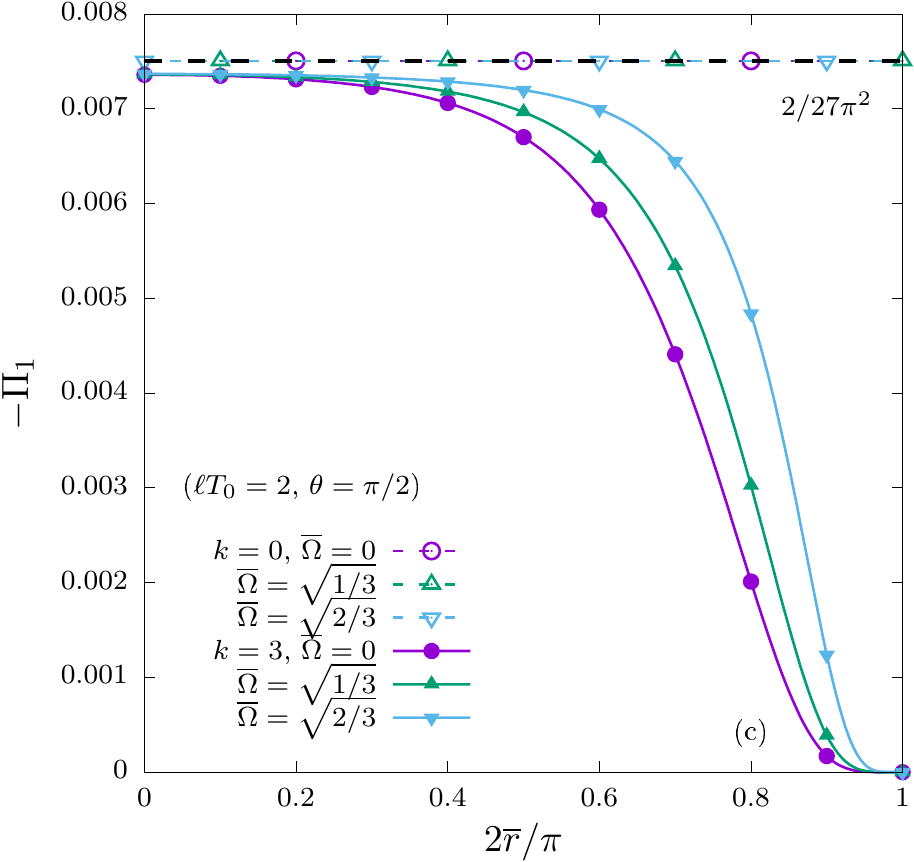} & 
 \includegraphics[width=0.45\linewidth]{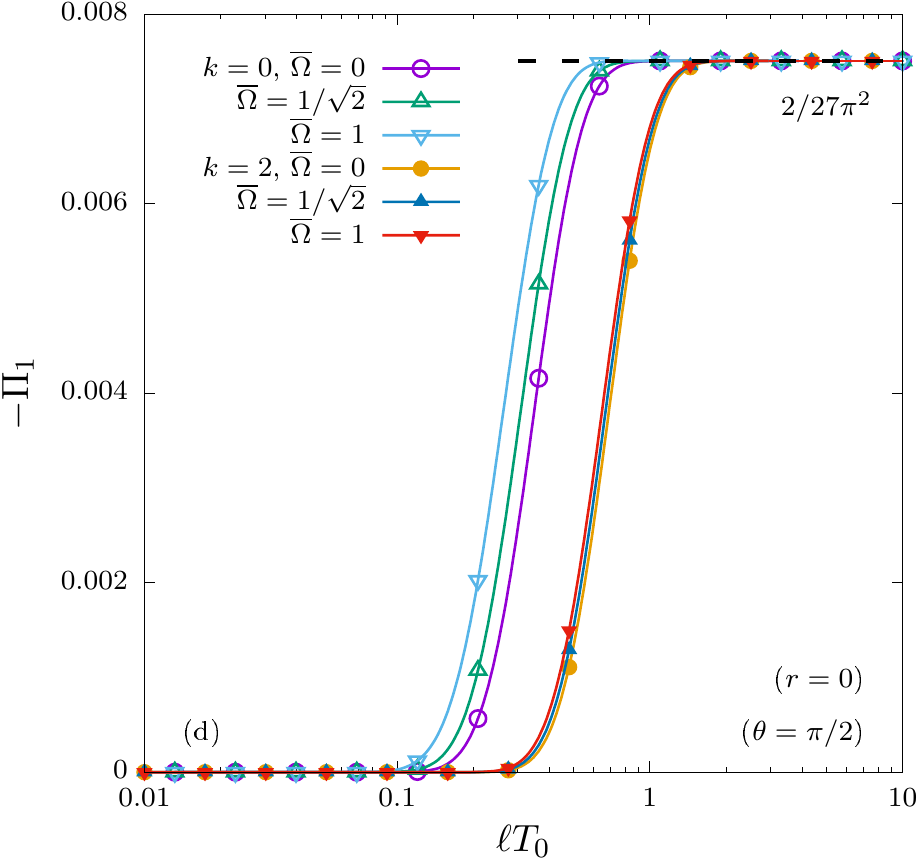}\\
\end{tabular}
\caption{(\textbf{a}--\textbf{c}) Profiles of the coefficient $\Pi_1$, taken with negative sign, in the equatorial plane ($\theta = \pi/2$). (\textbf{a}) Results for massless quanta ($k = 0$) at various values of $\wO$ and $T_0$; (\textbf{b}) Static ($\wO = 0$) states at low ($T_0 = 0.5 \ell^{-1}$) and high ($T_0 = 2 \ell^{-1}$) temperatures, for various values of  $k$; (\textbf{c}) High-temperature ($T_0 = 2\ell^{-1}$) results for various values of $k$ and $\wO$.
(\textbf{d}) Log plot of $\Pi_1$ at the origin $r = 0$ as a function of $T$ for $\wO = 0$, $1/\sqrt{2}$ and $1$ at $k = 0$ and $2$. The black dotted lines represent the high-temperature result in Equation~\eqref{eq:SET_highT}.
\label{fig:Pi1}}

\end{figure}

\subsection{Energy Density and Vacuum Regularisation}
\label{sec:energy}

The energy density can be obtained by adding $\frac{3}{4}(E + P)$ and $\frac{1}{4} M\, SC$,
where the scalar condensate SC is given in Equation~\eqref{eq:SC_highT}, with the result:
\begin{multline}
 E = \frac{7\pi^2 T^4}{60} +  \frac{T^2}{24}\left(3\bm{\omega}^2 + \bm{a}^2 + \frac{R}{12} - 2M^2\right) - \frac{M^2}{8\pi^2}\left(M^2 + \frac{R}{12}\right) \ln \ell \pi T 
  \\ + 
 \frac{1}{2880\pi^2} \Bigg[45 \bm{\omega }^4 + 46 \bm{\omega}^2\left(\bm{a}^2 + \frac{R}{12}\right) - 
 51\left(\bm{a}^2 + \frac{R}{12}\right)^2 + 44 (\bm{\omega} \cdot \bm{a})^2
 + 44 \bm{\omega}^2 \frac{R}{12}\Bigg]\\
 + \frac{M}{16\pi^2 \ell^3} \left(1 - \frac{5k}{6} - k^2 -k^3 - 
 2k(1 - k^2) [\psi(k + 1) + {\mathcal {C}}]\right) 
 \\ 
 + \frac{M^2}{48\pi^2}\left(\bm{a}^2 + \frac{R}{2} - 3 \bm{\omega}^2 + \frac{9M^2}{2}\right)
 + O(T^{-1}).\label{eq:E_highT}\tag{172}
\end{multline}
The above formula is validated in Figure~\ref{fig:E} by comparison with the numerical results obtained by computing the sum in Equation~\eqref{eq:EpP}. Panels (a) and (b) show the profiles of $\ell^4 E$ in the equatorial plane at vanishing mass ($k = 0$) with various values of $T_0$ and $\wO$; and at high temperature ($\ell T_0 = 2$) and various values of $k$ and $\wO$. Excellent agreement between the numerical (continuous lines and points) and analytical (dashed black lines) results can be seen, even at $\ell T_0 = 0.5$ and $k = 2$. Panels (c) and (d) show $\ell^4 E$ and $\ell^4(E - E_{\rm an})$, respectively, as functions of $\ell T_0$. Good agreement with the asymptotic result in Equation~\eqref{eq:E_highT} can be seen in panel (c) for $\ell T_0 \gtrsim 0.5$, while panel (d) confirms the validity of the logarithmic and temperature-independent terms.

\begin{figure}[H]

\begin{tabular}{cc}
 \includegraphics[width=0.45\linewidth]{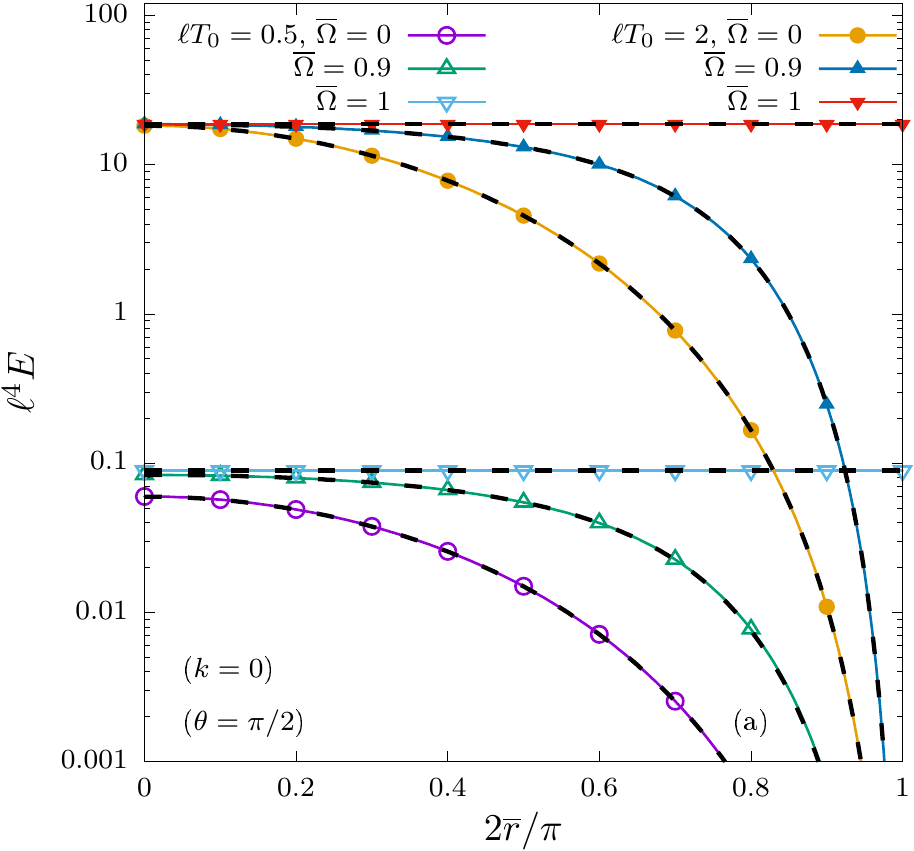} & 
 \includegraphics[width=0.45\linewidth]{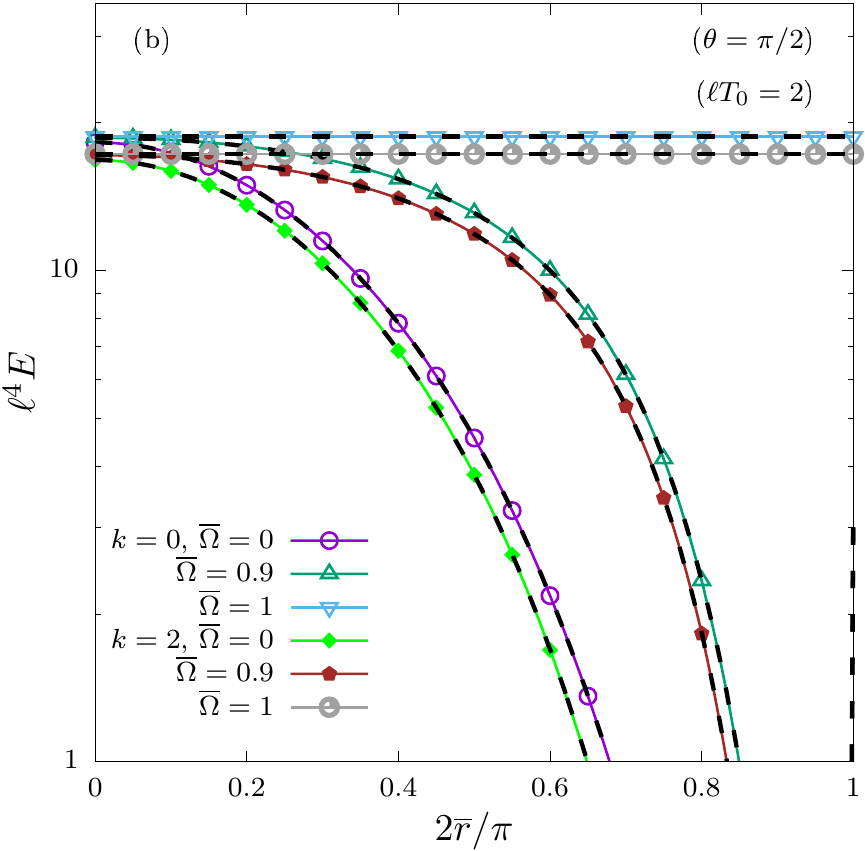}\\
 \includegraphics[width=0.45\linewidth]{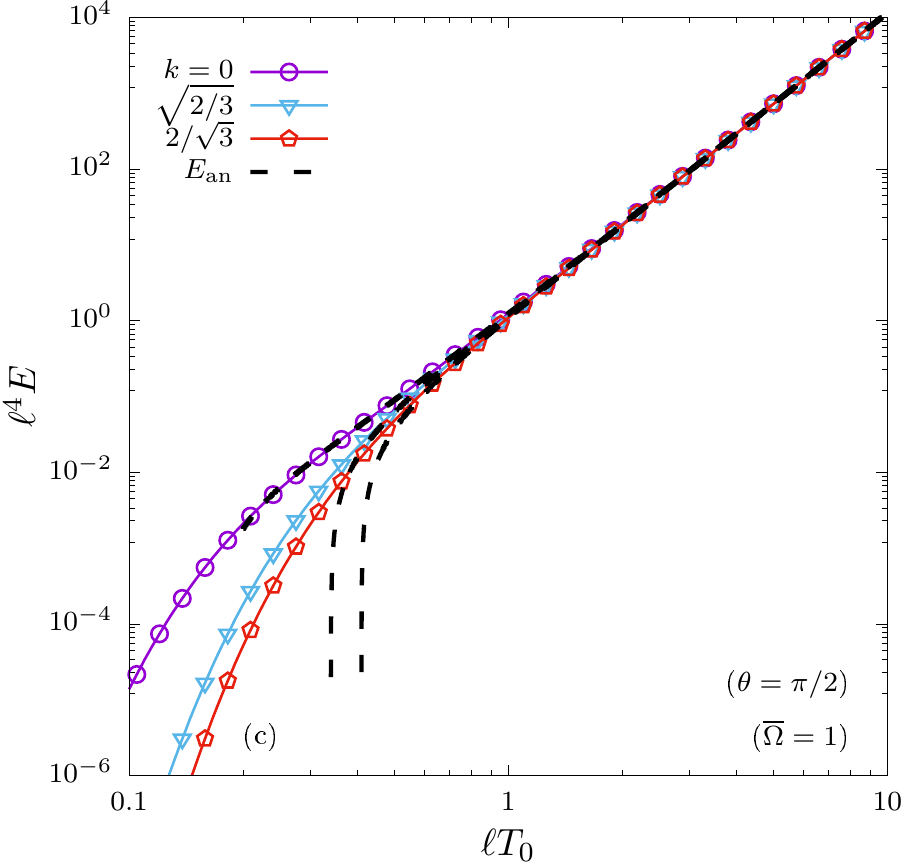} & 
 \includegraphics[width=0.45\linewidth]{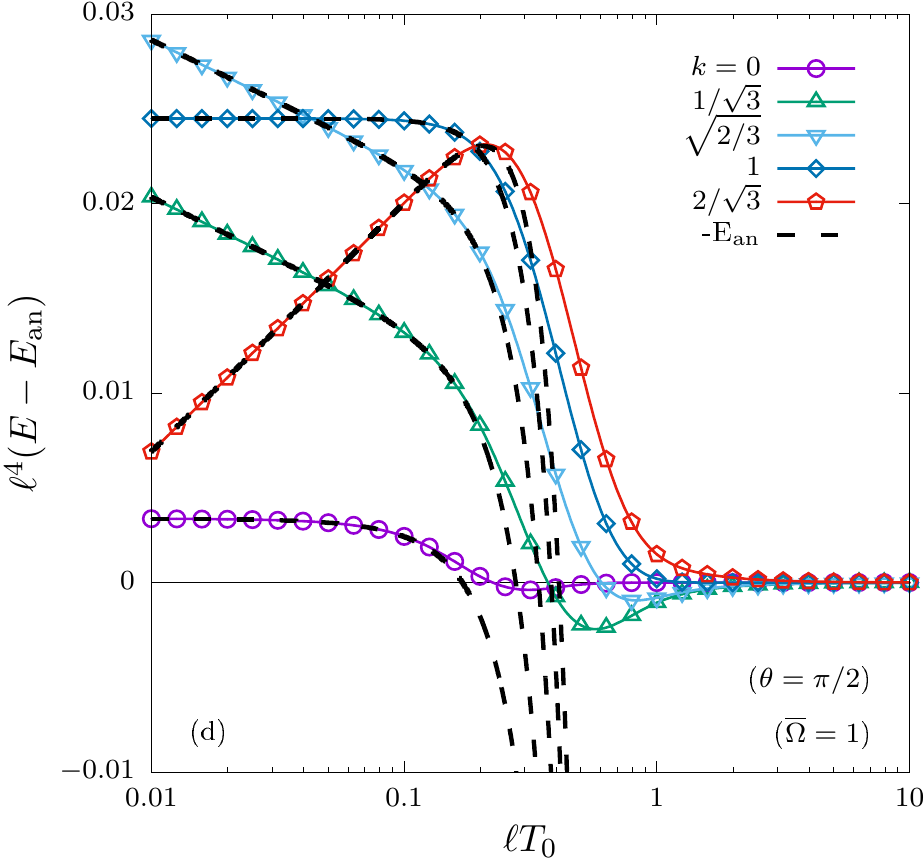} \\
\end{tabular}
\caption{(top line) Profiles of the energy density $E$ in the equatorial plane ($\theta = \pi/2$). (\textbf{a}) Massless ($k = 0$) quanta at low ($T_0 = 0.5 \ell^{-1}$) and high ($T_0 = 2 \ell^{-1}$) temperatures, for various values of  $\overline{\Omega}$. (\textbf{b}) High-temperature results for $k = 0$ and $k = 2$ at various $\overline{\Omega}$. (\textbf{c}) 
Log-log plot of $E$ in the equatorial plane ($\theta = \pi/2$) as a function of $\ell T_0$ for $\overline{\Omega} = 1$ and various values of $k$. (\textbf{d}) Linear-log plot of the difference $E - E_{\rm an}$ between the numerical result and the high temperature analytical expression in Equation~(\ref{eq:E_highT}). The black dotted lines represent the high-temperature result in Equation~\eqref{eq:E_highT}.
\label{fig:E}}

\end{figure}

The term on the penultimate line of Equation~\eqref{eq:E_highT} is in principle compensated by the vacuum expectation value of the stress-energy tensor, which we reproduce here, based on the Hadamard regularisation scheme~\cite{Ambrus:2015mfa}:

\end{paracol}
\nointerlineskip
\begin{align}
 E^{\rm Had}_{\rm vac} =& -P^{\rm Had}_{\rm vac}
 \nonumber \\ =& \frac{1}{16 \pi^2 \ell^4}\left\{\frac{11}{60} 
 + k - \frac{7k^2}{6} - k^3 + \frac{3k^4}{2} - 
 2k^2(k^2 - 1)\left[\psi(k) + \ln \frac{e^{\mathcal {C}}}{\ell \nu_{\rm Had} \sqrt{2}}\right]\right\} \nonumber\\
 =& \frac{1}{16 \pi^2 \ell^4}\left\{\frac{11}{60} 
 - k - \frac{7k^2}{6} + k^3 + \frac{3k^4}{2} - 
 2k^2(k^2 - 1)\left[\psi(1 + k) + \ln \frac{e^{\mathcal {C}}}{\ell \nu_{\rm Had} \sqrt{2}}\right]\right\}.\tag{173}
\end{align}
\begin{paracol}{2}
\switchcolumn

\noindent
It is interesting to note that, since $E^{\rm Had}_{\rm vac} = -P^{\rm Had}_{\rm vac}$, there is 
no vacuum contribution to the sum $E + P$, in other words $E^{\rm Had}_{\rm total} + P^{\rm Had}_{\rm total} = E +P$. 
Also, the vacuum contributions to the energy density and pressure are imperfectly balanced by those 
coming from the \mbox{SC, since}:
\begin{equation}
 E^{\rm Had}_{\rm vac} - \frac{M}{4} SC^{\rm Had}_{\rm vac} = 
 -P^{\rm Had}_{\rm vac} - \frac{M}{4} SC^{\rm Had}_{\rm vac} = \frac{11}{960\pi^2} \left(\frac{R}{12}\right)^2 + \frac{M^2}{16\pi^2} \frac{R}{12} + \frac{M^4}{32\pi^2}.\tag{174}
\end{equation}
The last term $M^4 / 32\pi^2$ survives in the limit $\ell \rightarrow \infty$, hence introducing 
a discrepancy between the flat space limits of $(E_{\rm total}^{\rm Had} - 3P_{\rm total}^{\rm Had}) / M$ and $SC_{\rm total}^{\rm Had}$. The results are given below for definiteness:
\begin{align}
 E^{\rm Had}_{\rm total} =& \frac{7\pi^2 T^4}{60} +  \frac{T^2}{24}\left(3\bm{\omega}^2 + \bm{a}^2 + \frac{R}{12} - 2M^2\right)
 \nonumber \\ & \qquad - \frac{M^2}{8\pi^2}\left(M^2 + \frac{R}{12}\right) \ln \frac{\pi T}{\nu_{\rm Had} \sqrt{2}} + \frac{M^2}{48\pi^2}\left(\bm{a}^2 + R - 3 \bm{\omega}^2 + 6M^2\right) \nonumber\\ 
 &+ \frac{1}{2880\pi^2} \left[45 \bm{\omega }^4 + 46 \bm{\omega}^2\left(\bm{a}^2 + \frac{R}{12}\right) - 
 51\left(\bm{a}^2 + \frac{R}{12}\right)^2 
 \right. \nonumber \\ & \qquad  \left.  + 44 (\bm{\omega} \cdot \bm{a})^2
 + 44 \bm{\omega}^2 \frac{R}{12} + 33\left(\frac{R}{12}\right)^2\right],\nonumber\\
 P^{\rm Had}_{\rm total} =& \frac{7\pi^2 T^4}{180} +  \frac{T^2}{72}\left(3\bm{\omega}^2 + \bm{a}^2 + \frac{R}{12} - 6M^2\right) \nonumber\\
 & \qquad + \frac{M^2}{8\pi^2}\left(M^2 + \frac{R}{12}\right) \ln \frac{\pi T}{\nu_{\rm Had}\sqrt{2}} + \frac{M^2}{48\pi^2}\left(\bm{a}^2 - \frac{R}{3} - 3 \bm{\omega}^2\right) \nonumber\\ 
 &+ \frac{1}{8640\pi^2} \left[45 \bm{\omega }^4 + 46 \bm{\omega}^2\left(\bm{a}^2 + \frac{R}{12}\right) - 
 51\left(\bm{a}^2 + \frac{R}{12}\right)^2 \right. \nonumber \\ & \qquad \left. + 44 (\bm{\omega} \cdot \bm{a})^2
 + 44 \bm{\omega}^2 \frac{R}{12} - 99\left(\frac{R}{12}\right)^2\right] .\label{eq:EP_highT_total}\tag{175}
\end{align}
Now sending $R, \bm{\omega}, \bm{a} \rightarrow 0$, agreement with the RKT result in Equation~\eqref{eq:RKT_highT} can be obtained when the Hadamard regularisation constant takes the value
\begin{equation}
 \nu_{\rm Had}^{\rm SET} = e^{-1/4} \nu_{\rm Had}^{SC} = \frac{M}{\sqrt{2}} e^{{\mathcal {C}} - \frac{3}{4}},
 \label{eq:SET_nuHad}\tag{176}
\end{equation}
which is not the same as the value $\nu_{\rm Had}^{SC}$ in Equation~\eqref{eq:SC_nuHad} required to match 
the RKT prediction for the SC.

\subsection{Comparison with Previous Results}
\label{sec:previous}

We now discuss the above results in connection with those previously obtained in the literature.
In the expression for $E + P$ in Equation~\eqref{eq:SET_highT}, the leading order contribution (proportional to $T^4$), as well 
as the terms involving $M^2 T^2$ and $M^4$, are identical 
to the ones derived within RKT in Equation~\eqref{eq:RKT_highT} and can be regarded as ``classical''. 
The first type of quantum corrections are due to the acceleration 
$\bm{a}$ and vorticity $\bm{\omega}$ of the medium. These corrections can be identified 
by setting the Ricci scalar $R = 0$ and are consistent with previous calculations performed on 
Minkowski space (with rotation and/or with acceleration). 
For example, Equation~(\hl{44}) in Ref.~\cite{Prokhorov:2019sss} gives the energy density 
for a uniformly \mbox{accelerating state,}
\begin{equation}
 E = \frac{7 \pi^2 T^4}{60} + \frac{T^2 \bm{a}^2}{24} - \frac{17 \bm{a}^4}{960\pi^2} 
 +M^2\left(-\frac{T^2}{12} + \frac{\bm{a}^2}{48\pi^2}\right),
 \label{eq:E_acc}\tag{177}
\end{equation}
where the mass correction was presented in Equation~(\hl{6.11}) in Ref.~\cite{Prokhorov:2019yft}.
The above result is consistent with the limit $\bm{\omega} = R = 0$ of Equation~\eqref{eq:E_highT}. 
Equation~\eqref{eq:E_acc} has the remarkable property that, for massless fermions 
($M = 0$), we have $E = 0$ at the Unruh temperature, when $T = |\bm{a}| / 2\pi$ (see also
Ref.~\cite{Becattini:2017ljh} for the scalar field case), however this property 
relies on taking into account the temperature-independent term, $-17 \bm{a}^4 / 960\pi^2$. 
On adS space-time, $E= 0$ can be achieved only when $T = 0$ (since $E>0$ when $T>0$, as can be seen from panels (c) and (d) in Figure~\ref{fig:E}), but the vacuum part $E_{\rm vac}^{\rm Had}$ makes a positive 
contribution when $M = 0$, so $E_{\rm total}^{\rm Had}$ remains positive 
for all temperatures. 
Due to this property and to the 
fact that the acceleration vector $a^\mu$ \eqref{eq:acc} on adS is not uniform even in 
the absence of rotation, an analogy with the Unruh effect on Minkowski space seems difficult 
to make, so we no longer pursue this issue in the present work.

At finite vorticity, both the $\bm{\omega}^2$ and the $\bm{a}^2$ terms were known 
from Minkowski space-time calculations~\cite{Ambrus:2014uqa}. The $\bm{\omega} \cdot \bm{a}$ term 
is not accessible though, since on Minkowski, the vorticity and acceleration 
are perpendicular. According to Ref.~\cite{Ambrus:2014uqa}, the components of the 
thermal expectation value of the SET, including the vacuum contributions due to the 
difference between the rotating (Iyer) and nonrotating (Vilenkin) vacua, are expressed in Equations (\hl{25c)--(25f}) with respect to the co-rotating coordinates ($t_{\rm r} = t, \varphi_{\rm r} = \varphi - \Omega t$) as~follows:
\begin{align}
 T_{t_{\rm r}t_{\rm r}} =& \frac{7\pi^2}{60\beta_0^4 \varepsilon} + \frac{\Omega^2}{8 \beta_0^2 \varepsilon^2}
 \left(\frac{4}{3} - \frac{\varepsilon}{3} \right) + \frac{\Omega^4}{64\pi^2 \varepsilon^3} \left(\frac{8}{9} + \frac{56}{45} \varepsilon - \frac{17}{15} \varepsilon^2\right),\nonumber\\
 -\frac{T_{\varphi_{\rm r} t_{\rm r}}}{\rho^2 \Omega} =& \frac{7\pi^2}{60\beta_0^4 \varepsilon^2} + 
 \frac{13 \Omega^2}{72 \beta_0^2 \varepsilon^3} \left(\frac{16}{13} - \frac{3 \varepsilon}{13}\right) + \frac{119 \Omega^4}{960 \pi^2 \varepsilon^4} \left(\frac{200}{119} - \frac{64}{119} \varepsilon - \frac{\varepsilon^2}{7} \right),\nonumber\\
 \frac{1}{\rho^2} T_{\varphi_{\rm r}\varphi_{\rm r}} =& \frac{7\pi^2}{180\beta_0^4 \varepsilon^3}(4 - 3\varepsilon) + \frac{\Omega^2}{24 \beta_0^2 \varepsilon^4}
 \left(8 -8\varepsilon + \varepsilon^2\right) 
 \nonumber \\ & \qquad + \frac{\Omega^4}{192\pi^2 \varepsilon^5} \left(64 - \frac{456}{5}\varepsilon + \frac{124}{5} \varepsilon^2 + \frac{17}{5} \varepsilon^3\right), \nonumber\\
 T_{\rho\rho} = T_{zz} =& \frac{7\pi^2}{180\beta_0^4 \varepsilon^2} + \frac{\Omega^2}{24 \beta_0^2 \varepsilon^3}
 \left(\frac{4}{3} - \frac{\varepsilon}{3}\right) + \frac{\Omega^4}{192\pi^2 \varepsilon^4} \left(8 - \frac{88}{15} \varepsilon - \frac{17}{15} \varepsilon^2\right),
 \label{eq:SET_vort}\tag{178}
\end{align}
where $\varepsilon = 1 - \rho^2 \Omega^2 = \Gamma^{-2}$, $\beta_0 = \Gamma / T$ and $\rho$ is the distance from the rotation axis measured in the transverse plane.
The energy density can be calculated from the formula $E = u^\mu T_{\mu\nu} u^\nu$ with 
$u^{t_{\rm r}} = \Gamma$ and $u^{\varphi_{\rm r}} = 0$ (the covariant components are 
$u_{t_{\rm r}} = -\Gamma^{-1}$ and \mbox{$u_{\varphi_{\rm r}} = \rho^2 \Omega \Gamma$), giving}
\begin{equation}
 E = \frac{7\pi^2 T^4}{60} +  \frac{T^2}{24}(3\bm{\omega}^2 + \bm{a}^2) 
 + \frac{1}{2880\pi^2} (45 \bm{\omega }^4 + 46 \bm{\omega}^2 \bm{a}^2 - 
 51 \bm{a}^4 ),\tag{179}
\end{equation}
in exact agreement with Equation~\eqref{eq:E_highT} when $M = R = 0$. 
The $O(T^2)$ correction terms were also derived in Refs.~\cite{buzzegoli17,buzzegoli18}. An alternative derivation of the above result, based on the Wigner function formalism, can be found in recent work in Ref.~\cite{Palermo:2021hlf}. The mass corrections appearing in Equation~\eqref{eq:E_highT} are consistent with those derived in Ref.~\cite{buzzegoli17} (see the $\rho$, $U_\alpha$ and $D_\alpha$ entries in \hl{Table 2}). 

Concerning the energy density at high temperatures, one may speculate that the additional term 
involving the Ricci scalar, which appears in the $O(T^2)$ term of 
Equation~\eqref{eq:E_highT} may trace its origin to 
the TTT gravitational (conformal) anomaly~\cite{Chernodub:2019tsx}, however establishing
this connection requires an explicit calculation, which is beyond the scope of the current 
paper. The  $O(T^0)$ term is revealed on Minkowski space as a vacuum term, which arises due 
to the difference between the rotating and static vacua. On adS, this term survives only in 
the large temperature limit, since $\lim_{\beta_0 \rightarrow \infty} E = 0$.

In the case of the circular heat conductivity $\sigma_\varepsilon^\tau$ appearing 
in Equation~\eqref{eq:SET_highT}, the Minkowski expression can be recovered from the results 
quoted in Equation~\eqref{eq:SET_vort} using the definition \eqref{eq:Walpha_def}, namely 
$\sigma_\varepsilon^\tau = \tau_\mu W^\mu / \tau^2$ with 
$W^\mu = -\Delta^{\mu\nu} T_{\nu\lambda} u^\lambda$. On Minkowski space with respect to the co-rotating coordinates, 
$\tau_{t_{\rm r}} = 0$, $\tau_{\varphi_{\rm r}} = -\rho^2 \Omega^3 \Gamma^5$ and $\tau^2 = \rho^2 \Omega^6 \Gamma^8$ (the contravariant components are $\tau^{t_{\rm r}} = -\rho^2 \Omega^4 \Gamma^5$ and $\tau^{\varphi_{\rm r}} = \Omega^3 \Gamma^3$),
which lead to the result
\begin{equation}
 \sigma^\tau_\varepsilon({\rm Minkowski}) = 
 \frac{1}{\rho^2 \Omega^3 \Gamma^4} (\rho^2 \Omega \Gamma^2 T_{t_{\rm r}t_{\rm r}} + T_{\varphi_{\rm r} t_{\rm r}})
 = -\frac{T^2}{18} - \frac{1}{360\pi^2}(39\bm{\omega}^2 + 31\bm{a}^2),\tag{180}
\end{equation}
in perfect agreement with the $R = M = 0$ limit of the appropriate line in  Equation~\eqref{eq:SET_highT}. 
The first term (proportional to $T^2$) was also found in 
Refs.~\cite{buzzegoli17,buzzegoli18,Ambrus:2019ayb}, while the second term (independent of $T$) was also derived in Ref.~\cite{Palermo:2021hlf}.   
The mass corrections appearing in Equation~\eqref{eq:SET_highT} are consistent with those derived in Ref.~\cite{buzzegoli17} (see the $G$ entry in \hl{Table 2}).

Moving now to the shear stress coefficient, $\Pi_1$, we compute this quantity from 
\mbox{Equation~\eqref{eq:SET_vort}} by noting that 
$\Pi_1 = \frac{2}{\bm{a}^2 \bm{\omega}^2}(P - T_{zz})$:
\begin{equation}
\Pi_1 = -\frac{2}{27\pi^2},\tag{181}
\end{equation}
in agreement with the adS result in Equation~\eqref{eq:SET_highT}. 

\subsection{Total Energy}
\label{sec:total}

We end this section by evaluating the total energy density contained within 
the boundaries of adS. For this purpose, we proceed by integrating the 
quantity $E + P$ over the whole adS volume:
\begin{multline}
 V^{E+P}_{\beta_0, \Omega} = \frac{(2+k)\Gamma_k}{6\pi^2 \ell^4} 
 \sum_{j = 1}^\infty (-1)^{j+1} \cosh \frac{j \beta_0}{2\ell} 
 \cosh \frac{\Omega j \beta_0}{2} 
\\ \times \int d^3x \sqrt{-g} \,\zeta_j^{2+k} 
 {}_2F_1(k,3+k;1+2k;-\zeta_j) \mathcal{P}, 
 \label{eq:SET_vol_aux}\tag{182}
 \end{multline}
 where we have defined the quantity 
 \begin{align}
 \mathcal{P} =& \frac{2(\sinh \frac{j\beta_0}{\ell} - \wrho^2 \wO \sinh \Omega j \beta_0)
 (\tanh \frac{j \beta_0}{2\ell} - \wrho^2 \wO \tanh\frac{\Omega j \beta_0 }{2})}
 {(1 - \wrho^2 \wO^2)(\sinh^2 \frac{j \beta_0}{2\ell} - \wrho^2 \sinh^2\frac{\Omega^2 j\beta_0}{2})} - 1 \nonumber\\
 =& \frac{4 [(1 - \wrho^2 \wO \tO)^2 - 2 \wrho^2 \wO \tO \mathcal{S}]}{(1 - \wrho^2 \wO^2)(1 - \wrho^2 \tO^2)} - 1,
 \label{eq:calPdef}\tag{183}
\end{align}
and the following notation was introduced:
\begin{align}
 \tO &= \frac{\sinh(\wO j \beta_0 / 2\ell)}{\sinh(j \beta_0 / 2 \ell)}, \nonumber \\ 
 \mathcal{S} &= \frac{1}{2}\left(\frac{\cosh \frac{j \beta_0}{2\ell}}{\cosh \frac{\wO j \beta_0}{2\ell}} + \frac{\cosh \frac{\wO j \beta_0}{2\ell}}{\cosh \frac{j \beta_0}{2\ell}}\right) - 1 = \frac{2 (\sinh^2 \frac{j \beta_0}{4\ell} - \sinh^2 \frac{\wO j \beta_0}{4\ell})^2}{\cosh \frac{j \beta_0}{2\ell} \cosh \frac{\wO j \beta_0}{2\ell}}.\tag{184}
\end{align}
In order to perform the $d^3x$ integration in Equation~\eqref{eq:SET_vol_aux}, we employ the strategy used for the computation of the flux of axial charge $F_A$ discussed in Section~\ref{sec:CC} and make use of the $(\wrho, \wz)$ coordinates introduced in Equation~\eqref{eq:wOrhoz}. With respect to these coordinates, Equation~\eqref{eq:SET_vol_aux} can be written as:
\begin{multline}
 V^{E+P}_{\beta_0, \Omega} = \frac{(2+k)\Gamma_k}{3\pi \ell} 
 \sum_{j = 1}^\infty (-1)^{j+1} \frac{\cosh \frac{j \beta_0}{2\ell} \cosh \frac{\Omega j \beta_0}{2}}{\sinh^4 \frac{j \beta_0}{2\ell}}
 \int_0^1 \frac{d\wrho\,\wrho \alpha^k}{(1 - \wrho^2 \tO^2)^2} \mathcal{P}\\
 \times 
 \int_{-\infty}^\infty \frac{d\wz}{(1 + \wz^2)^{1+k}} {}_2F_1\left(k,3+k;1+2k;-\frac{\alpha}{1 + \wz^2}\right),
 \label{eq:SET_vol_aux2}\tag{185}
\end{multline}
where
\begin{equation}
 \alpha = \frac{1}{\sinh^2 \frac{j \beta_0}{2\ell}} \frac{1 - \wrho^2}{1 - \wrho^2 \tO^2}.\label{eq:SET_vol_alpha}\tag{186}
\end{equation}
It is convenient to change the argument of the hypergeometric function using \linebreak \mbox{Equation~\eqref{eq:hyp_smallz}}:
\vspace{+12pt}
\end{paracol}
\nointerlineskip
\begin{equation}
 {}_2F_1\left(k, 3+k; 1+2k; -\frac{\alpha}{1+\wz^2}\right) = \left(\frac{1 + \wz^2}{1 + \wz^2 + \alpha}\right)^{3+k} {}_2F_1\left(1+k, 3+k; 1+2k; \frac{\alpha}{1 + \wz^2 + \alpha}\right).\tag{187}
\end{equation}
\begin{paracol}{2}
\switchcolumn

\noindent
Next, replacing the hypergeometric function by its series representation \eqref{eq:hyp_ser}, the $\wz$ integral can be performed analytically using 
\begin{multline}
 \int_{-\infty}^\infty \frac{d\wz \, (1 + \wz^2)^2}{(1 + \alpha + \wz^2)^{3+k+n}} \\ = \frac{ \sqrt{\pi} \Gamma(\frac{1}{2} + k + n)}{(1 + \alpha)^{\frac{5}{2} + k + n} \Gamma(3 + k + n)} \left[(1+k+n)(2+k+n) + (2+k+n)\alpha + \frac{3\alpha^2}{4}\right].\tag{188}
\end{multline}
Furthermore, the series can be resummed, leading to
\begin{multline}
 \int_{-\infty}^\infty \frac{d\wz}{(1 + \wz^2)^{1+k}} {}_2F_1\left(k,3+k;1+2k;-\frac{\alpha}{1 + \wz^2}\right) 
 \\ = \frac{\sqrt{\pi} \Gamma(\frac{1}{2} + k)}{(1+\alpha)^{\frac{5}{2}+k} \Gamma(3+k)}  \Bigg[
 \alpha \left(\frac{3\alpha}{4} + 1\right) {}_2F_1\left(1 + k, \frac{1}{2} + k; 1+2k; \frac{\alpha}{1+\alpha}\right)
 \\ + \alpha(1+k) {}_2F_1\left(2 + k, \frac{1}{2} + k; 1+2k; \frac{\alpha}{1+\alpha}\right) \\
 + (1+k)(2+k) {}_2F_1\left(3 + k, \frac{1}{2} + k; 1+2k; \frac{\alpha}{1+\alpha}\right)\Bigg].
 \label{eq:SET_vol_intz_aux}\tag{189}
\end{multline}
The hypergeometric functions can be expressed in terms of elementary functions using Equation~\eqref{eq:hyp_an_SET}, leading to
\begin{multline}
 \int_{-\infty}^\infty \frac{d\wz}{(1 + \wz^2)^{1+k}} {}_2F_1\left(k,3+k;1+2k;-\frac{\alpha}{1 + \wz^2}\right) \\
 = \frac{\pi}{2\Gamma_k(2+k)(1+\sqrt{1+\alpha})^{2k}} 
 \left[4 + \frac{k(6 + 5\alpha)}{(1+\alpha)^{3/2}} + \frac{2k^2}{1+\alpha}\right].\tag{190}
\end{multline}
Substituting the above result into Equation~\eqref{eq:SET_vol_aux2}, we arrive at
\begin{multline}
 V^{E+P}_{\beta_0, \Omega} = \sum_{j = 1}^\infty \frac{(-1)^{j+1}}{6\ell} \frac{\cosh \frac{j \beta_0}{2\ell} \cosh \frac{\Omega j \beta_0}{2}}{ \sinh^4 \frac{j \beta_0}{2\ell}} 
 \\ \times \int_0^1 \frac{d\wrho\, \wrho}{(1 - \wrho^2 \tO^2)^2} \left(\frac{\sqrt{\alpha}}{1 + \sqrt{1 + \alpha}}\right)^{2k}\left[4 + \frac{k(6+5\alpha)}{(1 + \alpha)^{3/2}} + \frac{2k^2}{1+ \alpha}\right] \mathcal{P}.
 \label{eq:SET_vol_rho}\tag{191}
\end{multline}

In order to extract the high temperature limit of $V_{\beta_0,\Omega}^{E+P}$, first of all we must account for its divergence when $\wO \rightarrow 1$. This can be achieved by changing the integration variable in Equation~\eqref{eq:SET_vol_rho} from $\wrho$ to 
\begin{equation}\tag{192}
\chi = \alpha \sinh^2 \frac{j \beta_0}{2\ell},
\end{equation} 
 where $\alpha$ was introduced in Equation~\eqref{eq:SET_vol_alpha}. We obtain
\begin{multline}
 V^{E+P}_{\beta_0, \Omega} = \sum_{j = 1}^\infty \frac{(-1)^{j+1}}{12\ell} \frac{\cosh \frac{j \beta_0}{2\ell} \cosh \frac{\Omega j \beta_0}{2}}{ ( 1- \tO^2) \sinh^4 \frac{j \beta_0}{2\ell}} 
 \int_0^1 d\chi \,\left(\frac{\sqrt{\chi}}{\sinh\frac{j \beta_0}{2\ell} +\sqrt{\sinh^2\frac{j \beta_0}{2\ell} + \chi}}\right)^{2k}\\\times
 \left[4 + \frac{6 \sinh^2\frac{j \beta_0}{2\ell} +5\chi}{(\sinh^2\frac{j \beta_0}{2\ell} + \chi)^{3/2}} k \sinh\frac{j \beta_0}{2\ell} + \frac{2k^2 \sinh^2\frac{j \beta_0}{2\ell}}{\sinh^2\frac{j \beta_0}{2\ell} + \chi}\right] \mathcal{P}.
 \label{eq:SET_vol_chi}\tag{193}
\end{multline}
Since we are interested only in the small $\beta_0$ behaviour, it is tempting to perform a series expansion of the integrand with respect to $\beta_0$ and then integrate order by order. We require terms up to $\beta_0^4$ to balance the $\sinh^4 \frac{j \beta_0}{2\ell}$ factor in the denominator of the factor appearing in front of the integral. However, the higher order powers of $\beta_0$ also introduce negative powers of $\sqrt{\chi}$, making this procedure invalid. While the full integral cannot be computed analytically, we first note that $\mathcal{P}$ (\ref{eq:calPdef}) admits the following series expansion with respect to $\beta_0$:
\begin{equation}\tag{194}
 \mathcal{P} = 3 - \frac{8}{9} \left(\frac{j \beta_0}{2\ell}\right)^4 \wO^2 (1 - \chi)(1 - \wO^2 \chi) + O(\beta_0^6).
\end{equation}
The above expansion shows that the dependence of $\mathcal{P}$ on $\chi$ comes in only at the fourth order with respect to $\beta_0$. It is not too difficult to observe that the second and third terms in Equation~\eqref{eq:SET_vol_chi} contribute terms of order $\sinh \frac{j \beta_0}{2\ell}$, thus for these terms, the $O(\beta_0^4)$ contribution to $\mathcal{P}$ can be neglected. We thus perform the computation in two steps. First, we approximate $\mathcal{P} \simeq 3$ (corresponding to its $\beta_0 = 0$ limit). The $\chi$ integral can be performed by switching the integration variable to $\alpha = \chi / \sinh^2\frac{j \beta_0}{2\ell}$ and noting that
\begin{multline}
 \sinh^2 \frac{j \beta_0}{2\ell} \int_0^{(\sinh\frac{j \beta_0}{2\ell})^{-2}} d\alpha \,\left(\frac{\sqrt{\alpha}}{1 +\sqrt{1 + \alpha}}\right)^{2k}
 \left[4 + \frac{k(6+5\alpha)}{(1 + \alpha)^{3/2}} + \frac{2k^2}{1+ \alpha}\right]
 \\ = 4 e^{-j M \beta_0} \left(1 + \frac{k}{2} \tanh \frac{j \beta_0}{2\ell}\right).\tag{195}
\end{multline}
The contribution of the fourth order term in $\mathcal{P}$ can be estimated by setting $\beta_0 = 0$ in the integrand, giving 
\begin{multline}
 \int_0^1 d\chi \,\left(\frac{\sqrt{\chi}}{\sinh\frac{j \beta_0}{2\ell} +\sqrt{\sinh^2\frac{j \beta_0}{2\ell} + \chi}}\right)^{2k}
 \\ \times
 \left[4 + \frac{6 \sinh^2\frac{j \beta_0}{2\ell} +5\chi}{(\sinh^2\frac{j \beta_0}{2\ell} + \chi)^{3/2}} k \sinh\frac{j \beta_0}{2\ell} + \frac{2k^2 \sinh^2\frac{j \beta_0}{2\ell}}{\sinh^2\frac{j \beta_0}{2\ell} + \chi}\right] (\mathcal{P} - 3) \\
 \simeq -\frac{32}{9}  \wO^2\left(\frac{j \beta_0}{2\ell}\right)^4 \int_0^1 d\chi (1 - \chi) (1 - \wO^2 \chi)
 \\ = -\frac{16}{27} \left(\frac{j \beta_0}{2\ell}\right)^4 \wO^2 (3 - \wO^2).\tag{196}
\end{multline}
Substituting the above results into Equation~\eqref{eq:SET_vol_chi} yields
\begin{multline}
 V^{E+P}_{\beta_0, \Omega} \simeq \sum_{j = 1}^\infty \frac{(-1)^{j+1}}{\ell} \frac{\cosh \frac{j \beta_0}{2\ell} \cosh \frac{\Omega j \beta_0}{2}}{ ( 1- \tO^2) \sinh^4 \frac{j \beta_0}{2\ell}} 
 \\ \times \left[e^{-j M \beta_0}\left(1 + \frac{k}{2} \tanh\frac{j \beta_0}{2\ell}\right) - \frac{4}{81} \left(\frac{j \beta_0}{2\ell}\right)^4 \wO^2 (3 -\wO^2)\right],\tag{197}
\end{multline}
where the $O(\beta_0^6)$ contributions were ignored inside the square brackets.
We now perform a small $\beta_0$ expansion inside the summand and then calculate the sum term by term to find
\begin{multline}
 V^{E+P}_{\beta_0, \Omega} = \frac{\ell^3}{1 - \wO^2} \Bigg[\frac{7 \pi^4 T_0^4}{45} - 9 M T_0^3 \zeta(3) + \frac{\pi^2 T_0^2}{18}\left(6M^2 + \Omega^2 + \frac{R}{12}\right)
 \\ - \frac{M  T_0 \ln 2}{6}\left(4M^2 + 3\Omega^2 + \frac{R}{12}\right) \\
 + \frac{1}{6480}\left(-29\Omega^4 + 540 M^2 \Omega^2 + \frac{177}{6} \Omega^2 R - 45 M^2 R - \frac{17}{16} R^2\right) + O(T_0^{-1})\Bigg].\label{eq:SET_vol}\tag{198}
\end{multline}
Using  the volume integral of the SC given in Equation~\eqref{eq:SC_V_highT}, it can be seen that the volume integral $V^E_{\beta_0,\Omega} = \frac{3}{4} V^{E+P}_{\beta_0,\Omega} + \frac{M}{4} V^{SC}_{\beta_0, \Omega}$ of the energy density can be written as
\begin{multline}
 V^E_{\beta_0, \Omega} = V^{E_{\rm RKT}}_{\beta_0, \Omega} + 
 \frac{\ell^3}{1 - \wO^2}\Bigg[\frac{\pi^2 T_0^2}{24} \left(\Omega^2 + \frac{R}{12}\right) - \frac{M T_0 \Omega^2}{3} \ln 2 \\
 + \frac{1}{8640}\left(-29 \Omega^4 + 360 M^2 \Omega^2 + \frac{177}{6} \Omega^2 R - 90 M^2 R - \frac{17}{16} R^2\right)+ O(T_0^{-1})\Bigg].
 \label{eq:SET_VE_highT}\tag{199}
\end{multline}

\begin{figure}[H]

\begin{tabular}{cc}
 \includegraphics[width=0.45\linewidth]{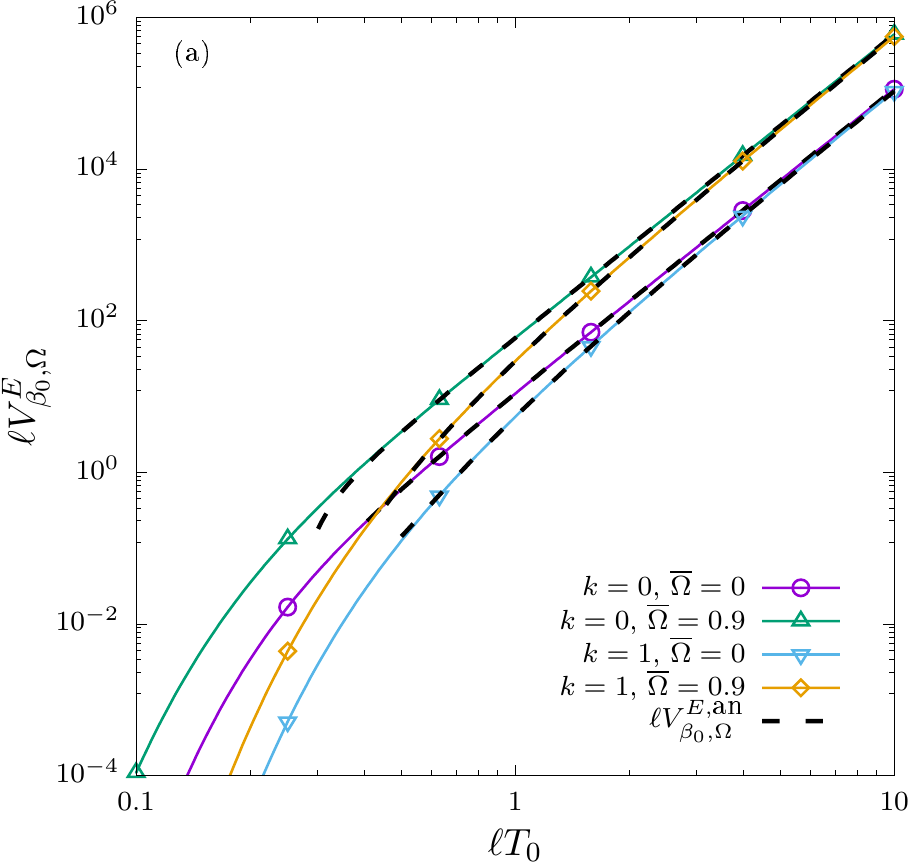} &
 \includegraphics[width=0.45\linewidth]{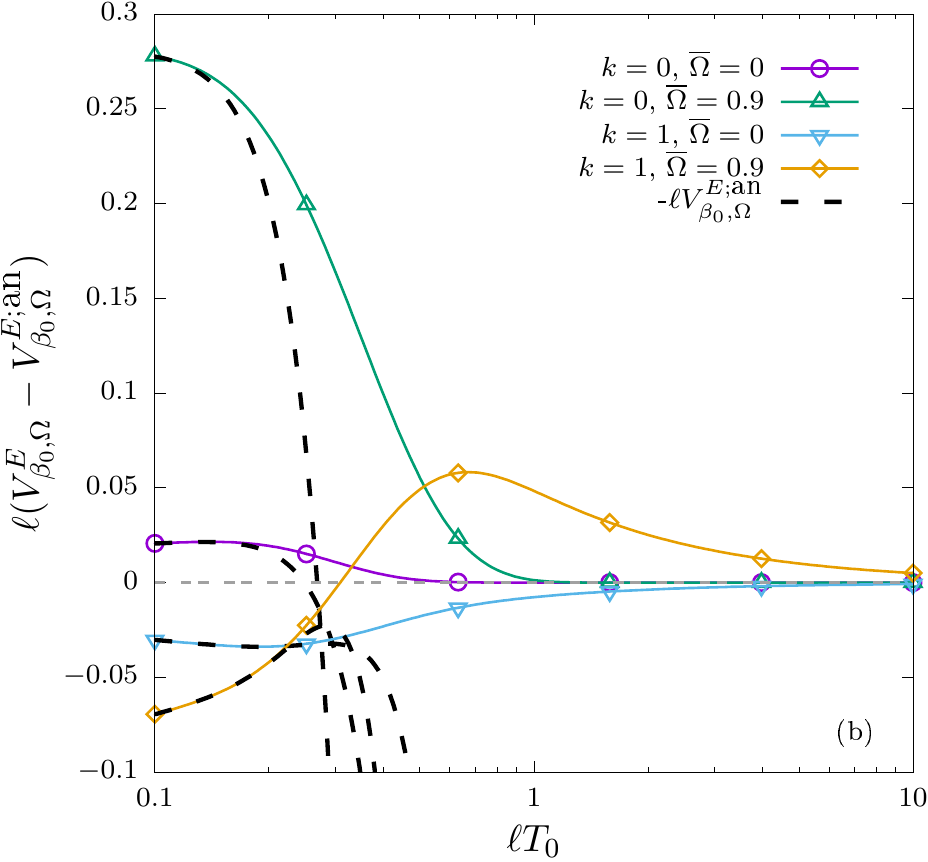}\\
\end{tabular}



\begin{center}
\includegraphics[width=0.45\linewidth]{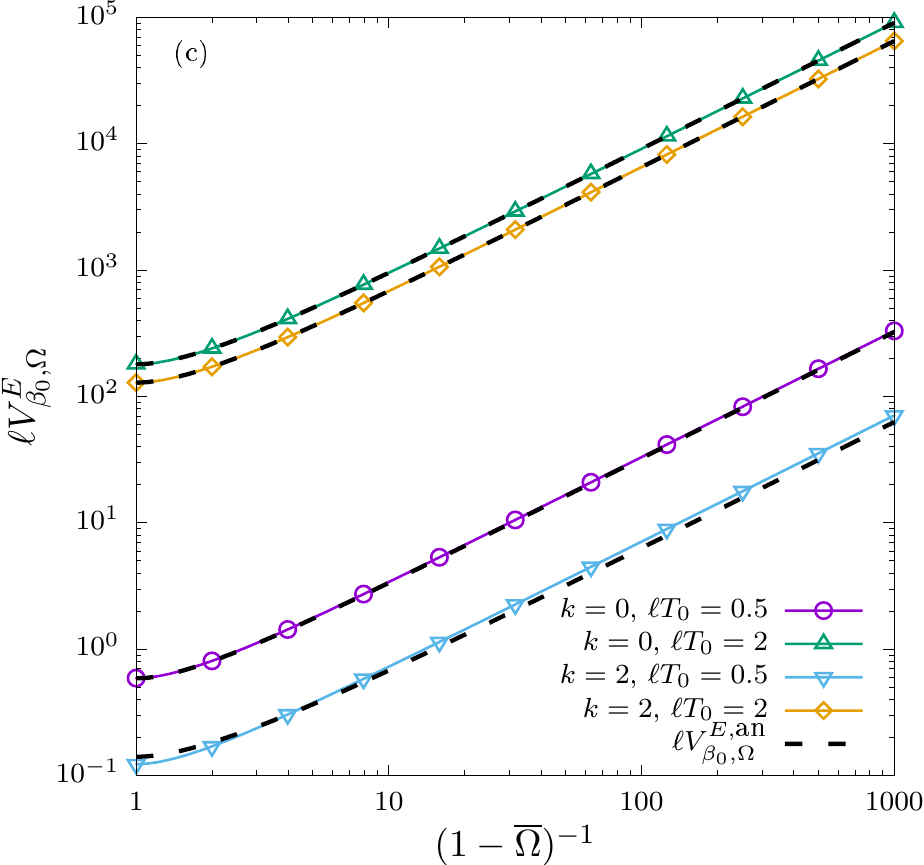}
\end{center}
\caption{(\textbf{a}) Log-log plot of $V^E_{\beta_0,\Omega} = \frac{3}{4} V^{E+P}_{\beta_0,\Omega} + \frac{M}{4} V^{\rm SC}_{\beta_0, \Omega}$, computed using Equations~\eqref{eq:SC_V} and \eqref{eq:SET_vol_rho}, as a function of $\ell T_0$. (\textbf{b}) Difference $V^E_{\beta_0,\Omega} - V^{E; {\rm an}}_{\beta_0,\Omega}$ as a function of  $\ell T_0$. (\textbf{c}) $V_{\beta_0,\Omega}^E$ as a function of $(1 - \overline{\Omega})^{-1}$ for various values of $k$ and $\ell T_0$. The dashed lines represent the high temperature limit given in Equation~\eqref{eq:SET_VE_highT}.
\label{fig:SET_VE}}

\end{figure}

Compared to the RKT prediction $V^{E_{\rm RKT}}_{\beta_0,\Omega}$ derived in Equation~\eqref{eq:RKT_vol_highT}, there are corrections appearing already at order $O(T_0^2)$. The high temperature limit obtained in \mbox{Equation~\eqref{eq:SET_VE_highT}} is validated in Figure~\ref{fig:SET_VE} by comparison with  numerically obtained results for $V^E_{\beta_0,\Omega} = \frac{3}{4} V^{E+P}_{\beta_0,\Omega} + \frac{M}{4} V^{\rm SC}_{\beta_0,\Omega}$, computed from Equations~\eqref{eq:SC_V} and \eqref{eq:SET_vol_rho}. Panel (a) validates the large temperature asymptotic, while panel (b) shows that the analytical expression in Equation~\eqref{eq:SET_VE_highT} recovers all leading order terms down to and including the $O(T_0^0)$ term. Finally, panel (c) confirms the divergence with respect to $\wO$ as the limit of critical rotation $\wO \rightarrow 1$ is approached.


\section{Discussion and Conclusions}
\label{sec:conc}

In this paper we have studied the properties of rotating vacuum and thermal states for free fermions on adS.  
We restricted our attention to the case when the rotation rate is sufficiently small that no SLS forms. This enabled us to exploit the
maximal symmetry of the underlying space-time and use a geometric approach to find the vacuum and thermal two-point functions.
We have investigated the properties of thermal states by computing the expectation values of the SC, PC, VC, AC and SET.
Our analysis concerns only the case of vanishing chemical potential, leaving the study of finite chemical potential effects for future work.

At the beginning of our work we put forward three questions, which we now address in turn.
\begin{enumerate}
    \item Is the rotating fermion vacuum state distinct from the global static fermion vacuum on adS?
\end{enumerate}

For a quantum scalar field, the rotating and static vacua coincide irrespective of the angular speed, both on Minkowski and on adS \cite{Kent:2014wda}.
For a fermion field on unbounded Minkowski space-time, the rotating and static vacua do not coincide. In the situation of small rotation rate $\Omega$ considered here, there is no SLS, and the rotating fermion vacuum coincides with the global static vacuum, as we have shown on the basis of the quantisation of energy derived in Ref.~\cite{Cotaescu:1998ts}.
\begin{enumerate}
\setcounter{enumi}{1}
    \item Can rigidly-rotating thermal states be defined for fermions on adS? 
    \end{enumerate}
    
    This question has a simple answer (at least when there is no SLS): yes, and we have constructed these states in this paper. 
    For a quantum scalar field, this question is yet to be explored in the literature, although one might anticipate, in analogy with the situation in Minkowski space-time, that rigidly-rotating thermal states can be defined only when there is no SLS.
    Similarly, we expect that rigidly-rotating states for fermions can be constructed on adS even when there is an SLS, and plan to address this in future work. 
    \begin{enumerate}
        \setcounter{enumi}{2}
    \item What are the properties of these rigidly-rotating states?
\end{enumerate}

Answering this question has been the main focus of our work in this paper. 
We have considered the situation when the angular speed $|\Omega \ell| \le 1$ and there is no SLS.
In this case there are two competing factors at play.
First, static thermal radiation in adS tends to clump away from the boundary~\cite{Allen:1986ty,Ambrus:2017cow,Ambrus:2018olh}.
Second, in Minkowski space-time, the energy density $E$ of rotating thermal radiation increases as the distance from the axis of rotation increases~\cite{Ambrus:2014uqa}. Our results indicate that at any distance $r$ from the origin, $E$ increases monotonically with the angular velocity $\Omega$. In the limit of critical rotation $\Omega = \ell^{-1}$, the energy density becomes constant in the equatorial plane. A similar conclusion holds for the other quantities considered in this paper, namely the scalar condensate (SC), the axial vortical conductivity $\sigma^\omega_A$, the circular heat conductivity $\sigma^\tau_\varepsilon$ and the pressure deviator coefficient $\Pi_1$. 

Thus, on adS, we find that the properties of rotating vacuum and thermal states mirror those on Minkowski space-time.
In particular: 
\begin{enumerate}[label=(\alph*)]
    \item The rotating and nonrotating vacua are identical for both scalar and fermion fields in the case where there is no SLS;
    \item If there is no SLS, rigidly-rotating thermal states can be defined for fermion fields (and presumably also for scalar fields), and these states are regular everywhere in the space-time;
    \item For sufficiently large temperatures, all quantities (SC, $\sigma^\omega_A$, $E$, $P$, $\sigma^\tau_\varepsilon$, $\Pi_1$) reproduce the corresponding Minkowski expressions, 
    plus corrections proportional to the Ricci scalar $R$ due to the space-time curvature.
\end{enumerate}

We conjecture that if the angular speed $\Omega \ell > 1$, then rigidly-rotating thermal states cannot be defined for scalar fields, and can be defined for fermion fields, but further analysis is required to answer this definitively.

Taking advantage of the bounded nature of adS, we were able to evaluate the total SC and energy contained inside the adS boundary. Compared to estimates based on relativistic kinetic theory, we highlighted quantum corrections to the high temperature $T$ limit,  appearing at next-to-next-to-leading order. We also considered the axial flux $F_A(\wz)$ through two-dimensional slices of adS which are orthogonal to the local vorticity vector $\omega^\mu$, corresponding to constant values of the effective vertical coordinate $\wz$. By analogy with Minkowski space, the chiral vortical effect induces a nonvanishing axial flux through the equatorial plane. We have shown that in the case of massless fermions,  conservation of the axial current requires the axial flux to penetrate the adS boundaries, originating from the southern hemisphere and leaving adS through its northern hemisphere (defined with respect to the orientation of $\bm{\Omega}$). We were also able to show that for nonvanishing fermion mass, $F_A$ is  zero on the adS boundary, which can be understood by noting that timelike geodesics require an infinite time to reach the adS boundary. In this case, the axial flux appearing in the equatorial plane due to the axial vortical effect is converted into a nonvanishing distribution of the pseudoscalar condensate $PC$, which is antisymmetric with respect to the equatorial plane and integrates to zero over the whole adS volume.

In the introduction to this work, we outlined the analogy between the Unruh effect and the definition of quantum states on static black hole space-times~\cite{Sewell:1982zz,Kay:1985zs,Kay:1988mu}.
We close our discussion with some thoughts on the analogy between the definition and properties of rigidly-rotating quantum states in Minkowski and adS space-times and those on the corresponding rotating black hole space-times.
We consider only the space-time exterior to the black hole event horizon, and posit that the Boulware state~\cite{Boulware:1974dm,Boulware:1975pe} is analogous to the rigidly-rotating vacuum, while the Hartle-Hawking state~\cite{Hartle:1976tp} will correspond to a rigidly-rotating thermal state.

Let us first consider asymptotically flat rotating Kerr black holes~\cite{Kerr:1963ud} (for which there is always an SLS) and a quantum scalar field.
In this case the Hartle-Hawking state does not exist~\cite{Kay:1988mu,Ottewill:2000qh,Ottewill:2000yr}, in accordance with the nonexistence of  rigidly-rotating thermal states.
Since there is no rigidly-rotating vacuum state distinct from the nonrotating vacuum, one would expect that there is also no Boulware state on Kerr. 
Indeed, no vacuum state exists on Kerr black holes which is as empty as possible at both future and past null infinity~\cite{Ottewill:2000qh}. 

The situation is markedly different for fermion fields on Kerr black holes~\cite{Casals:2012es}.
In particular, a ``Hartle-Hawking''-like state exists for fermions, and this state is regular close to the event horizon but divergent on the SLS~\cite{Casals:2012es}, as observed for rigidly-rotating thermal states in unbounded Minkowski space-time~\cite{Ambrus:2014uqa}.
There is also a ``Boulware''-like state. 
This is a vacuum state far from the black hole, which is regular there, but diverges at the stationary limit surface~\cite{Casals:2012es}, which is the surface at which, due to the rotation of the black hole, an observer can no longer remain at rest relative to the infinity. 
There is no equivalent to the stationary limit surface for unbounded Minkowski space-time, but nonetheless the analogy between rigidly-rotating states and the situation on the black hole space-time \mbox{remains pertinent}.

With this in mind, what do the results presented here for rotating states in adS imply for asymptotically adS rotating black holes?
Although further research is required before we have a complete picture of scalar fields on pure adS, our results for fermion fields are nontheless suggestive.
Rotating Kerr-adS black holes~\cite{Carter:1968ks,Plebanski:1976gy} do not necessarily have an SLS, depending on their angular speed. 
If the angular speed of the black hole is sufficiently small so there is no SLS, our pure adS results lead us to conjecture that there will be a well-defined Hartle-Hawking state regular throughout the space-time.
This conjecture is in line with studies of the thermodynamics of Kerr-adS black holes in the context of the adS/CFT correspondence (see, for example,~\cite{Caldarelli:1999xj,Hawking:1998kw,Hawking:1999dp}), when it is shown that if there is no SLS, the Kerr-adS black hole can be in thermal equilibrium with a bath of radiation at the Hawking temperature, and there is a associated state in the boundary CFT. 
We await future work to examine whether the above conjecture holds. 

\vspace{6pt}

\authorcontributions{Conceptualization, V.E.A.~and E.W.; methodology, V.E.A.; software, V.E.A.; validation, V.E.A.; formal analysis, V.E.A.; investigation, V.E.A.; resources, N/A; data curation, N/A; writing---original draft preparation, V.E.A. and E.W.; writing---review and editing, V.E.A. and E.W.; visualization, N/A; supervision, E.W.; project administration, N/A; funding acquisition, V.E.A.~and E.W. All authors have read and agreed to the published version of the manuscript.}

\funding{The work of V.E.A.~is supported by a grant from the Romanian 
National Authority for Scientific Research and Innovation, CNCS-UEFISCDI,
project number PN-III-P1-1.1-PD-2016-1423. The work of E.W.~is supported by the Lancaster-Manchester-Sheffield Consortium for Fundamental
Physics under STFC grant ST/T001038/1 and partially supported by the 
H2020-MSCA-RISE-2017 Grant No. FunFiCO-777740.}

\acknowledgments{We thank Stephen A.~Fulling for the invitation to contribute to this topical collection. We also thank the organizers and participants in the virtual conference ``Acceleration and Radiation:  Classical and Quantum,
Electromagnetic and Gravitational'' for the stimulating presentations and discussion.}

\conflictsofinterest{The authors declare no conflicts of interest.}

\appendixtitles{yes}

\appendixstart
\appendix

\renewcommand{\theequation}{\thesection.\arabic{equation}}

\section{Thermal Spinor Traces Involving the Bispinor of Parallel Transport}\label{app:traces}

The following traces are useful for the computation of the scalar and pseudoscalar condensates in Section~\ref{sec:SCPC}:
\begin{align}
 {\rm tr}[\Lambda(x,x')] =& \frac{4 \sec\frac{s}{2\ell } \cos\frac{\Delta \wt}{2}}
 {\sqrt{\cos\wr \cos\wr'}} \left(\cos\frac{\wr}{2} \cos\frac{\wr'}{2}
 - \cos\Upsilon \sin\frac{\wr}{2} \sin\frac{\wr'}{2}\right),\nonumber \\
 {\rm tr}[\gamma^\hatt \gamma^\hatz \Lambda(x,x')] =& 
 -\frac{4 \sec\frac{s}{2\ell } \sin\frac{\Delta \wt}{2}}
 {\sqrt{\cos\wr \cos\wr'}} \left(\sin\frac{\wr}{2} \cos\frac{\wr'}{2} \cos\theta 
 + \cos\frac{\wr}{2} \sin\frac{\wr'}{2} \cos\theta'\right),\nonumber \\
 {\rm tr}[\gamma^5 \gamma^\hatt \gamma^\hatz \Lambda(x,x')] =& 
 \frac{4 i \sec\frac{s}{2\ell } \cos\frac{\Delta \wt}{2}}
 {\sqrt{\cos\wr \cos\wr'}} \sin\frac{\wr}{2} \sin\frac{\wr'}{2} 
 \sin\theta \sin\theta' \sin \Delta \varphi,
 \label{eq:tr_FC}\tag{A1}
\end{align}
while ${\rm tr}[\gamma^5 \Lambda(x,x')] = 0$.

For the computation of the vector (VC) and axial (AC) currents 
in Section~\ref{sec:CC}, the following traces are required:
\begin{align}
 {\rm tr}[\gamma^\hatt \slashed{n} \Lambda(x,x')] =& 
 \frac{4 \, {\rm cosec}\frac{s}{2\ell } \sin\frac{\Delta \wt}{2}}
 {\sqrt{\cos\wr\cos\wr'}} \left( 
 \cos\frac{\wr}{2} \cos\frac{\wr'}{2} + 
 \cos\Upsilon \sin\frac{\wr}{2} \sin\frac{\wr'}{2}\right),\nonumber\\
 {\rm tr}[\gamma^\hati \slashed{n} \Lambda(x,x')] =& 
 \frac{4\,  {\rm cosec}\frac{s}{2\ell } \cos\frac{ \Delta \wt}{2}}
 {\sqrt{\cos\wr\cos\wr'}} 
 \frac{x^\hati}{r}\left(\sin\frac{\wr}{2} \cos\frac{\wr'}{2}  - 
 \frac{x'^\hati}{r'} \cos\frac{\wr}{2} \sin\frac{\wr'}{2}  \right),\nonumber\\
 {\rm tr}[\gamma^5 \slashed{n} \Lambda(x,x') \gamma^\hati] =& 
 \frac{4 i\, {\rm cosec}\frac{s}{2\ell} \sin\frac{\Delta \wt}{2}}
 {\sqrt{\cos\wr\cos\wr'}} 
  \left(\frac{\vx \times \vx'}{rr'}\right)^\hati
 \sin\frac{\wr}{2} \sin\frac{\wr'}{2} 
,\nonumber\\
 {\rm tr}[\gamma^5 \gamma^\hatt \gamma^\hati \slashed{n} \Lambda(x,x') \gamma^\hatz] =& 
 \frac{4 i \, {\rm cosec}\frac{s}{2\ell } \cos\frac{\Delta \wt}{2}}
 {\sqrt{\cos\wr\cos\wr'}} 
 \varepsilon^{\hatt\hati\hatj\hatz}\left(
  \frac{x_\hatj}{r} \sin\frac{\wr}{2} \cos\frac{\wr'}{2} -
 \frac{x'_\hatj}{r'} \cos\frac{\wr}{2} \sin\frac{\wr'}{2} \right),\nonumber\\
 {\rm tr}[\gamma^\hatt \gamma^\hati \slashed{n} \Lambda(x,x') \gamma^\hatz] =& 
 \frac{4 \, {\rm cosec}\frac{s}{2\ell } \sin\frac{\Delta \wt}{2}}
 {\sqrt{\cos\wr\cos\wr'}} 
 \left[
 \left(\frac{x^\hati z' - z x'^\hati}{rr'} + 
 \eta^{\hati\hatz} \cos\Upsilon \right)\sin\frac{\wr}{2} \sin\frac{\wr'}{2}   \right. \nonumber\\
 & \left. + 
 \eta^{\hati\hatz} \cos\frac{\wr}{2} \cos\frac{\wr'}{2}\right],
 \label{eq:tr_CC}\tag{A2}
\end{align}
while ${\rm tr}[\gamma^\hatt \gamma^5 \slashed{n} \Lambda(x,x')] = 0$.

Finally, the computation of the stress-energy tensor (SET) 
in Section~\ref{sec:SET}, requires the following formulae:
\begin{align}
 {\rm tr}(e^{-\Omega j \beta_0 S^\hatz} \Lambda_j) =& 
 \frac{4 \cosh \frac{j \beta_0}{2\ell } \cosh \frac{\Omega j \beta_0}{2}}
 {\cos \frac{s_j}{2\ell }}, \nonumber\\
 R_z(i \Omega j \beta_0)^\hlambda{}_{\halpha} 
 {\rm tr}(e^{-\Omega j \beta_0 S^\hatz} \gamma_\hlambda \slashed{n} \Lambda_j) =& 
 -\frac{4i \cosh \frac{j \beta_0}{2\ell } \sinh\frac{\Omega j \beta_0}{2}}{\cos\wr \sin\frac{s_j}{2\ell }} 
 \begin{pmatrix}
  \tanh \frac{j \beta_0}{2\ell } \coth\frac{\Omega j \beta_0}{2} \\
   \wrho \sin \varphi \\
  -\wrho \cos \varphi \\ 
  0
 \end{pmatrix}_\halpha,
 \label{eq:tr_SET}\tag{A3}
\end{align}
where $\wrho = \sin\wr \sin \theta$.

\section{High Temperature Summation Formulae}\label{app:summ}

In this paper, the computation of thermal expectation values involves
a summation over the index $j$ which labels the thermal contour images $x_j$ of the 
space-time coordinate $x$. At high 
temperatures, the summation over this index reduces to expressions
of the following type:
\begin{equation}
 \sum_{j = 1}^\infty \frac{(-1)^{j + 1}}{j^\nu} = \left(1 - \frac{1}{2^{\nu - 1}}\right) \zeta(\nu),
 \label{eq:sumj_zeta}\tag{A4}
\end{equation}
where the Riemann zeta function $\zeta(\nu)$ is defined by
\begin{equation}
 \zeta(\nu) = \sum_{j = 1}^\infty \frac{1}{j^\nu}.
 \label{eq:zeta_def}\tag{A5}
\end{equation}
Clearly, Equation~\eqref{eq:zeta_def} is valid only for ${\rm Re}(\nu) > 1$, while elsewhere $\zeta(\nu)$ is defined by analytic continuation. In the same spirit, Equation~\eqref{eq:sumj_zeta} is strictly valid only when $\nu > 1$, however the limit for $\nu = 1$ can be obtained using 
\begin{equation}\tag{A6}
 \lim_{\nu \rightarrow 1}\left(1 - \frac{1}{2^{\nu - 1}}\right) \zeta(\nu) = \ln 2,
\end{equation}
which follows after noting that $\zeta(1 + \delta) = \delta^{-1} + O(1)$.

We now compile a number of summation formulae which are 
useful for performing the calculations presented in the main text.
At finite temperature, the summation over $j$ typically involves the
function $\zeta_j$ introduced in Equation~\eqref{eq:zetaj_def},
which admits the following large temperature expansion:
\begin{equation}\tag{A7}
 \zeta_j = \frac{4 T^2\ell ^{2}}{j^2 } - \frac{\Gamma^4 \cos^2\wr}{3} \left(1 - \wrho^2 \wO^4\right) + O(T^{-1}),\label{eq:zetaj_largeT}
\end{equation}
where $T = \Gamma \cos\wr / \beta_0$ is the local temperature. For the computation of the scalar (SC) and pseudoscalar (PC) condensates in Section~\ref{sec:SCPC}, the following summation formulae \mbox{are required}:
\begin{align}
 \sum_{j = 1}^\infty (-1)^{j+1} \zeta_j \cosh\frac{ j \beta_0}{2\ell } \cosh\frac{\Omega j \beta_0}{2} =& 2\pi^2\ell ^{2} \left[\frac{T^2}{6} +
 \frac{1}{24\pi^2}\left(3\bm{\omega}^2 - \bm{a}^2 - \frac{R}{12}\right)\right] 
 \nonumber \\ &  \qquad + O(T^{-1}),\nonumber\\
 \sum_{j = 1}^\infty (-1)^{j+1} \cosh\frac{j \beta_0}{2\ell } \cosh\frac{\Omega j \beta_0}{2} =& \frac{1}{2} + O(T^{-1}), \nonumber\\
 \sum_{j = 1}^\infty (-1)^{j+1} \ln(\zeta_j^{-1/2}) \cosh\frac{ j \beta_0}{2\ell } \cosh\frac{\Omega j \beta_0}{2} =& -\frac{1}{2} \ln \pi T\ell.
 \label{eq:SC_summation}\tag{A8}
\end{align}

For the computation of the large temperature behaviour of the 
axial vortical conductivity $\sigma_A^\omega$ considered in
Section~\ref{sec:CC}, the following summation formulae are useful:
\begin{align}
 \sum_{j = 1}^\infty (-1)^{j+1} \frac{\zeta_j^2 \sinh\frac{ j \beta_0}{2\ell } \sinh\frac{\Omega j \beta_0}{2}}{2\pi^2 \ell^3 \Omega \Gamma^2 \cos^2 \wr} =& \frac{T^2}{6} + \frac{1}{24\pi^2}\left(\bm{\omega}^2 + 3\bm{a} + \frac{R}{4}\right) + O(T^{-1}),\nonumber\\
 \sum_{j = 1}^\infty (-1)^{j+1} \frac{\zeta_j \sinh\frac{ j \beta_0}{2\ell } \sinh\frac{\Omega j \beta_0}{2}}{2\pi^2 \ell^3 \Omega \Gamma^2 \cos^2 \wr} =& \frac{1}{4\ell^2 \pi^2} + O(T^{-1}).
 \label{eq:sA_summation}\tag{A9}
\end{align}
The computation of the high temperature expansion of the axial flux $F_A(\bar{z})$ through slices of constant $\bar{z}$, discussed also in Section~\ref{sec:CC}, requires the following formulae:
\begin{align}
 \sum_{j = 1}^\infty \frac{(-1)^{j + 1} \sinh\frac{\Omega j \beta_0}{2}}{(\sinh^2\frac{j \beta_0}{2\ell } - \sinh^2\frac{\Omega j \beta_0}{2})\sinh\frac{j \beta_0}{2\ell}} =& 
 \frac{\wO}{1 - \wO^2}\left[\frac{1}{3}(\pi \ell T_0)^2 - \frac{3 + \wO^2}{12} + O(T_0^{-1})\right],\nonumber\\
 \sum_{j = 1}^\infty \frac{(-1)^{j + 1} \sinh\frac{j \beta_0}{2\ell} \sinh\frac{\Omega j \beta_0}{2}}{\sinh^2\frac{j \beta_0}{2\ell } - \sinh^2\frac{\Omega j \beta_0}{2}} =& 
 \frac{\wO}{2(1 - \wO^2)} + O(T_0^{-1}),\nonumber\\
 \sum_{j = 1}^\infty \frac{(-1)^{j + 1} \sinh\frac{j \beta_0}{2\ell} \sinh\frac{\Omega j \beta_0}{2}}{\sinh^2\frac{j \beta_0}{2\ell} - \sinh^2\frac{\Omega j \beta_0}{2}} \ln j =& 
 \frac{\wO}{2(1 - \wO^2)} \ln \frac{\pi}{2} + O(T_0^{-1}).
 \label{eq:sumj_FA}\tag{A10}
\end{align}

\section{Special Functions}\label{app:hyp}

The Gauss hypergeometric function ${}_2F_1(a,b;c;Z)$ 
admits the following series representation~\cite{NIST:DLMF},
\begin{equation}
 {}_2F_1(a, b; c; Z) = \sum_{n = 0}^\infty \frac{(a)_n (b)_n Z^n}{n! (c)_n},\label{eq:hyp_ser}\tag{A11}
\end{equation}
valid when $|Z| < 1$. At finite mass, the vacuum propagator
\eqref{eq:geom_SF_def} and 
its thermal analogue~\eqref{eq:SF_beta} involve the hypergeometric
functions ${}_2F_1(k,2+k;1+2k; Z)$ and ${}_2F_1(1+k,2+k;1+2k; Z)$
[see Equation~\eqref{eq:SF_AB_gen}]. In this section of the Appendix,
we compile a handful of properties of the
hypergeometric functions, which are useful for the
calculations presented in
Sections~\ref{sec:SCPC}--\ref{sec:SET}.

First, noting that the first two arguments $k$ (or $k+1$) and $k+2$
differ by an integer, the expansion for large argument can be 
performed using the formula~\cite{NIST:DLMF}
\begin{multline}
 \frac{{}_2F_1(a, a+m; c; Z)}{\Gamma(c)} = \frac{(-Z)^{-a}}{\Gamma(a+ m)} \sum_{{\mathsf {k}} = 0}^{m-1} \frac{(a)_{\mathsf {k}} (m - {\mathsf {k}} - 1)!}{{\mathsf {k}}! \Gamma(c - a - {\mathsf {k}})} Z^{-{\mathsf {k}}}
\\ + \frac{(-Z)^{-a}}{\Gamma(a)} \sum_{{\mathsf {k}} = 0}^\infty \frac{(a + m)_{\mathsf {k}} (-1)^{\mathsf {k}} Z^{-{\mathsf {k}}-m}}{{\mathsf {k}}!({\mathsf {k}}+m)! \Gamma(c - a - {\mathsf {k}} - m)} \\\times \left[\ln(-Z) + \psi({\mathsf {k}} + 1) + \psi({\mathsf {k}} + m + 1) - \psi(a + {\mathsf {k}} + m) - \psi(c - a - {\mathsf {k}} - m)\right],
 \label{eq:hyp_largez}\tag{A12}
\end{multline}
where $(a)_{\mathsf {k}} = \Gamma(a + {\mathsf {k}}) / \Gamma(a)$ is the Pochhammer symbol and the digamma function \mbox{$\psi(Z)$ satisfies }
\begin{equation}
 \psi(Z) = \frac{d \ln\Gamma(Z)}{dZ}, \qquad 
 \psi(1 + Z) = \psi(Z) + Z^{-1}, \qquad 
 \psi(1) = -\mathcal{C}, \qquad 
 \mathcal{C} \simeq 0.577,
 \label{eq:digamma}    \tag{A13}
\end{equation}
where we used ${\mathcal {C}}$ to denote the 
Euler-Mascheroni constant. 

Next, the computation of volume integrals requires a representation
of the hypergeometric functions which is convergent for both
small and large values of $Z$. In this case, the following formula 
is useful~\cite{NIST:DLMF}:
\begin{equation}
 {}_2F_1(a, b; c; Z) = (1 - Z)^{-a} {}_2F_1\left(a, c-b; c; \frac{Z}{Z - 1}\right),
 \label{eq:hyp_smallz}\tag{A14}
\end{equation}
taking into account that $Z = -\zeta_j < 0$ in the applications 
relevant to this paper.

In the computation of the volume integral $V^{\rm SC}_{\beta_0, \Omega}$
of the SC in Section~\ref{sec:SCPC}, the following identity
is useful~\cite{NIST:DLMF}:
\begin{equation}
{}_2F_1\left(a, \frac{1}{2}+a; 1 + 2a; Z\right) = \left(\frac{1}{2} +\frac{1}{2}\sqrt{1 - Z}\right)^{-2a}.
\label{eq:hyp_an}\tag{A15}
\end{equation}
Furthermore, the computation of $V^{E+P}_{\beta_0, \Omega}$ 
in Section~\ref{sec:SET} requires the identities 
\end{paracol}
\nointerlineskip
\begin{align}
 {}_2F_1\left(1 + k, \frac{1}{2} + k; 1+2k; \frac{\alpha}{1+\alpha}\right) =& 4^k \frac{(1 + \alpha)^{\frac{1}{2} + k}}{(1 + \sqrt{1 + \alpha})^{2k}},\nonumber\\
 {}_2F_1\left(2 + k, \frac{1}{2} + k; 1+2k; \frac{\alpha}{1+\alpha}\right) =& \frac{4^k}{2(1+k)} \frac{(1 + \alpha)^{1 + k}}{(1 + \sqrt{1 + \alpha})^{2k}}\left(\frac{2+\alpha}{\sqrt{1 +\alpha}} + 2k\right),\nonumber\\
 {}_2F_1\left(3 + k, \frac{1}{2} + k; 1+2k; \frac{\alpha}{1+\alpha}\right) =& \frac{4^k}{4(1+k)(2+k)} \frac{(1 + \alpha)^{\frac{3}{2} + k}}{(1 + \sqrt{1 + \alpha})^{2k}}
 \nonumber \\ &  \qquad \qquad \qquad \times \left[\frac{8 + 8\alpha + 3\alpha^2}{1+\alpha} + \frac{6k(2+\alpha)}{\sqrt{1 + \alpha}} + 4k^2 \right],
 \label{eq:hyp_an_SET}\tag{A16}
\end{align}
\begin{paracol}{2}
\switchcolumn
\noindent which can be derived from Equation~\eqref{eq:hyp_an} by applying the 
the following recurrence relation~\cite{NIST:DLMF}:
\begin{equation}
 \left(Z \frac{d}{dZ} Z\right)^n [Z^{a-1}{}_2F_1(a,b;c;Z)] = (a)_n Z^{a+n-1} {}_2F_1(a+n,b;c;Z).\tag{A17}
\end{equation}

In the RKT approach presented in Section~\ref{sec:adS:RKT}, 
the following small argument expansion of the modified Bessel
functions $K_n(Z)$ of integer argument $n$ is useful~\cite{NIST:DLMF}:
\begin{multline}
 K_n(Z) = \frac{1}{2} \left(\frac{Z}{2}\right)^{-n} \sum_{{\mathsf {k}} = 0}^{n - 1} \frac{(n -{\mathsf {k}} -1)!}{{\mathsf {k}}!}\left(-\frac{Z^2}{4}\right)^{\mathsf {k}} + (-1)^{n+ 1} I_n(Z) \ln\frac{Z}{2} \\
 + \frac{1}{2}\left(-\frac{Z}{2}\right)^n \sum_{{\mathsf {k}} = 0}^\infty [\psi({\mathsf {k}} + 1) + \psi(n + {\mathsf {k}} + 1)] 
 \frac{(Z^2 / 4)^{\mathsf {k}}}{{\mathsf {k}}! (n + {\mathsf {k}})!},
 \label{eq:Kn_smallZ}\tag{A18}
\end{multline}
where the modified Bessel function of the first kind $I_\nu(Z)$ is defined in terms of the usual Gamma function $\Gamma (Z)$ via
\begin{equation}
 I_\nu(Z) = \left(\frac{Z}{2}\right)^\nu \sum_{{\mathsf{k}} = 0}^\infty \frac{(Z^2 / 4)^{\mathsf {k}}}{{\mathsf {k}}! \Gamma(\nu + {\mathsf {k}} + 1)},
 \label{eq:Inu_def}\tag{A19}
\end{equation}
while the digamma function $\psi(Z)$ is introduced in Equation~\eqref{eq:digamma}.

\end{paracol}

\reftitle{References}

\end{document}